\documentclass[twocolumn]{aastex63}

\usepackage{graphicx}
\usepackage{wrapfig}
\usepackage{natbib}
\usepackage{color}
\graphicspath{{./figures/}} 
\usepackage{mathtools}
\usepackage{epstopdf}
\usepackage[autostyle]{csquotes}
\usepackage{hyperref} 
\def    \apjl  		{\rm {ApJL}}

\def    \apjl  		{\rm {ApJL}}

\def	\cm		{\,{\rm {cm}}}
\def	\K		{\,{\rm K}}
\def	\g		{\,{\rm {g}}}
\def	\mum	{\,{\mu \rm{m}}}
\def	\erg		{\,{\rm {erg}}}
\def	\s		{\,{\rm {s}}}
\def	\H		{\,{\rm {H}}}

\def \bea {\begin{eqnarray}}
\def \ena {\end{eqnarray}}                  

\begin{document}
\shorttitle{ }
\title{On planet formation around supermassive black holes and the grain disruption barriers by radiative torques}

\author{Nguyen Chau Giang}
\affil{Korea Astronomy and Space Science Institute, Daejeon 34055, Republic of Korea; \href{mailto:thiemhoang@kasi.re.kr}{thiemhoang@kasi.re.kr}}
\affil{Korea University of Science and Technology, Daejeon 34113, Republic of Korea}

\author{Thiem Hoang}
\affil{Korea Astronomy and Space Science Institute, Daejeon 34055, Republic of Korea; \href{mailto:thiemhoang@kasi.re.kr}{thiemhoang@kasi.re.kr}}
\affil{Korea University of Science and Technology, Daejeon 34113, Republic of Korea}

\author{Le Ngoc Tram}
\affil{Max Planck Institute for Radio Astronomy, Auf dem Hügel 69, 53121 Bonn, Germany}

\author{Nguyen Duc Dieu}
\affil{Department of Physics, International University, Ho Chi Minh City, Vietnam}
\affil{Vietnam National University, Ho Chi Minh City, Vietnam}

\author{Pham Ngoc Diep}
\affil{Vietnam National Space Center, Vietnam Academy of Science and Technology, 18 Hoang Quoc Viet, Hanoi, Vietnam}

\author{Nguyen Thi Phuong}
\affil{Korea Astronomy and Space Science Institute, Daejeon 34055, Republic of Korea; \href{mailto:thiemhoang@kasi.re.kr}{thiemhoang@kasi.re.kr}}
\affil{Vietnam National Space Center, Vietnam Academy of Science and Technology, 18 Hoang Quoc Viet, Hanoi, Vietnam}

\author{Bui Van Tuan}
\affil{University of Science and Technology of Hanoi, Vietnam Academy of Science and Technology, 18 Hoang Quoc Viet, Hanoi, Vietnam}

\author{Bao Truong}
\affil{Department of Physics, International University, Ho Chi Minh City, Vietnam}
\affil{Vietnam National University, Ho Chi Minh City, Vietnam}

\begin{abstract}

It has recently been suggested that planets can form by dust coagulation in the torus of active galactic nuclei (AGN) with low luminosity of $L_{\rm bol}\lesssim 10^{42}\erg\s^{-1}$, constituting a new class of exoplanets orbiting the supermassive black hole called \textit{blanets}. However, large dust grains in the AGN torus may be rotationally disrupted by the Radiative Torque Disruption (RATD) mechanism due to AGN radiation feedback, which would prevent the blanet formation. To test this scenario, we adopt the simple smooth and clumpy dust/gas distribution inside the torus region to study the effect of RATD on the evolution of composite dust grains in the midplane of the torus.  We found that grain growth and then blanet formation are possible in the smooth torus model. However, in the clumpy torus model, grain growth will be strongly constrained by RATD, assuming the gas density distribution as adopted in Wada et al. We also found that icy grain mantles inside clumps are quickly detached from the grain core by rotational desorption, reducing the sticking coefficient between icy grains and coagulation efficiency. The grain rotational disruption and ice desorption occur on timescales much shorter than the growth time up to a factor of $\sim 10^{4}$, which are the new barriers that grain growth must overcome to form blanets. Further studies with more realistic AGN models are required to better constrain the effect of RATD on grain growth and blanet formation hypothesis around low luminosity AGN.
  
\end{abstract}
\keywords{active galactic nuclei, dust, planets}

\section{Introduction}\label{sec:intro}
Planets are thought to form from dust grains in the protoplanetary disk (PPD) around young stars. Although the precise mechanism of planet formation remains elusive, one of the leading theories is {\it core-accretion} which is induced by dust coagulation and gravitational instability (\citealt{Gold_1973}, \citealt{Kobuko}). According to the core-accretion theory, small dust grains covered by ice mantles collide and stick together to form the large dust aggregates, followed by the formation of planetesimals of km-size (\citealt{Kataoka_2012}, \citealt{Okuzumi}). Subsequently, gravity acts on planetesimals to form planets (see \citealt{Chiang:2010} for a review). Recently, \cite{Wada2019,Wada2021} studied the possibility of planet formation in the dusty torus around supermassive black holes (SMBHs) and suggested that planets can be formed by coagulation of icy grain mantles beyond the snow line, provided that the luminosity of Active Galactic Nuclei (AGN) is relatively low of $L_{\rm bol} \sim 10^{42}\erg ~\rm s^{-1}$. Such a proposed scenario is expected to present a new class of exoplanets orbiting SMBHs, termed as \textit{blanets}.

While dust grains in the midplane of PPDs are significantly shielded from the stellar radiation due to the high gas density, dust grains in the torus around SMBHs are subject to intense radiation feedback from AGN. Thus, dust grains can be spun up to extremely fast rotation by RAdiative Torques (RATs) such that the centrifugal force can disrupt micron-sized grains to smaller fragments (\citealt{Hoang_2019b}). This Radiative Torque Disruption (RATD) mechanism is an essential dynamical constraint for the grain size distribution in the interstellar medium (\citealt{Hoang_2019b}), which would be a critical barrier for blanet formation around SMBHs because large aggregates have a lower tensile strength (\citealt{Tatsuuma:2019}; \citealt{Kimura:2020}) and will be disrupted more easily than small ones.

The effect of RATD on grain growth and planet formation in protoplanetary disks (PPDs) is studied in detail by \cite{Tung_Hoang}. The authors found that the RATD effect is inefficient in the disk interior thanks to the shielded stellar radiation and very high gas density of $n_{\rm H}> 10^{10}\cm^{-3}$. Compared to PPDs, the environment condition for blanet formation around SMBHs is radically different, with a lower gas density of $n_{\rm H}\sim 10^{4}-10^{6}\cm^{-3}$ at $r = 1$ pc in the midplane of torus (\citealt{Wada_16,Wada2019,Wada2021}) and an intense radiation field of $L_{\rm bol} = 10^{42}\erg~\rm s^{-1}$. Recently, \cite{Giang_2021} studied the influence of RATD on composite grains around high AGN luminosity of $L_{\rm bol} = 10^{46}\erg~\rm s^{-1}$. They found that large grains of $a\geq 0.1\mum$ are significantly disrupted to smaller sizes up to distance $r \sim 100$ pc in the polar cone and $r \sim 10$ pc in the midplane of the torus. Therefore, it is important to study the strength of RATD around low-luminosity AGN to understand whether RATD suppresses grain growth via coagulation or not.

On the other hand, \cite{Wada2019} studied grain growth beyond the snow line of the torus where ice can form on the surface of the solid grain core. The presence of ice mantles increases the sticking collision between icy monomers and reduces the fragmentation due to the dust collision (\citealt{Chokshi_93}, \citealt{Gundlach_2011}). Large icy dust aggregates with highly porous structures thus can grow quickly and overcome the \enquote{radial drift barrier} to form the planetesimals of $km-$ size (\citealt{Okuzumi}). However, \cite{Hoang_Tram} showed that the fast rotating icy grain mantles by RATs could cause the separation between two components following the fragmentation of detaching ice mantles and then the quick sublimation of icy fragments to the gas phase. This mechanism is named rotational desorption. Considering this effect on PPDs, \cite{Tung_Hoang} found the significant removal of micron-sized icy grain mantle beyond the \enquote{original} snow line on the dense midplane region. The formation of thick ice mantles is more strongly suppressed due to efficient rotational desorption, that icy dust with size $a \sim 0.5\mum$ only can be formed at the outer boundary of the disc of $r \sim 300$ au. Therefore, we expect a similar effect on the evolution of icy grain mantles in the midplane of low-luminosity AGN and rotational desorption may be another barrier that prevents the blanet formation. 

To model the effect of rotational disruption and desorption around low AGN luminosity, one needs to know in detail the morphology and the distribution of gas and dust grains within the torus region. However, this problem has not yet been well answered now. Numerical studies of the circumnuclear region showed that both the smooth torus model, in which dust and gas distributes smoothly in the flared or tapered geometry of the torus (\citealt{Fritz_06}, \citealt{Schart}), and the clumpy torus model, in which dust and gas concentrate into the dense clumps that distribute randomly inside the diffuse torus region (\citealt{Nenkova2002}, \citealt{Nenkova08b}, \citealt{Honig10}), generally can reproduce the observed properties of the spectral energy distribution (SED) in infrared (IR) range, the feature of the line emission/absorption, and also the broad X-ray spectrum observed from AGN (\citealt{Buchner_2015}). However, the high resolution observations toward the circumnuclear region tend to reveal the inhomogeneous distributions of gas and dust grains instead of the homogeneous or discrete gas/dust distribution as proposed in the smooth and clumpy torus model (\citealt{Shi_2006}, \citealt{Hicks_2009}, \citealt{Izumi_2018}). This picture is consistent with the multi-gas phase structure driven by the radiation-driven fountain model proposed by \cite{Wada_09} (\citealt{Schartmann_2014}, \citealt{Wada_16}). This model also can reproduce many photometric observations and spectroscopy observed from AGN (\citealt{Buchner}. However, since the main purpose of our paper is to give the first insight of the effect of rotational disruption and desorption on grain growth around low luminosity AGN, we adopt two simple models of the smooth torus model and the clumpy model and solve the radiative transfer only in one dimension. We first perform numerical calculations of the disruption size of dust grains with composite structures and the desorption size of icy grain mantles for the low-luminosity AGN with $L_{\rm bol} = 10^{42}\erg ~\rm s^{-1}$. Next, we will compare the timescale between grain growth via coagulation and rotational destruction and desorption to understand how spinning dust affects grain evolution.

The structure of the paper is as follows. The spectral energy distribution of AGN and the model of the torus are described in Section \ref{sec:AGN_model}. In Section \ref{sec:disruprtion}, we briefly describe the mechanism of RATD, the radiative transfer model, and numerical calculations of the grain disruption size by RATD for a smooth and a clumpy gas/dust density model in the torus. We then study the destruction of icy grain mantles beyond the snow line under the effect of rotational desorption in Section \ref{sec:desorption}. We next compare the disruption and desorption timescale with the growth time by dust coagulation in Section \ref{sec:timescale}. The discussion of the rotational effect on the blanet formation around SMBHs will be presented in Sections \ref{sec:discuss}, and a summary of our main findings is presented in Section  \ref{sec:sum}.

\section{Model of AGN radiation field and dusty torus}\label{sec:AGN_model}
\subsection{Spectral energy distribution}
We adopt the spectral energy distribution of an unobscured AGN from the study of \cite{Nenkova08} and  \cite{Stalevski_2012}, which follows:
\bea 
\small
\lambda L_{\lambda}= A \left\{
\begin{array}{l l}    
    0.158 ~ (\lambda/1\mum)^{1.2} ~  {\rm if~ } 0.001 \leq \lambda  \leq 0.01 \mum\\
    6.3\times10^{-4} ~  ~~~~~~~~{\rm ~ if~ } 0.01  < \lambda \leq 0.1\mum \\
    2\times10^{-4} ~ (\lambda/1\mum)^{-0.5}   {\rm ~ if~ } 0.1  < \lambda \leq 5\mum\\
    0.011 ~ (\lambda/1\mum)^{-3} ~ ~ ~ ~ ~  ~ ~ {\rm ~ if~ } 5  < \lambda \leq 50 \mum
\end{array} \right \},
\label{eq:u_AGN}
\ena

where $A$ is a normalization constant determined by the bolometric luminosity $L_{\rm bol}$ of AGN. For convenience, we consider that AGN is a time-invariant source.
 
In order to model the effect of RATD, we only focus on the main radiation spectrum emitting from the accretion disk from $\lambda_{\rm min} = 0.1\mum$ to $\lambda_{\rm max} = 20\mum$ at which the lower limit is determined by the Lyman absorption of neutral hydrogen atoms. To describe the strength of the radiation field, we use the dimensionless parameter $U = u_{\rm rad}/u_{\rm ISRF}$ where $u_{\rm rad}$ is the radiation energy density of the local radiation field and $u_{\rm ISRF}=8.64\times10^{-13} \erg \cm^{-3}$ is the energy density of the average interstellar radiation field (ISRF) in the solar neighborhood \citep{Mathis83}. 

\subsection{Physical model of the AGN torus}
\subsubsection{Smooth torus model}
\begin{figure*}
\centering
        \includegraphics[height=8cm,width=18cm]{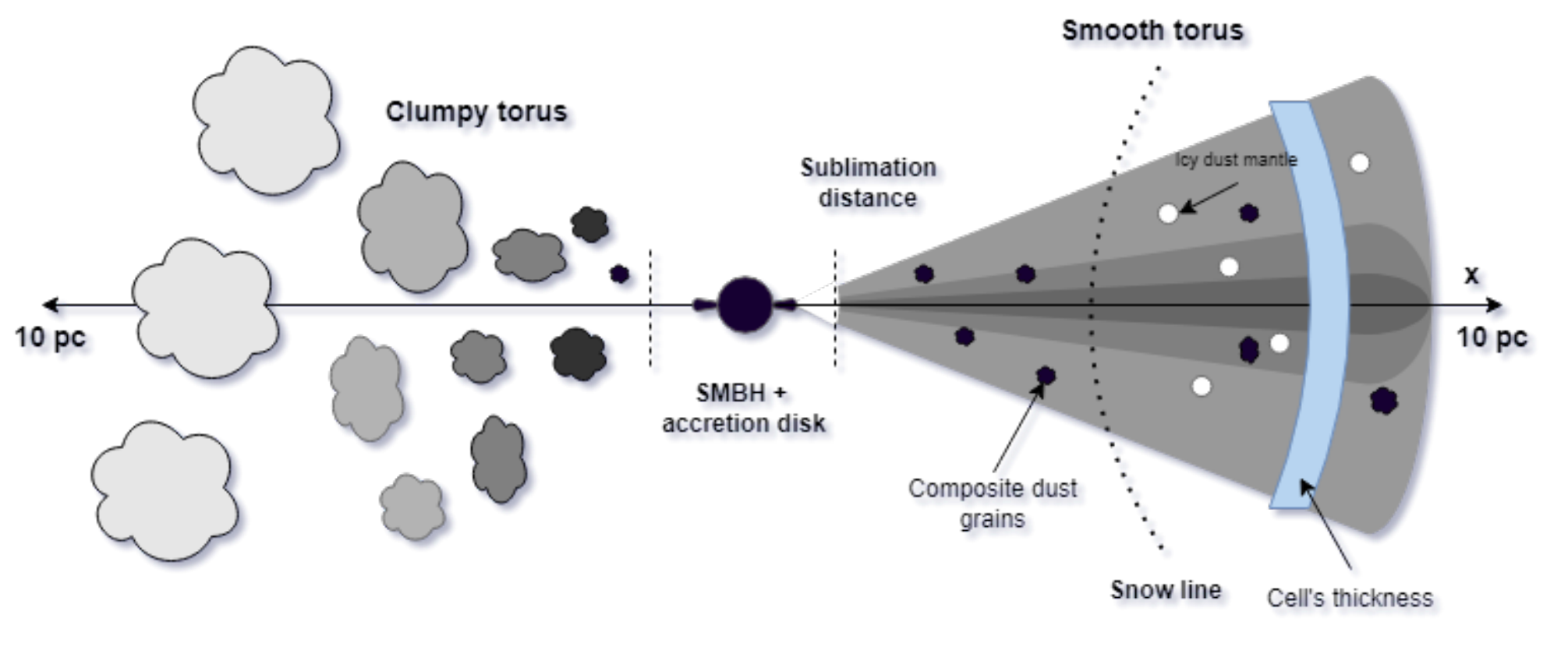}
     \caption{Illustration of an AGN torus in two dimensions with a smooth gas/dust density distribution (e.g., smooth torus model, right part), and a clumpy gas/dust distribution (clumpy torus model, left part), around a central SMBH (black circle), and an accretion disc. Composite dust grains (small black circles) and icy grain mantles (small white circles) are present beyond the sublimation front determined by dust temperature $T_{\rm sub} = 1200$ K and the snow line defined by the ice temperature $170$ K, respectively. 
     Small, dense clumps (dark clumps) locate near the central region, while large, dilute clumps (light clumps) are located far from the center. To model the effect of dust reddening, we divide the torus (in the case of the smooth torus model) and clumps (in the case of the clumpy torus model) into the equally small spherical shell of thickness $d_{\rm cell}$ (blue cell).}
          \label{fig:AGN_model}
\end{figure*}

The right part of Figure \ref{fig:AGN_model} shows the two-dimensional illustration of an AGN with the smooth distribution of dust and gas in the torus (hereafter smooth torus model). In the dense environment as the midplane of torus region, dust and gas are likely in thermal equilibrium, so that the gas temperature can be inferred from the dust temperature (i.e., $T_{\rm gas} \approx T_{\rm dust}$) determined by the balance between grain heating by UV-optical radiation from the central AGN and grain cooling by IR re-emission. The equilibrium temperature of large grains of size $a \geq 0.01\mum$ can be approximately given by \citep{Draine11}:
\bea
T_{\rm dust}(a) \simeq 16.4U^{1/6} \Bigg(\frac{a}{0.1\mum}\Bigg)^{-1/15} \K.
\label{eq:Tgas_torus}
\ena

Dust grains near the center region exposed to higher $U$ have higher temperatures and can be sublimated when its temperature exceeds the sublimation threshold, $T_{\rm sub}$. The distance where dust starts to survive from the thermal sublimation is called the sublimation distance and denoted as $r_{\rm sub}$. Here, we assume that carbonaceous grains have the same sublimation temperature of $T_{\rm sub} =1200$ K as silicate grains. Therefore, the sublimation distance $r_{\rm sub}$ of grain size $a$ can be found by setting $T_{\rm dust}(a) = T_{\rm sub}$,  yielding (\citealt{Draine11}):
\bea 
r_{\rm sub}  &\simeq & 2\times10^{-3} ~\rm pc ~ \Bigg(\frac{L_{\rm bol}}{10^{42}\erg ~\rm s^{-1}}\Bigg)^{1/2} \Bigg(\frac{T_{\rm sub}}{1200 ~ \rm K}\Bigg)^{-3} \nonumber \\
 &~~& \times \Bigg(\frac{a}{0.01 \mum}\Bigg)^{-1/5}.
\label{eq:r_sub}
\ena
We consider the sublimation distance of small grains of $a = 0.01\mu m$ is the inner boundary of dusty torus region, which yields $r_{\rm sub} = 0.002$ pc for $L_{\rm bol} = 10^{42}\erg ~ \rm s^{-1}$, assuming $T_{\rm sub}=1200\K$.
 
We assume the simple flared disk geometry for the torus from \cite{Fritz_06} and the gas density profile from \cite{Wada2021} which is given by

\bea
n_{\rm H} = \left\{ 
\begin{array}{l l}
    n_{\rm H, 0.05} ~~~~~~~~~~~~{\rm ~ for~ r} < 0.05~ \rm pc \\ 
    n_{\rm H, 1} (\rm r/\rm pc)^{-3/2} ~~ {\rm for ~ r} \geq 0.05 ~\rm pc    
\end{array}\right\},
\label{eq:n_H_smooth}
\ena
where $n_{\rm H, 0.05}$ and $n_{\rm H, 1}$ are the gas density at $r = 0.05$ pc and $r = 1$ pc, respectively.  

In our study, we only focus on studying the RATD effect on dust in the midplane of the torus where most of dust is present and the blanet formation is suggested to occur (\citealt{Wada2019}). 

\subsubsection{Clumpy torus model}
Dust and gas are also suggested to concentrate into dense clumps within the diffuse circumnuclear region (left part of Figure \ref{fig:AGN_model}). This clumpy torus model was first proposed by \cite{Krolik_1988} as a way to reduce dust destruction by intense UV-optical AGN radiation. It is then widely studied (e.g., \citealt{Nenkova2002,Nenkova08b}, \citealt{Netzer15}, \citealt{Dullemond}) and supported by observations (e.g., \citealt{Alonso}). For this configuration, small and dense clumps (dark color) are likely located near the AGN center, while large and dilute clumps (light color) are located in the middle and outer boundary of the torus (see also three-dimension (3D) structure of the clumpy torus model in \citealt{Schart}). The variation of the clump size $R_{\rm cl}$ with distance can be assumed as (\citealt{Honig10}):
\bea 
R_{\rm cl} = \Bigg(\frac{R_{\rm cl,0}}{r_{\rm sub}}\Bigg) \Bigg(\frac{r}{r_{\rm sub}}\Bigg)^{\beta_{\rm cl}} r_{\rm sub},
\label{eq:R_clump}
\ena
where $R_{\rm cl, 0}$ is the clump radius of at the sublimation front $r_{\rm sub}$, and $\beta_{\rm cl}$ is the power-law index of the distribution. We assume $R_{\rm cl, 0} = 0.01 ~ \rm r_{\rm sub}$. To simplify the model, we consider that the near side of the clump (closer to the AGN) is directly illuminated by AGN radiation and only consider the radiation attenuation inside the clump.

We adopt the same gas density profile as in the smooth torus model given by Equation (\ref{eq:n_H_smooth}) for comparison. The gas temperature is taken from Equation (\ref{eq:Tgas_torus}) due to the thermal equilibrium between gas and dust inside the dense clump.

\section{Rotational Disruption of Composite grains} \label{sec:disruprtion}
We first describe the main points of the RATD mechanism in Section \ref{sec:RATD} and the radiative transfer modeling in Section \ref{sec:radiative_transfer}. The numerical calculations of the disruption size of composite grains for the smooth and clumpy torus models are described in Section \ref{sec:a_disr}.

\subsection{The RATD Mechanism}\label{sec:RATD}
Following the principle of RATD (\citealt{Hoang_2019b}), a fast rotating irregular dust grain spun up by radiative torques (\citealt{Draine96}, \citealt{Laza07}) can be destroyed to smaller pieces if the induced tensile stress exceeds the maximum tensile strength of the dust grain. Assuming the constant AGN radiation field, the time-evolution of the grain angular velocity driven by RATs is given by
\bea 
\omega(t) = \omega_{\rm RAT}(1-e^{-t/\tau_{\rm damp}}),
\label{eq:omega(t)}
\ena
where $\tau_{\rm damp}$ is the gas damping timescale, which is:
\bea 
\tau_{\rm damp} = \frac{\tau_{\rm gas}}{1 + \rm FIR}
\ena
with $\tau_{\rm gas}$ is the typical timescale for the rotational damping by gas collisions, and $\rm FIR$ is the coefficient presenting for the rotational damping caused by re-IR emission (see \citealt{Hoang_2019b} for details). $\omega_{\rm RAT}$ is the maximum angular velocity that the dust grain can reach after a long time. For an irregular grain of effective size $a$ which is defined as the radius of a spherical grain with the same volume, the formula for $\omega_{\rm RAT}$ is described by (\citealt{Hoang_2019b} and \citealt{Giang_2021})
\bea 
\omega_{\rm RAT} ~&\approx&~ 1.2\times10^{11} \gamma_{\rm rad} a_{-5}^{0.7} \bar{\lambda}_{0.5}^{-1.7} \nonumber \\ &~~~~& \times \Bigg(\frac{U_{6}}{1.2 n_{5}T_{2}^{1/2}}\Bigg) \Bigg(\frac{1}{1+\rm F_{\rm IR}}\Bigg) ~\rm rad~ s^{-1},
\label{eq:omega_RAT_small}
\ena 
for grain of size $a \leq \bar{\lambda}/1.8$, and
\bea 
\omega_{\rm RAT} ~&\approx&~ 2.1\times10^{12} \gamma_{\rm rad} a_{-5}^{-2} \bar{\lambda}_{0.5} \nonumber \\
&~~~& \times \Bigg(\frac{U_{6}}{1.2 n_{5}T_{2}^{1/2}}\Bigg) \Bigg(\frac{1}{1+F_{\rm IR}}\Bigg)~\rm rad ~s^{-1},
\label{eq:omega_RAT_large}
\ena
for grain of size $a > \bar{\lambda}/1.8$. In these equations, $a_{\rm -5} = a/(10^{-5}\cm)$, $\bar{\lambda}_{0.5} = \bar{\lambda}/(0.5\mu m)$ with $\bar{\lambda}$ the mean wavelength of the radiation spectrum from $\lambda = 0.1-20\mum$, $U_{6} = U/(10^{6})$ is the radiation strength, $n_{5} = n_{\rm H}/(10^{5}\cm^{-3})$, and $T_{2} = T_{\rm gas}/(100~\rm K)$ is the gas temperature given by Equation (\ref{eq:Tgas_torus}). Also, $\gamma_{\rm rad}$ is the anisotropic degree of the radiation field ($0 \leq \gamma_{\rm rad} \leq 1$), and we adopt $\gamma_{\rm rad} = 1$ for the unidirectional radiation field of AGN. The  last term of $F_{\rm IR}$ is the dimensionless coefficient of rotational damping due to thermal dust emission (see \citealt{Hoang_2019b} for details). The change of $\omega_{\rm RAT}$ at $a =a_{\rm trans}= \bar{\lambda}/1.8$ is due to the change of RAT efficiency with the mean wavelength $\bar{\lambda}$ and the grain size $a$ (see \citealt{Hoang08}, \citealt{Hoang14}, \citealt{Hoang_2019b}, \citealt{Giang_2021} for details). 

The rotating dust grain will be disrupted to smaller sizes if its tensile stress, $S=\rho a^{2}\omega^{2}/4$, exceeds the maximum tensile strength of the grain material, $S_{\rm max}$.  The value of $S_{\rm max}$ depends on the grain material, grain structure, and grain size, but have not yet well constrained. Ideal material can have high maximum tensile strength, such as diamond with $S_{\rm max} \sim 10^{11}\erg\cm^{-3}$ (\citealt{Draine79}, \citealt{Burke_1974}), or polycrystalline bulk material with $S_{\rm max} \sim 10^{9}-10^{10}\erg\cm^{-3}$ (\citealt{Hoang_2019b}), while composite dust grains which are formed via sticking collisions between sub-micron grains in cold and dense environments such as molecular clouds and protoplanetary disks (\citealt{Tatsuuma:2019}, \citealt{Kimura:2020}) are expected to be porous with low maximum tensile strength of $S_{\rm max} \sim 10^{5}-10^{8}\erg\cm^{-3}$ (\citealt{Hoang_2019a}). 
 

The critical angular velocity for rotational disruption can be found by setting the centrifugal stress $S = S_{\rm max}$, which is given by:
\bea
\omega_{\rm disr} &=& \frac{2}{a}\Bigg(\frac{S_{\rm max}}{\rho}\Bigg)^{1/2} \nonumber \\
&=&  \frac{1.14\times10^{9}}{a_{-5}} S_{\rm max,8}^{1/2} \hat{\rho}^{-1/2}~\rm rad ~~ s^{-1},
\label{eq:omega_disr}
\ena
where $S_{\rm max,8} = S_{\rm max}/(10^{8}\erg \cm^{-3})$, and $\hat{\rho} = \rho/(3 ~\rm g \cm^{-3})$ with $\rho$ the grain mass density. The disruption threshold decreases with decreasing maximum tensile strengths and increasing grain sizes, implying the strong effect of RATD on large composite grains.

Comparing Equation (\ref{eq:omega_RAT_small}) with Equation (\ref{eq:omega_disr}) for grains of $a \leq a_{\rm trans}$, one can see that large grains are easier to be destroyed by RATD due to the increase of $\omega_{\rm RAT}$ and the decrease of $\omega_{\rm disr}$ with increasing grain sizes. The first intersection between $\omega_{\rm RAT}$ and $\omega_{\rm disr}$ determines the grain disruption size, denoted by $a_{\rm disr}$, above which dust will be disrupted by RATD. By comparing Equation (\ref{eq:omega_RAT_large}) with Equation (\ref{eq:omega_disr}) for grains of $a > a_{\rm trans}$, grains with larger size are hard to be destroyed due to the decrease of $\omega_{\rm RAT}$ with increasing grain sizes. Thus, the second intersection between $\omega_{\rm RAT}$ and $\omega_{\rm disr}$ determines the maximum size that dust still be affected by RATD, denoted by $a_{\rm disr,max}$. The range of disrupted grains of $a_{\rm disr} - a_{\rm disr,max}$ will be extended for stronger radiation field and weaker maximum tensile strength $S_{\rm max}$. 

\subsection{Radiative transfer model}\label{sec:radiative_transfer}
UV-optical radiation passing through the dusty torus will be attenuated due to scattering and absorption by dust grains. The spectral energy density at distance $r$ from the central source thus will be given by:
\bea 
u_{\rm \lambda} = u_{\rm \lambda,0} e^{-\tau_{\rm \lambda}},
\label{eq:u_redden}
\ena
where $u_{\rm \lambda,0} = L_{\rm \lambda,0}/ (4 \pi c r^{2})$ is the intrinsic energy density with $L_{\rm \lambda,0}$ the specific luminosity given by Equation (\ref{eq:u_AGN}), and $\tau_{\rm \lambda}$ is the optical depth at wavelength $\lambda$.

We adopt the popular interstellar dust model consisting of separate astronomical silicate and carbonaceous grains (see \citealt{Li_2001}, \citealt{Wein01}, \citealt{Drain07}) with the grain size spanning from the minimum size of $a_{\rm min} = 3.5~\rm \AA$ to the maximum size of $a_{\rm max} = 10\mum$.  We assume that dust grains initially follow the MRN distribution of $dn/da = n_{\rm H} C a ^{-3.5}$ (\citealt{Mathis:1977}) with $C$ the normalization constant determined by the dust-to-gas mass ratio $\eta$ (\citealt{Laor_93}). We consider the presence of micron-sized grains in the torus, thus, in order to keep the typical dust-to-gas mass ratio found in the interstellar medium $\eta = 0.01$, the normalization constant must be reduced by a factor of 3 from the values of \cite{Mathis:1977}, to $C_{\rm sil} = 1.16\times10^{-26}\cm^{2.5}$ and $C_{\rm carb} = 1.036\times10^{-26}\cm^{2.5}$, respectively.

Without RATD, AGN radiation is attenuated by all grains from $a_{\rm min}$ to $a_{\rm max}$. With RATD, grains of size from $a_{\rm disr}$ to $a_{\rm disr,max}$ are disrupted to smaller sizes, resulting in two separate populations of the grain size distribution. The first population includes grains of size from $a_{\rm min}$ to $a_{\rm disr}$, which are enhanced by the disruption of large grains by RATD. The second population includes large grains from $a_{\rm disr,max}$ to $a_{\rm max}$ which are not affected by RATD and still follow the MRN distribution. Due to the lack of information about the redistribution of disrupted grains, we assume that small grains of size between $a_{\rm min} - a_{\rm disr}$ still follow the power-law distribution with a steeper slope, denoted as $\epsilon$, implying the enhancement of very small grains. Assuming the constant values of $C_{\rm sil}$ and $C_{\rm carb}$, the new slope $\epsilon$ can be found from the mass conservation of dust from $a_{\rm min}$ to $a_{\rm disr,max}$ (see \citealt{Giang_2021} for details). Consequently, the optical depth at wavelength $\lambda$ produced by all dust grains from the sublimation front to distance $d$ from the AGN center is given by
\bea 
\tiny
\tau_{\rm \lambda}(d) &=& \sum_{j=\rm sil,carb} \int_{\rm 0}^{d} \int C_{\rm ext}^{j}(a,\lambda) \frac{dn^{j}}{da}(r) da  \nonumber\\
&=& \sum_{j =\rm sil,carb} \int_{0}^{d} \Bigg(\int_{a_{\rm min}}^{a_{\rm disr}(r)} C_{\rm ext}^{j}(a,\lambda) a^{\epsilon(r)} da \nonumber\\ 
&~~& + \int_{a_{\rm disr, max}(r)}^{a_{\rm max}} C_{\rm ext}^{j}(a,\lambda) a^{-3.5} da \Bigg) n_{\rm H}(r) C^{j} dr, ~~~ \label{eq:sigma_ext}
\ena
where $a_{\rm disr}(r)$, $a_{\rm disr,max}(r)$, $\epsilon(r)$, and $n_{\rm H}(r)$ are the grain disruption size, maximum grain disruption size, new power-index of the distribution of grain size of $a\leq a_{\rm disr}(r)$, and the gas density at distance $r$ given by Equation (\ref{eq:n_H_smooth}), respectively. Above, $C_{\rm ext}(a,\lambda) = Q_{\rm ext}(a,\lambda) \pi a^{2}$ is the extinction cross section of grain size $a$ with wavelength $\lambda$ with $Q_{\rm ext}(a,\lambda)$ the extinction efficiency. The values of $Q_{\rm ext}(a, \lambda)$ are adopted from \cite{Hoang13} for sub-micron sized grains and calculated from the public Discrete Dipole Scattering code (DDSCAT) (\citealt{Draine:1994}) for micron-sized grains (\citealt{Giang_2021}), assuming an oblate spheroidal shape with an axial ratio of 2. 

 \cite{Wada2019} suggested that micron-sized grains beyond the snow line can grow and form blanet in the midplane of the torus region around low luminousity AGN. Therefore, we focus to study the effect of RATD and rotational desorption in this region, and consider the radiative transfer of AGN radiation in one dimension, i.e., along x-direction (Figure \ref{fig:AGN_model}). \footnote{This choice neglects the angle-dependence of the UV-optical radiation emitting from the accretion disk around AGN.}

To numerically model the attenuation of AGN radiation by dust along x-direction, we divide the midplane of the flared disk (in case of the smooth torus model) and the clump (in case of the clumpy torus model) into $n$ thin cells of the same thickness $d_{\rm cell} = 0.01$ pc, respectively. The dust size distribution is considered uniform in each cell (see the thin blue cell in Figure \ref{fig:AGN_model}). The total optical depth produced by dust from the sublimation front to distance $d$ can be written as:
\bea 
\tau_{\rm \lambda,n} = \sum_{i = 1}^{n} \Delta \tau_{\rm \lambda,i},
\ena 
where $\Delta \tau_{\rm \lambda,i}$ is the optical depth produced by dust in the $i^{th}$ cell (by using Equation \ref{eq:sigma_ext} with $d$ replaced by $d_{\rm cell}$) and $i$ is the order of the cell. We denote $i = 1$ be the first cell at the sublimation front (in case of the smooth torus model) and the near side (closer to the AGN) of the clump (in case of the clumpy torus model) and $i = n$ be the last cell at distance $d$  on x-direction. 

\subsection{Grain disruption size}\label{sec:a_disr}
To study the effect of RATD on composite dust grains in the midplane of torus, we first solve the radiative transfer equation (Equation \ref{eq:u_redden}) to get the radiation strength $U$ with distances, then calculate the terminal angular velocity $\omega_{\rm RAT}$ of all grain sizes from $a_{\rm min}$ to $a_{\rm max}$ (Equations \ref{eq:omega_RAT_small} and \ref{eq:omega_RAT_large}). By comparing $\omega_{\rm RAT}$ with the critical angular velocity $\omega_{\rm disr}$ (Equation \ref{eq:omega_disr}), we can determine the range $a_{\rm disr}-a_{\rm disr,max}$ in which grains are destroyed by RATD. We consider the evolution of dust grains under the RATD effect for the typical case of luminosity of $L_{\rm bol} = 10^{42}\erg ~\rm s^{-1}$ in which blanet is suggested to form (\citealt{Wada2019,Wada2021}).

\subsubsection{Smooth torus model}\label{sec:adisr_smooth}

\begin{figure*}
        \includegraphics[width = 0.5\textwidth]{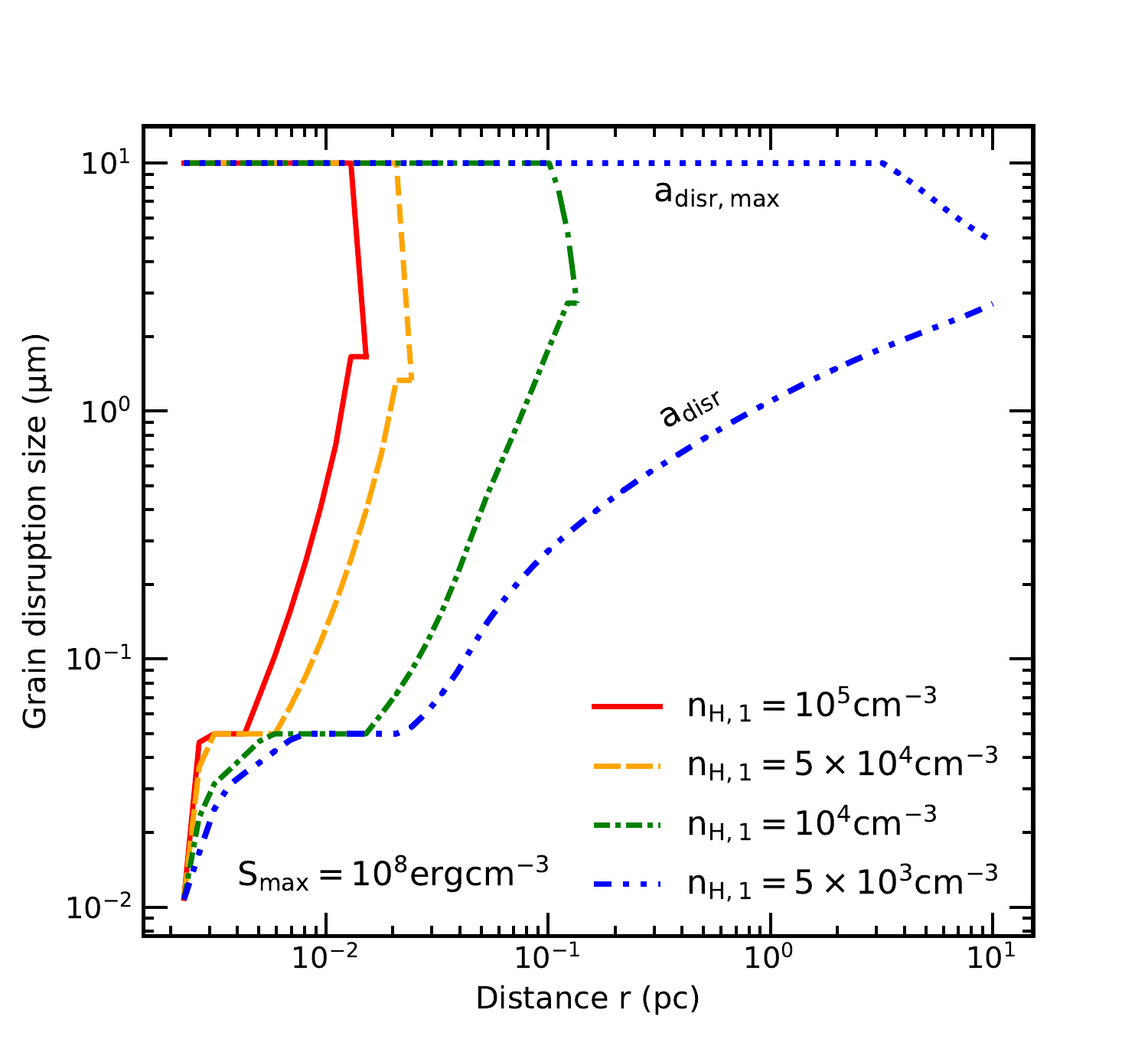}
        \includegraphics[width = 0.5\textwidth]{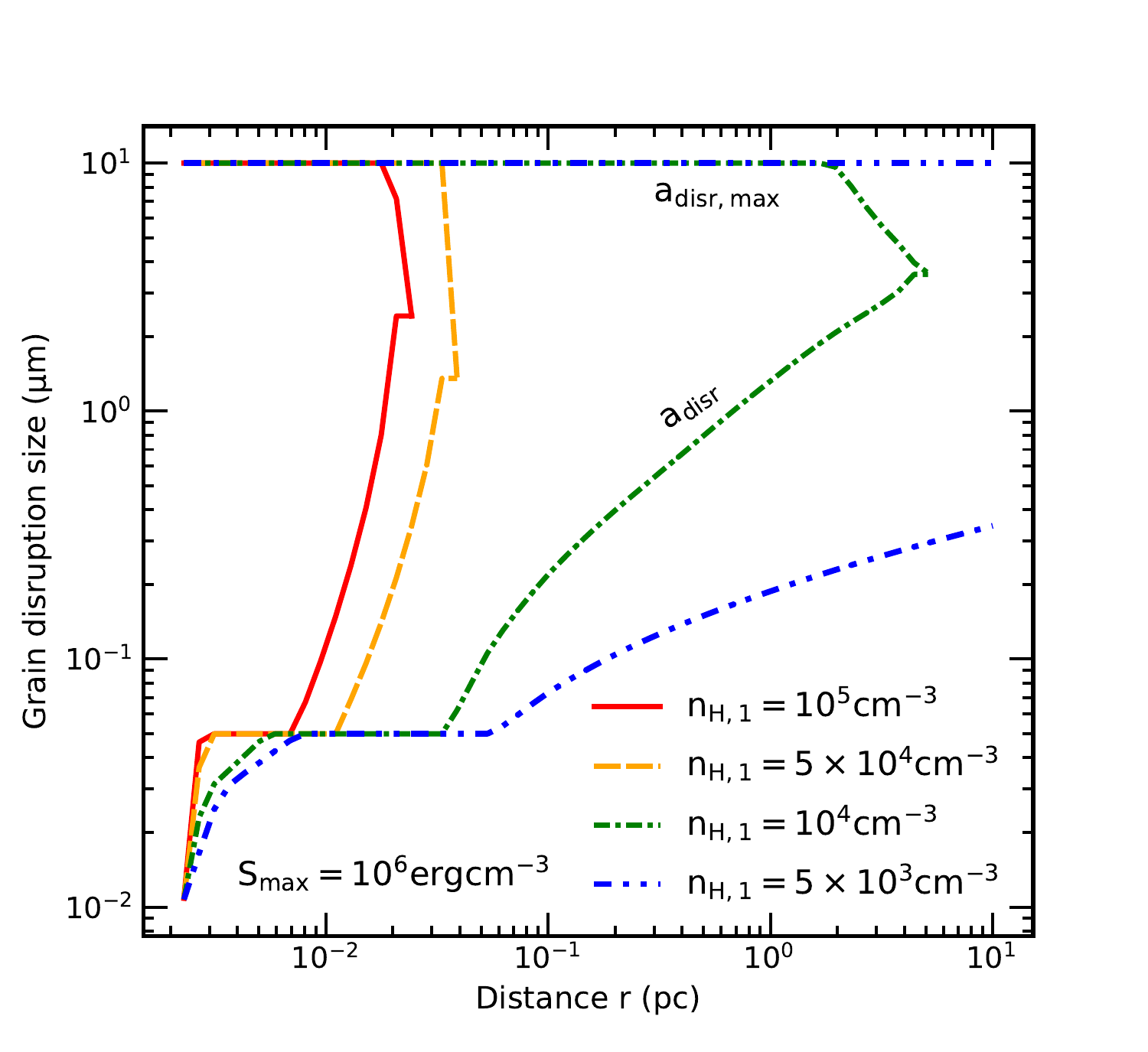}
     \caption{Dependence of the range of the disruption sizes of composite grains, $a_{\rm disr} - a_{\rm disr,max}$, on the distance to the AGN center for the smooth torus model, assuming different gas density profiles described by $n_{\rm H,1}$. Two values of $S_{\rm max} = 10^{8}\erg\cm^{-3}$ (typically for compact dust grains) and $S_{\rm max} = 10^{6}\erg\cm^{-3}$ (typically for porous dust grains) are considered in the left and right panels, respectively.} 
     \label{fig:a_disr_smooth}
\end{figure*}

\begin{figure*}
        \includegraphics[width = 0.5\textwidth]{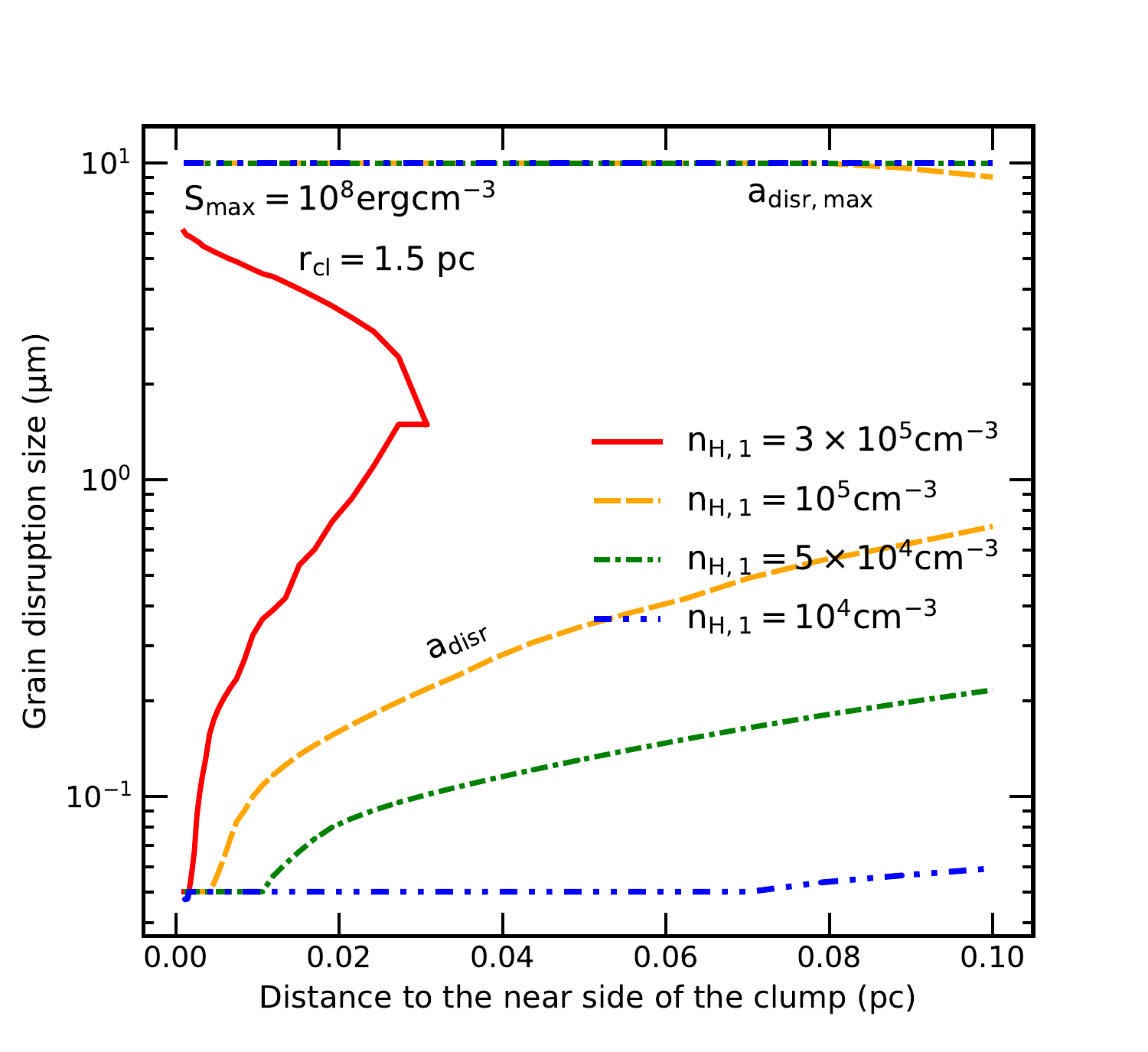}
        \includegraphics[width = 0.5\textwidth]{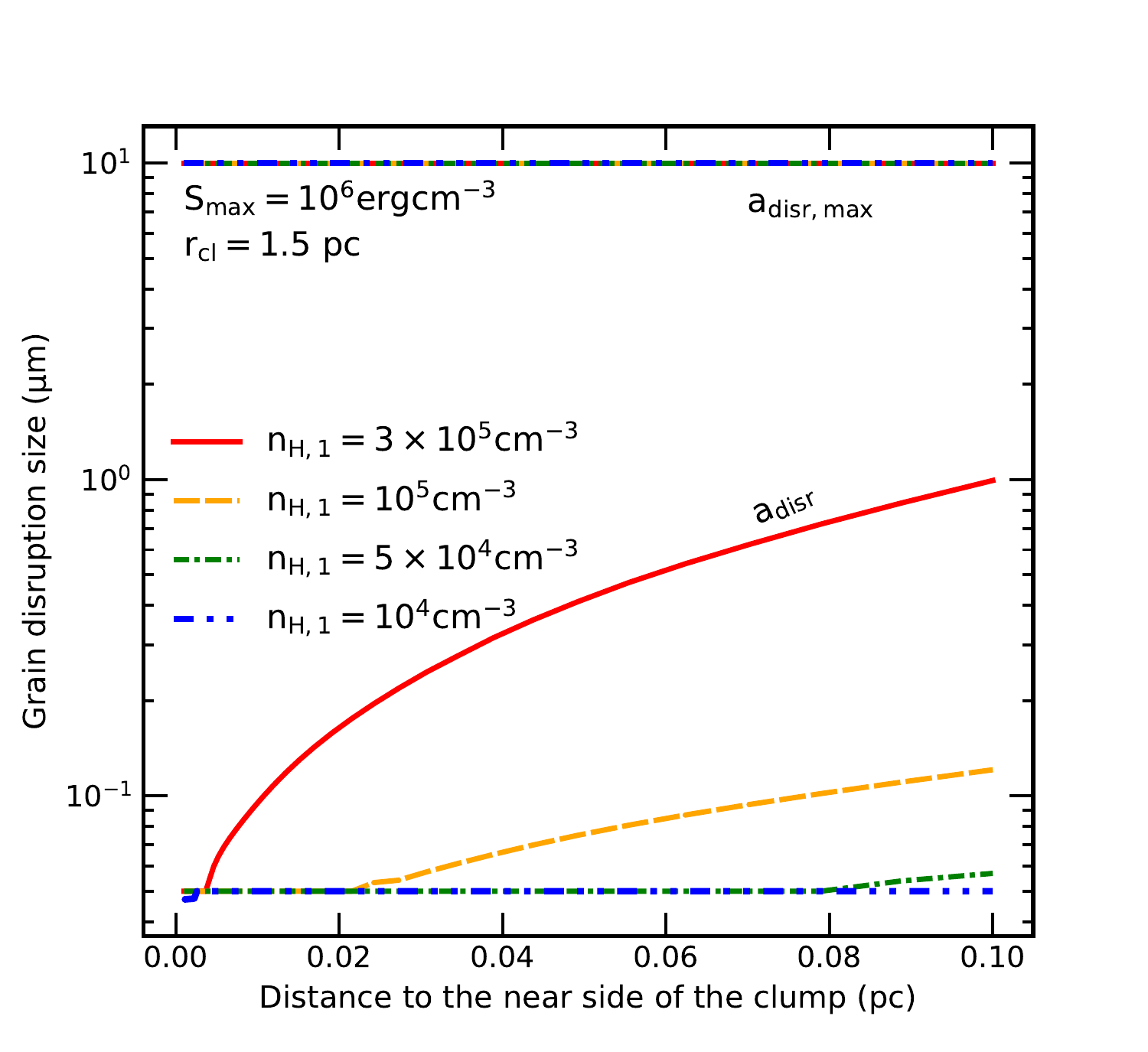}
     \caption{Similar to Figure \ref{fig:a_disr_smooth} but for composite grains inside the clumps of radius $R_{\rm cl} = 0.05$ pc at $r = 1.5$ pc. The RATD effect is more effective, as demonstrated by a broader range of the disruption sizes, due to the reduction of dust reddening (see the main text).}
     \label{fig:a_disr_clumpy}
\end{figure*}

\begin{figure*}[t]
\centering
        \includegraphics[width = 0.49\textwidth]{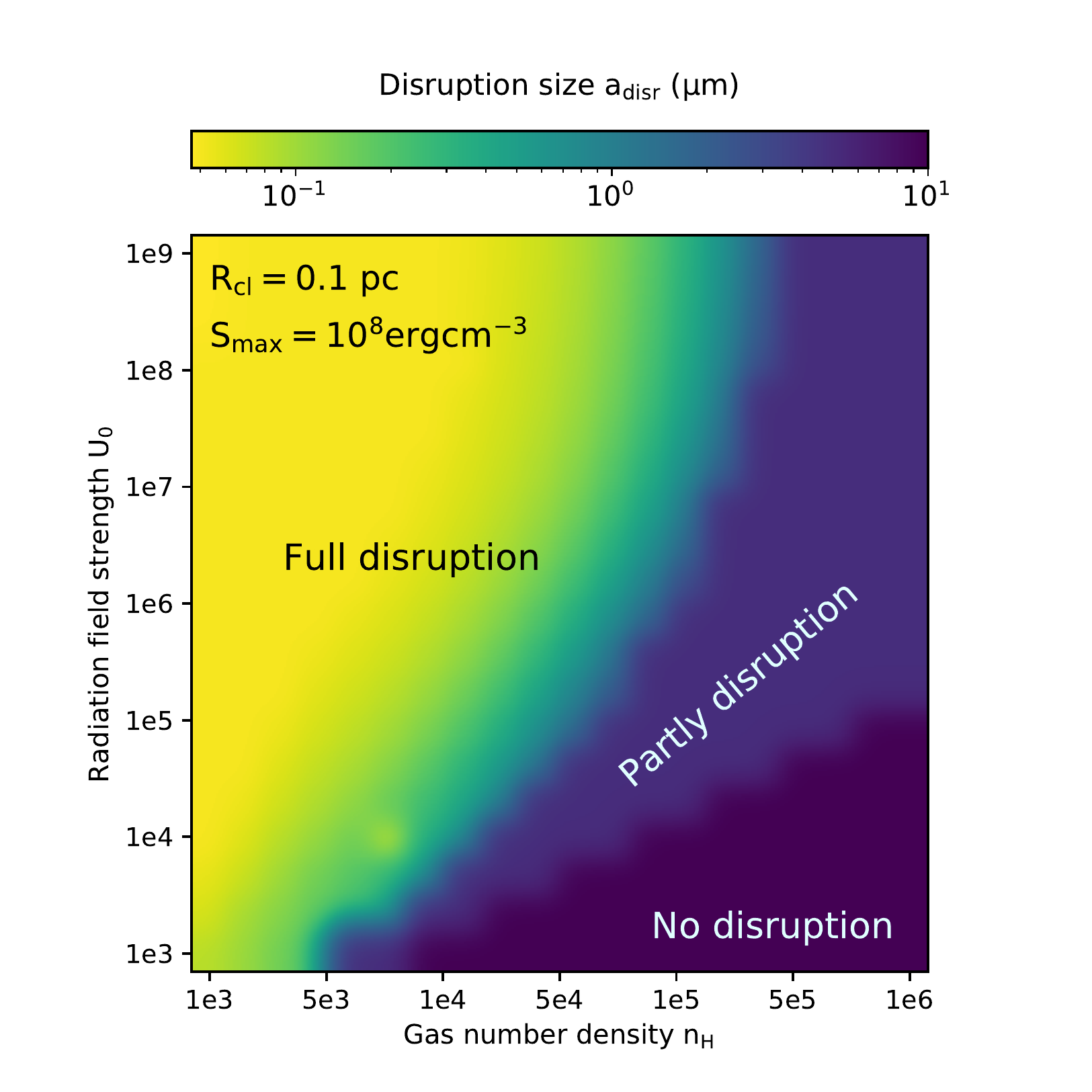}
        \includegraphics[width = 0.49\textwidth]{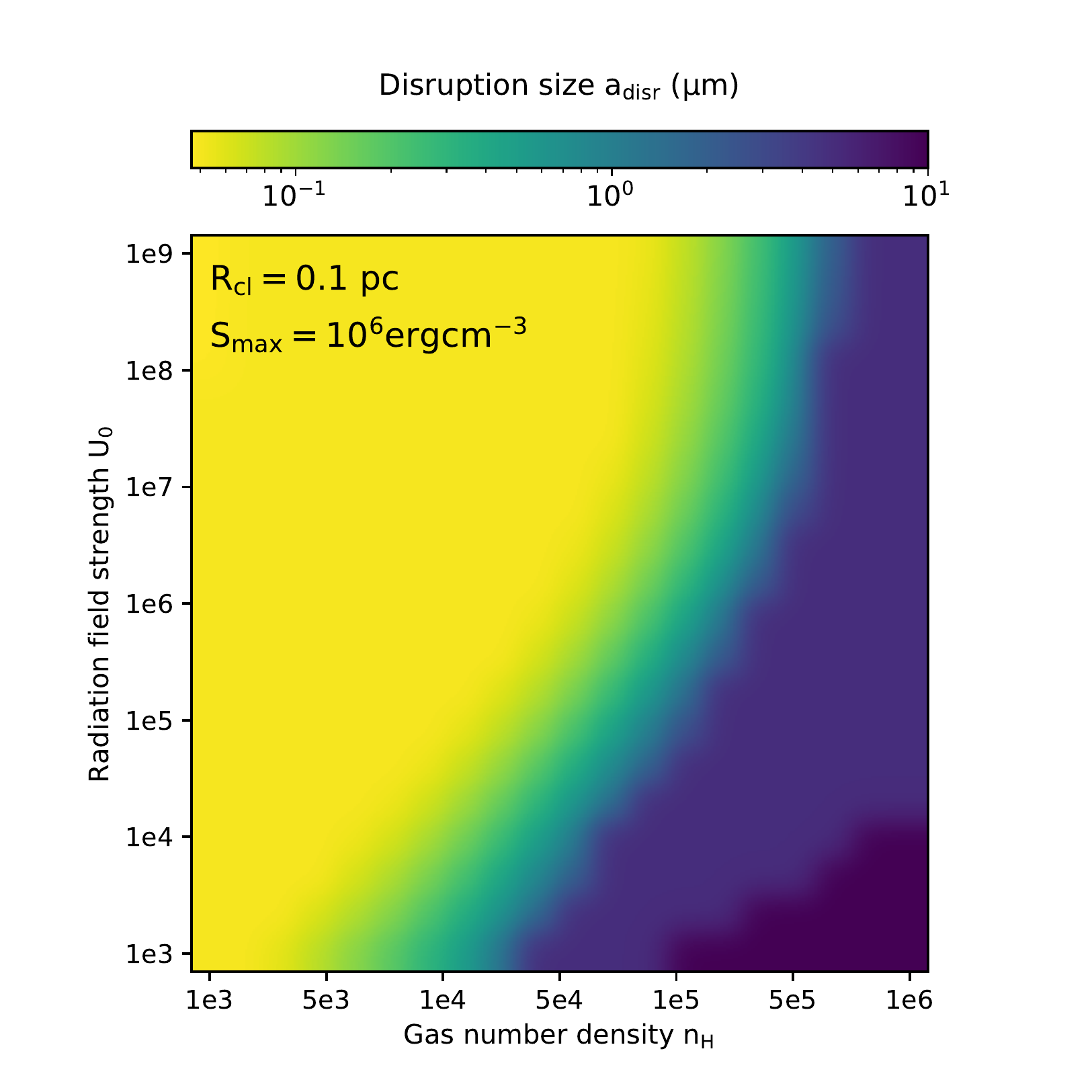}
     \caption{Variation of grain disruption size at the far side of clumps as the function of radiation field strength and gas density at the near side of clumps $U_{0}$ and $n_{\rm H}$ for the clump of size $R_{\rm cl} = 0.1$ pc, assuming $S_{\rm max} = 10^{8}\erg \cm^{-3}$ (left panel) and $S_{\rm max} = 10^{6}\erg\cm^{-3}$ (right panel). Yellow to blue implies the full dust disruption effect by RATD in clumps, darkblue and black imply the partly disruption and no disruption case.}
          \label{fig:adisr_clump_distance}
\end{figure*}

\begin{figure*}
        \includegraphics[width = 0.49\textwidth]{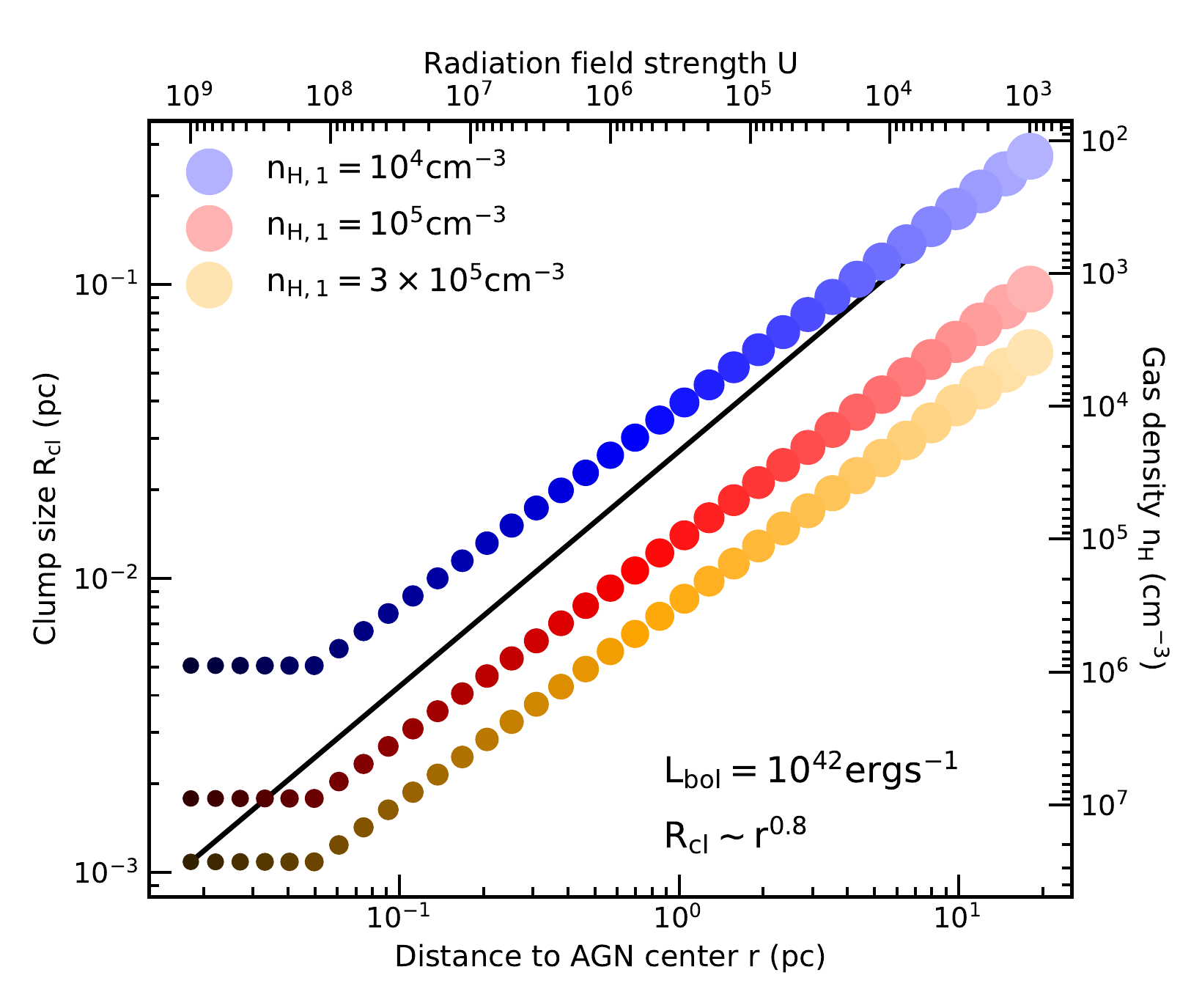}
        \includegraphics[width = 0.49\textwidth]{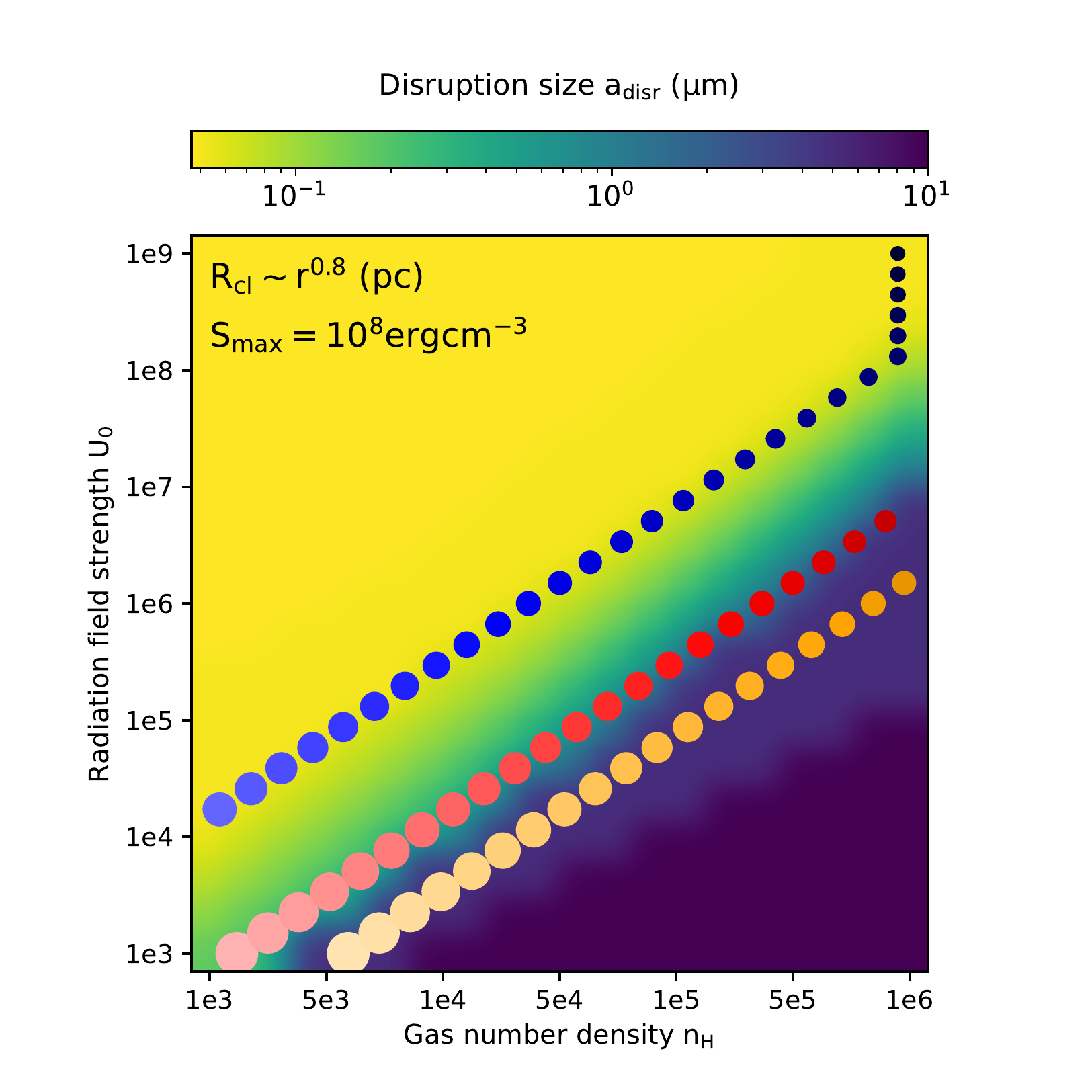}
        \includegraphics[width = 0.49\textwidth]{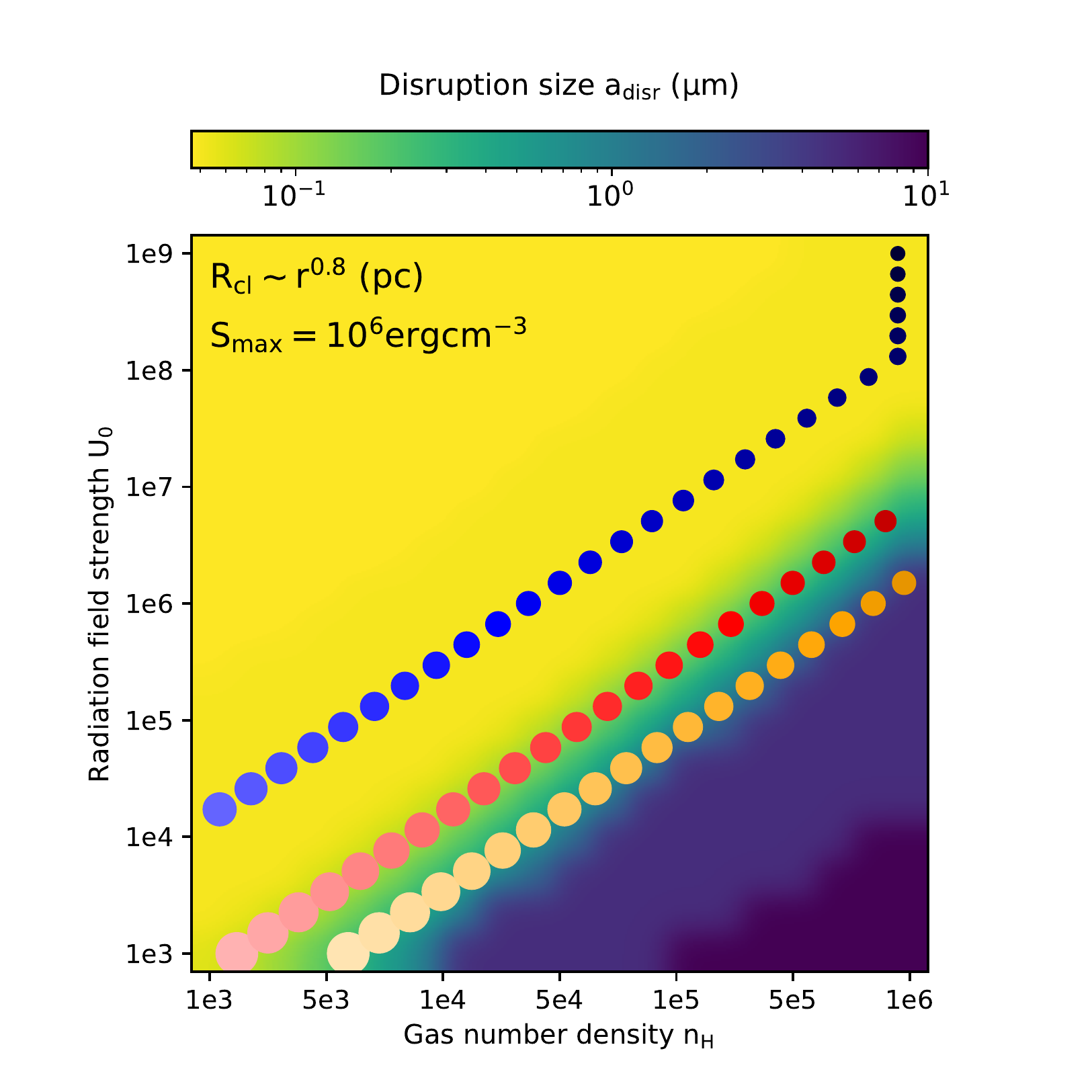}
     \caption{Upper left panel shows the space-dependence of gas density and clump size as the function of distances $r$ to the AGN center (or radiation field strength $U$). From $R_{\rm cl}-r$ relation (black line), we sample 35 points to study the effect of RATD inside clumps, in which small, dark circles present for small, dense clumps near the AGN center, while large, faint circles present for large, optically thin clumps near the outer region of torus. We note that each point corresponds to one couple value of $r-n_{\rm H}$ (horizontal and right vertical axis), and the point size only represents quantitatively the clump size. Upper right and lower left panels show the disruption size as the function of $U_{0}$ and $n_{\rm H}$, accounting for the variation of clump size with distances as in the upper left panel. Studied cases are plotted to see how RATD destroys composite dust grains with these conditions. }
          \label{fig:adisr_clump_distance_cloud_size}
\end{figure*}

The left panel of Figure \ref{fig:a_disr_smooth} shows the variation of the disruption size of composite dust grains as the function of distances, assuming different gas density profiles with $n_{\rm H, 1}$ varying from $5\times10^{3}\cm^{-3}$ to $10^{5}\cm^{-3}$. To account for the dependence of the maximum tensile strength on grain sizes, we adopt $S_{\rm max} = 10^{10}\erg\cm^{-3}$ for small grain of size $a < 0.05\mum$ and $S_{\rm max} = 10^{8}\erg\cm^{-3}$ for larger grains. Dust grains are strongly disrupted near the sublimation front due to strong radiation and are less affected by RATD at larger distances, as illustrated by the rise of $a_{\rm disr}$ and the decline of $a_{\rm disr, max}$. The intersection between $a_{\rm disr}$ and $a_{\rm disr,max}$ determines the distance where RATD ceases. We call this boundary the active region of RATD. The disruption of composite dust grains is stronger for lower gas densities due to smaller rotational damping and dust reddening effect. For example, micron-sized grains of $a > 1\mum$ can be disrupted by RATD up to $r = 1$ pc if the gas density at $r = 1$ pc decreases below $n_{\rm H, 1} \leq 5\times10^{3}\cm^{-3}$.

The right panel of Figure \ref{fig:a_disr_smooth} shows the similar results as the left panel, but assuming large grains of $a > 0.05\mum$ have $S_{\rm max} = 10^{6}\erg\cm^{-3}$. By decreasing the maximum tensile strength, large grains are easier to be destroyed by RATD due to lower disruption threshold (Equation \ref{eq:omega_disr}), resulting in the expansion of the active region of RATD. For example, with low gas density of $n_{\rm H, 1} = 5\times10^{3}\cm^{-3}$, all large grains of $a> 0.03\mum$ with $S_{\rm max} = 10^{6}\erg\cm^{-3}$ can be destroyed by RATD up to the outer boundary of torus at $r = 10$ pc. However, for low-luminosity AGN of $L_{\rm bol} = 10^{42}\erg~\rm s^{-1}$ with the sublimation distance at $r_{\rm sub} = 0.002$ pc, the smooth distribution of dust density in torus region produces very high dust reddening effect that significantly suppresses the RATD effect at the parsec scale. Dust grains at this distance are not affected by rotational disruption if $n_{\rm H, 1} > 5\times10^{3}\cm^{-3}$. 

\subsubsection{Clumpy torus model}\label{sec:adisr_clumpy}
The left panel of Figure \ref{fig:a_disr_clumpy} shows the variation of grain disruption size with the distance to the near side of the clump (i.e., closest to the AGN) for the clump of radius $R_{\rm cl} = 0.05$ pc located at distance $r = 1.5$ pc, assuming different values of $n_{\rm H, 1}$, and $S_{\rm max} = 10^{8}\erg\cm^{-3}$. Dust is destroyed stronger toward the near side, while weaker in the middle and far side of the clump. In contrast to the inefficient RATD at the parsec scale in the smooth torus model (see Figure \ref{fig:a_disr_smooth}), this mechanism now is strong enough to destroy all large grains of $a\geq 1\mum$ in the whole clump located at $r = 1.5$ pc with $n_{\rm H}  < 3\times10^{5}\cm^{-3}$. The higher efficiency of RATD comes from the configuration of the clumpy torus model such that dust near the front face of clumps is directly illuminated by the strong UV-optical AGN radiation and thus can be destroyed by RATD stronger. However, for a higher gas density of $n_{\rm H, 1} > 3\times10^{5}\cm^{-3}$, RATD is only efficient to disrupt large grains in the first half of the clumps near the AGN.

The right panel of Figure \ref{fig:a_disr_clumpy} shows the similar results as the left one but for $S_{\rm max} = 10^{6}\erg\cm^{-3}$. The RATD effect becomes stronger (i.e, smaller $a_{\rm disr}$) due to the smaller $S_{\rm max}$. For instance, micron-sized grains of $a > 0.1\mum$ with low $S_{\rm max} = 10^{6}\erg\cm^{-3}$ will be totally removed by RATD even if the clumps at $r = 1.5$ pc are optically thick with $n_{\rm H, 1} > 3\times10^{5}\cm^{-3}$.

We now study the effect of RATD for a given clump of $R_{\rm cl} = 0.1$ pc, assuming the different radiation strength ($U_{\rm 0}$) and density ($n_{\rm H}$) at the near side of the clump. The results for $S_{\rm max} = 10^{8}\erg\cm^{-3}$ and $S_{\rm max} = 10^{6}\erg\cm^{-3}$ are shown in the upper left and right panel of Figure \ref{fig:adisr_clump_distance}, respectively. The color from yellow to blue indicates the total disruption of large grains by RATD inside the clump (e.g., the right panel of Figure \ref{fig:a_disr_clumpy}), while the darkblue indicates the disruption in the front half of the clump (e.g., the curve with $n_{\rm H} = 3\times10^{5}\cm^{-3}$ in the left panel of Figure \ref{fig:a_disr_clumpy}). The black (e.g., $a_{\rm disr} = a_{\rm max} = 10\mum$) indicates no disruption effect.

For compact grains with $S_{\rm max} = 10^{8}\erg\cm^{-3}$ (left panel), they will be efficiently destroyed by RATD in the region of strong radiation field and/or low gas density (yellow region) and are less affected by RATD in the weak radiation field and/or high density region (darkblue and black region). The constrains of the maximum grain size by RATD is stronger, i.e., smaller $a_{\rm disr}$, for grains with lower maximum tensile strength (right panel for grains with $S_{\rm max} = 10^{5}\erg\cm^{-3}$). For example, with $U_{0} = 10^{5}$ and $n_{\rm H} = 10^{5}\cm^{-3}$, RATD only can remove large grains with $S_{\rm max} = 10^{8}\erg\cm^{-3}$ in the front half of the clump of size $0.1$ pc (darkblue region), while it can destroy all grains larger than $1\mum$ in the entire clump if grains have $S_{\rm max} = 10^{6}\erg\cm^{-3}$. 

The upper left panel of Figure \ref{fig:adisr_clump_distance_cloud_size} shows the variation of gas densities given by Equation (\ref{eq:n_H_smooth}) and cloud sizes given by Equation (\ref{eq:R_clump}) with $\beta_{\rm cl} = 0.8$ as the function of distances $r$ to the AGN center (or radiation field strength $U$). From the $R_{\rm cl}-r$ relation (black line), we sample 35 points from $r = r_{\rm sub}$ to $r = 20$ pc to study the effect of RATD inside clumps. Blue, red, and orange colors present for different gas density profiles with the gas density at 1 pc of $n_{\rm H,1} =10^{4}, 10^{5}$, and $3\times10^{5}\cm^{-3}$, respectively. Small, dark circles present for small, dense clumps near the AGN center while large, faint clumps present for large, less dense clumps near the outer region of the circumnuclear region (see the left part of Figure \ref{fig:AGN_model}). 

The upper right and lower left panels of Figure \ref{fig:adisr_clump_distance_cloud_size} show the similar disruption size map as the function of $U_{0}$ and $n_{\rm H}$ as Figure \ref{fig:adisr_clump_distance}, but accounting for the variation of clump size with distances from the upper left panel. RATD is more efficient in destroying composite dust grains inside the dense and compact clumps near the center region, but less efficient in the dilute and large clumps far from the center. Putting the sample of studied cases (red, blue, and orange circles) from the upper left panel to the map, one can see that with low gas density of $n_{\rm H, 1} = 10^{4}\cm^{-3}$, RATD is efficient enough to suppress the presence of large grains of $a \geq 0.1\mu m$ inside the torus. For a higher gas density of $n_{\rm H} = 3\times10^{5}\cm^{-3}$, grains of $a\geq 0.1\mu m$ with $S_{\rm max} = 10^{8}\erg\cm^{-3}$ at the far side of clumps can survive against RATD and grow to micron-sizes. However, the growth of micron-sized grains is likely coupled with the reduction of the material's strength  (\citealt{Dominik_1997}, \citealt{Wada_2008}, \citealt{Suyama_2012}). For newly formed large grains with $S_{\rm max} = 10^{6}\erg\cm^{-3}$ (the lower left panel), they will be quickly disrupted again to smaller sizes. In other words, the maximum grain size inside clumps will be constrained by the strength of RATD.

If clumps inside the torus are bigger and denser than our considered case, dust aggregates have more chance for growing around AGN. This sets the boundary region of RATD that should be taken into account when studying the formation of large grains inside the dusty torus.

\section{Rotational desorption of icy grain mantles} \label{sec:desorption}
In this section, we study the evolution of icy grain mantles beyond the snow line under the effect of rotational desorption (\citealt{Hoang_Tram}). The rotational desorption mechanism is described in Section \ref{sec:rotation_desp} and the results are shown in Section \ref{sec:a_desp}.

\subsection{Rotational desorption mechanism}
\label{sec:rotation_desp}
Similar to rotational disruption of composite dust grains in the intense radiation field described in Section \ref{sec:a_disr}, an ice mantle can be separated from a solid grain core if the tensile stress produced by the spinning icy grain exceeds the adhesive energy that holds the mantle and the core together (\citealt{Hoang_Tram}). 
 
Let $a_{\rm c}$ be the effective radius of a spherical grain that has the same volume as the irregular compact grain core, and $\Delta a_{\rm m}$ is the average thickness of the ice mantle. The effective radius of the icy grain can be considered as $a = a_{\rm c} + \Delta a_{\rm m}$. The average tensile stress $S$ produced by the centrifugal force of all ice layers applying on the interface of core-mantle is (\citealt{Hoang_Tram}):
\bea 
S = \frac{\rho_{\rm ice}\omega^{2} a^{2} }{4} \Bigg[1 - \Bigg(\frac{a_{\rm c}}{a}\Bigg)^{2}\Bigg],
\label{eq:tensile_stress}
\ena
where $\rho_{\rm ice} = 1~\rm g\cm^{-3}$ is the mass density of ice.
  
The detachment of the ice mantle from the grain surface will occur when $S$ exceeds the adhesive energy of the mantle. The adhesive energy depends on the surface properties of the solid core, i.e., rough surfaces induce larger adhesive strength up to $\sim 10^{9}\erg \cm^{-3}$ (\citealt{Work}) while smoother surfaces induce lower values. The tensile strength of ice $S_{\rm max,ice}$ depends on the temperature, that higher temperature, i.e., $200-300$ K, induces lower value of $S_{\rm max,ice} \sim 5\times10^{6}\erg \cm^{-3}$ (\citealt{Litwin}). In this paper, we consider that the adhesive strength has the same value as the maximum tensile strength of ice mantles.

The critical angular velocity for rotational desorption of the ice mantle can be obtained by setting $S = S_{\rm max,ice}$ (\citealt{Hoang_Tram}), which follows:
\bea 
\omega_{\rm desp} &=& \frac{2}{a(1 - a_{\rm c}^{2}/a^{2})^{1/2}}\Bigg(\frac{S_{\rm max}}{\rho_{\rm ice}}\Bigg)^{1/2} \nonumber \\
&\approx& \frac{6.3\times10^{8} \rho_{\rm ice}^{-1/2} }{a_{-5}(1 - a_{\rm c}^{2}/a^{2})^{1/2}} S_{\rm max,ice,7}^{1/2} ~\rm rad ~ s^{-1},
\label{eq:omega_desp}
\ena 
where $S_{\rm max,ice,7} = S_{\rm max,ice}/(10^{7} \erg \cm^{-3})$. 

By comparing the maximum angular velocity induced by RATs, $\omega_{\rm RAT}$, for all grain sizes (Equations \ref{eq:omega_RAT_small} and \ref{eq:omega_RAT_large}) with $\omega_{\rm desp}$, one can determine the range of desorption size $a_{\rm desp}-a_{\rm desp,max}$ in which the ice mantle is separated from the grain core.
 
\subsection{Grain desorption size}\label{sec:a_desp}
\begin{figure*}
        \includegraphics[width = 0.5\textwidth]{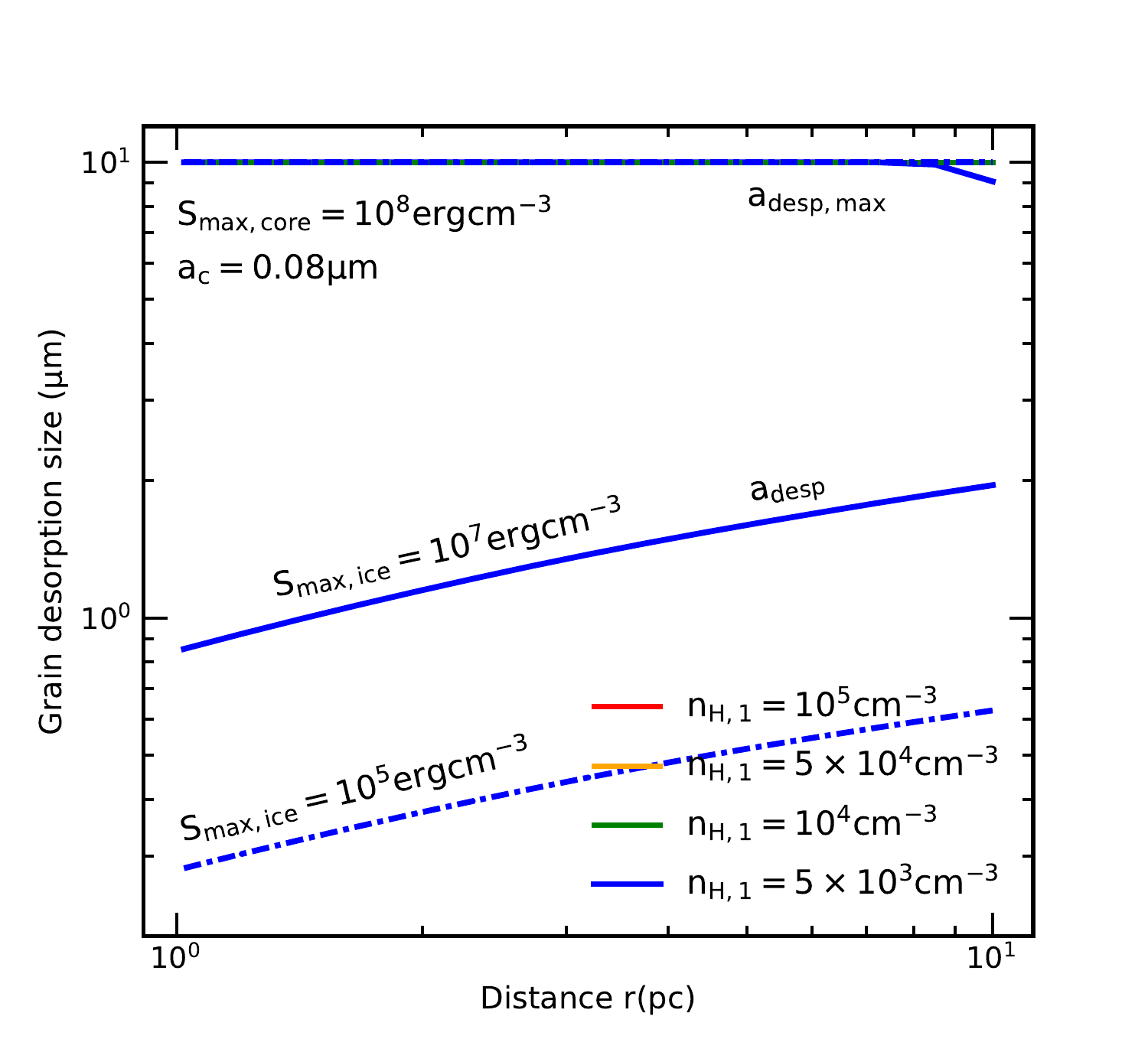}
        \includegraphics[width = 0.5\textwidth]{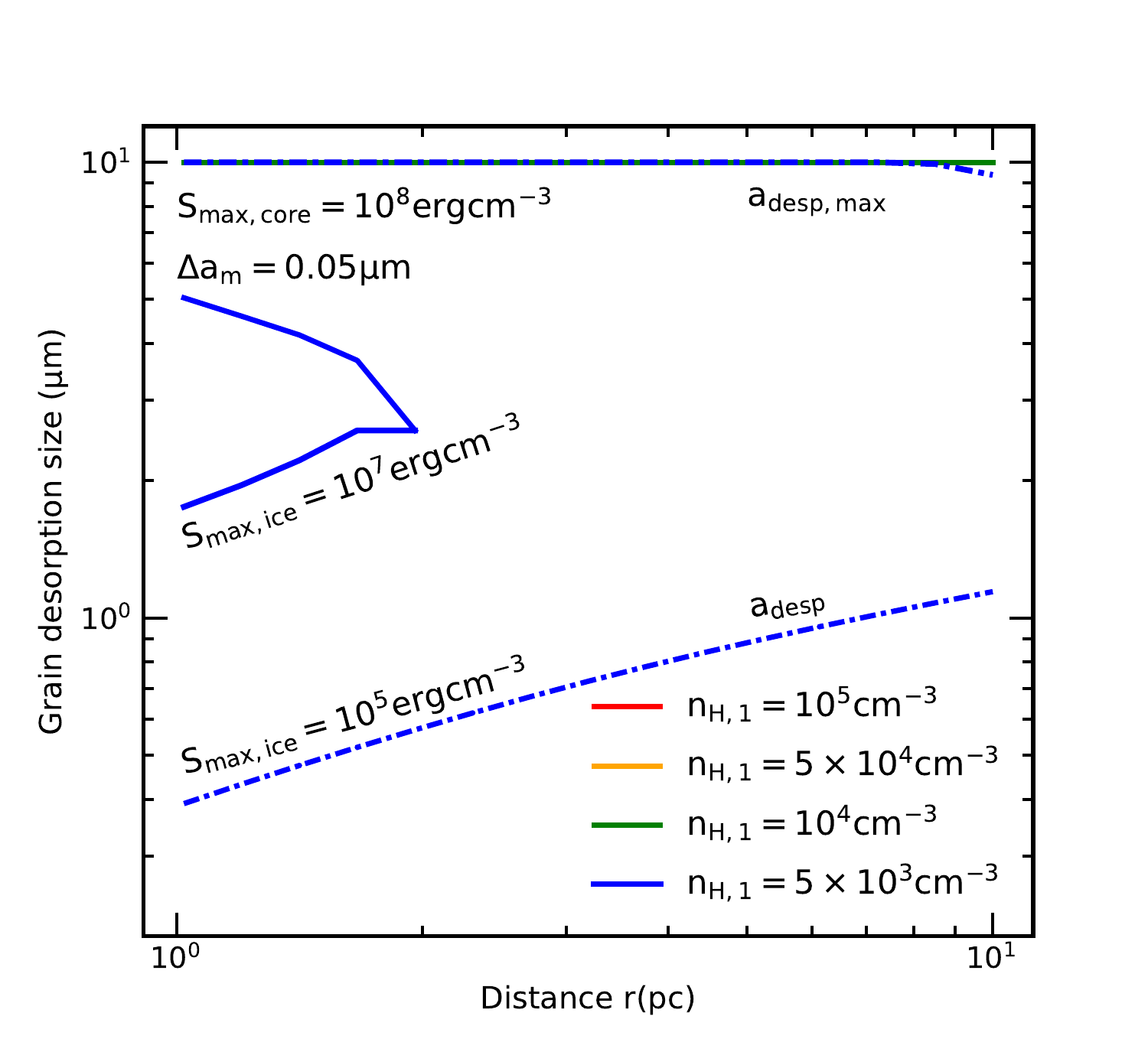}
     \caption{Dependence of the range of desorption sizes of icy grain mantles from minimum desorption size of $a_{\rm desp}$ to maximum desorption size of $a_{\rm desp,max}$ on the distance to the AGN center for the smooth torus model. Different gas density profiles $n_{\rm H, 1}$ and two values of the tensile strengths of ice $S_{\rm max, ice} = 10^{7}\erg\cm^{-3}$ (solid lines) and $S_{\rm max, ice} = 10^{5}\erg\cm^{-3}$ (dashed-dot lines) are assumed. For high gas density of $n_{\rm H,1} > 5\times10^{3}\cm^{-3}$ (green, orange, and red lines), ice mantles are not desorbed from the solid core due to the inefficient rotational desorption by weak radiation field, giving $a_{\rm desp} = a_{\rm desp,max} = 10\mum$, i.e., horizontal line at $10\mum$. Rotational desorption of ice mantles occurs only for low gas density $n_{\rm H, 1} \leq 5\times10^{3}\cm^{-3}$, with stronger desorption of ice mantles, i.e.,  smaller $a_{\rm desp,min}$, near the snow line at 1 pc. The left panel is for the grain mantle with the fixed grain core of $a_{\rm c} = 0.08\mum$, while the right panel is for the fixed ice mantle thickness of $\Delta a_{\rm m} = 0.05\mum$. The grain core is assumed to have $S_{\rm max} = 10^{8}\erg\cm^{-3}$.} 
     \label{fig:a_desp_smooth}
\end{figure*}

\begin{figure*}
   \begin{minipage}[t]{0.47\textwidth}
     \centering
        \includegraphics[width=1.05\textwidth]{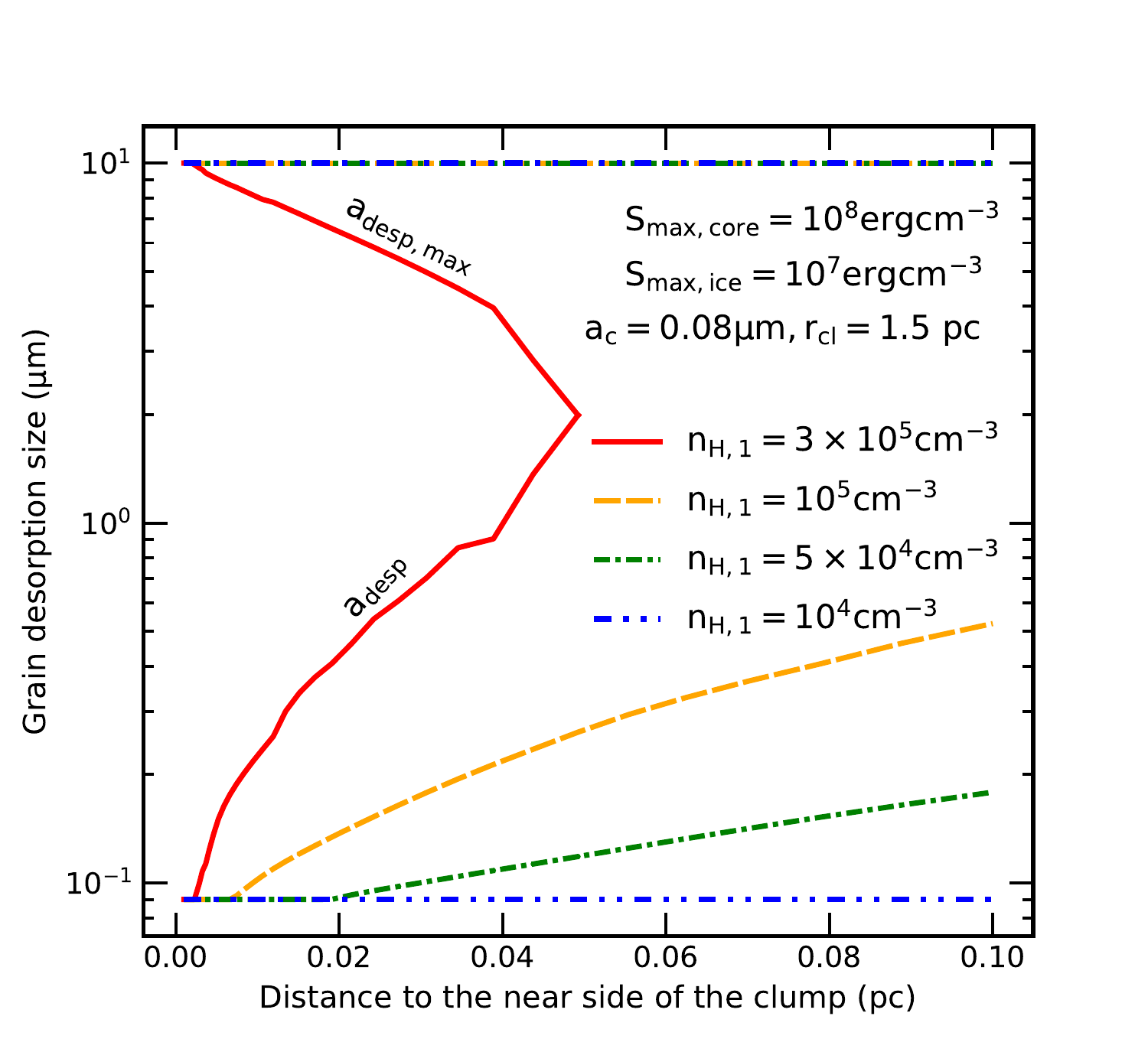}
        \includegraphics[width=1.05\textwidth]{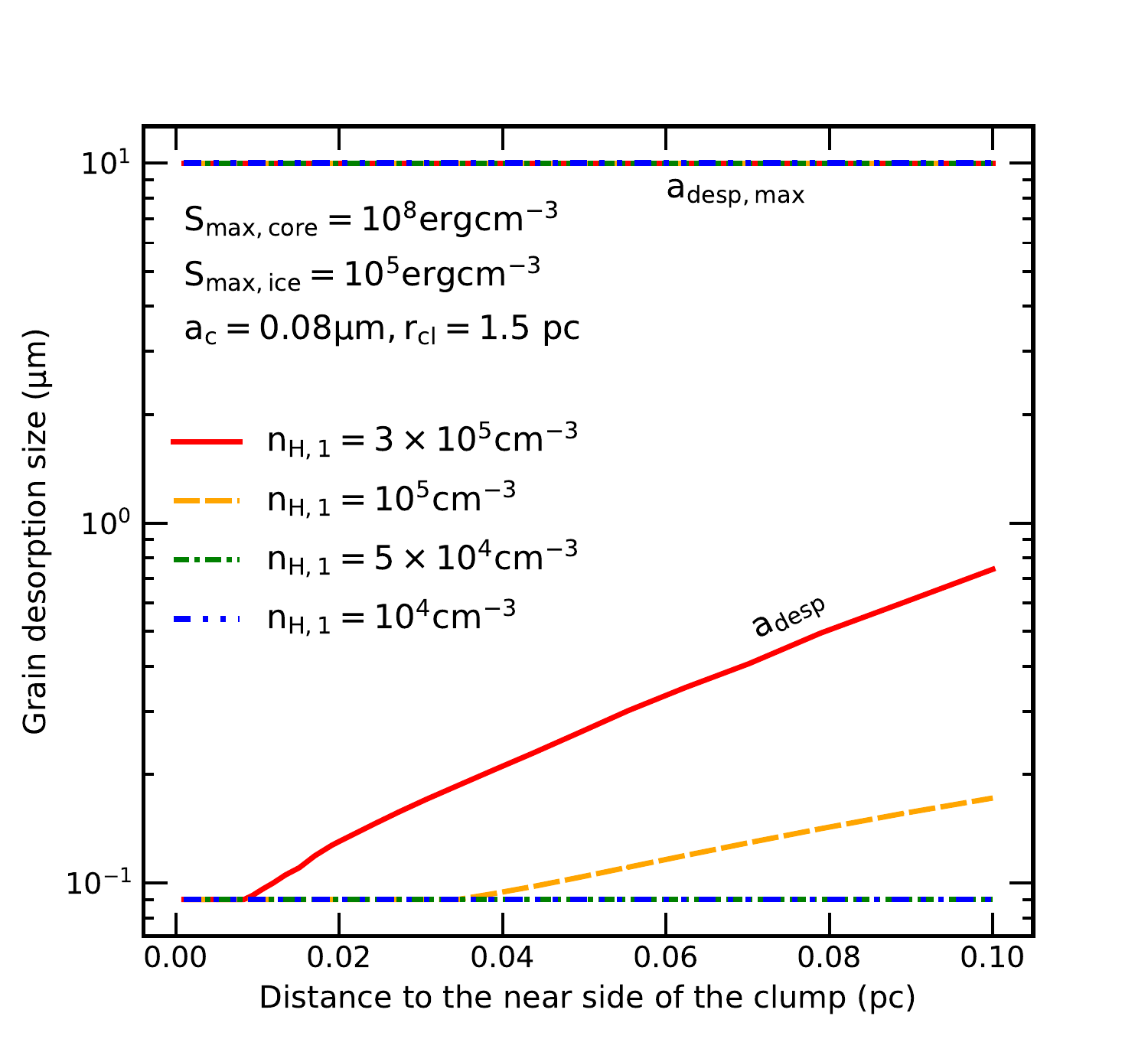}
     \caption{Variation of the range of desorption sizes of icy grain mantles inside the clump at $r = 1.5$ pc for different gas density profiles, $S_{\rm max, ice} = 10^{7}\erg\cm^{-3}$ (upper panel), and $S_{\rm max, ice} = 10^{5}\erg\cm^{-3}$ (lower panel). The grain core is assumed to have $a_{\rm c} = 0.08\mum$ and $S_{\rm max} = 10^{8}\erg \cm^{-3}$.}
     \label{fig:adesp_clumpy_fixcore}
   \end{minipage}\hfill
   \begin{minipage}[t]{0.47\textwidth}
     \centering
    \includegraphics[width=1.05\textwidth]{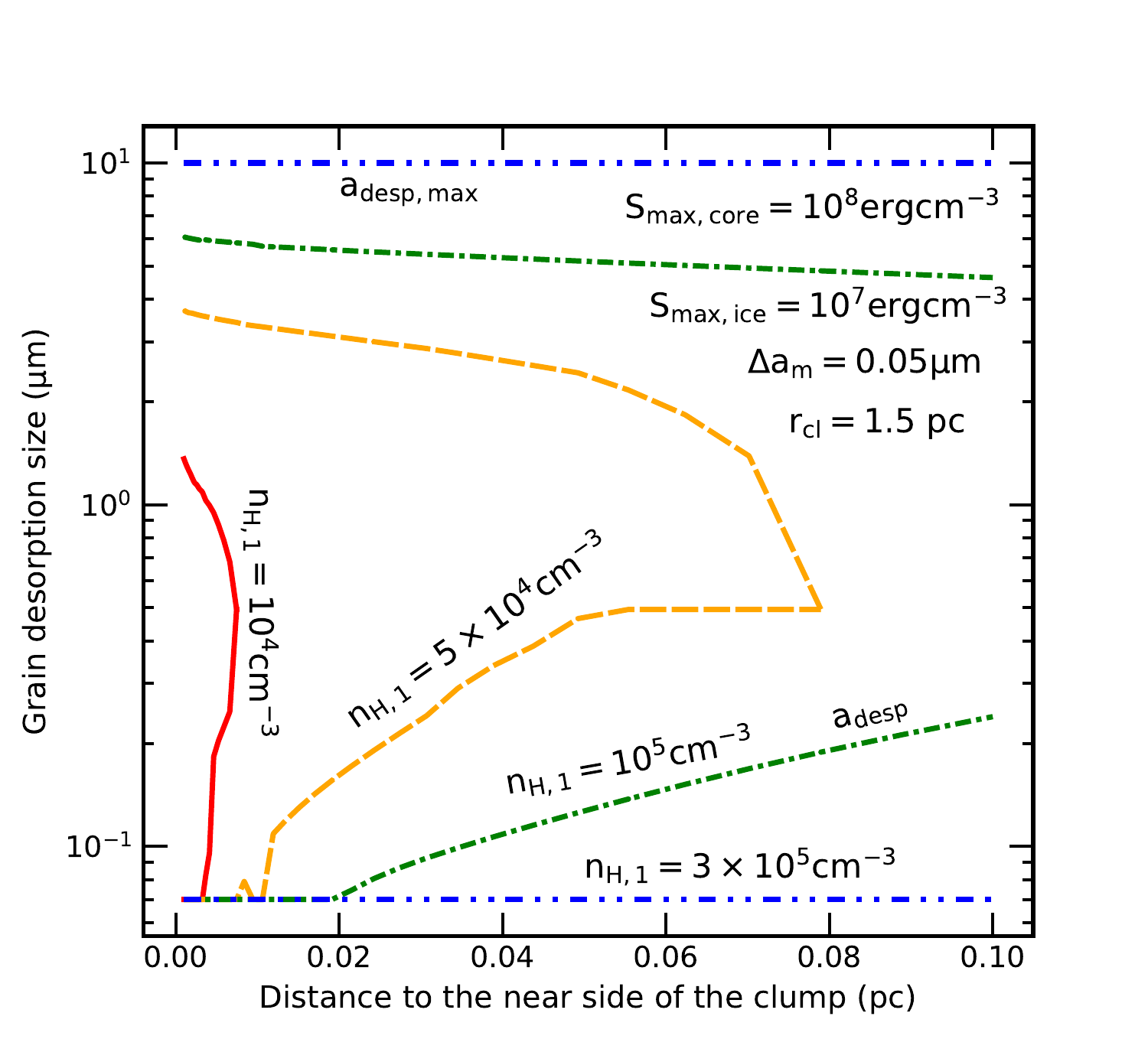}
    \includegraphics[width=1.05\textwidth]{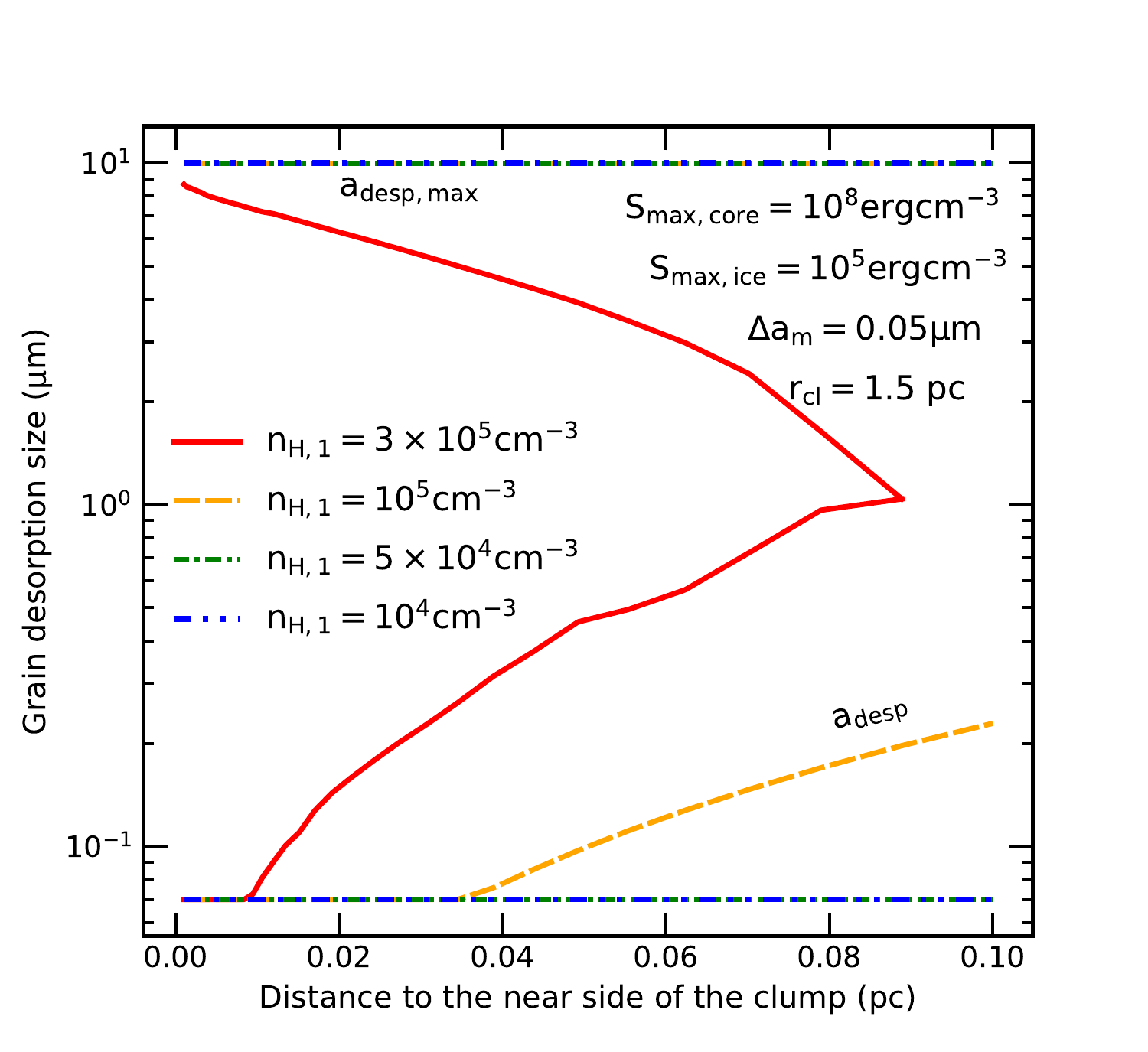}
   \caption{Similar results as Figure \ref{fig:adesp_clumpy_fixcore} but for icy grains with a fixed thickness of the ice mantle of $\Delta a_{\rm m} = 0.05\mum$. } 
   \label{fig:adesp_clumpy_fixmantle}
   \end{minipage}
\end{figure*}

\begin{figure*}
\centering
        \includegraphics[width=0.49\textwidth]{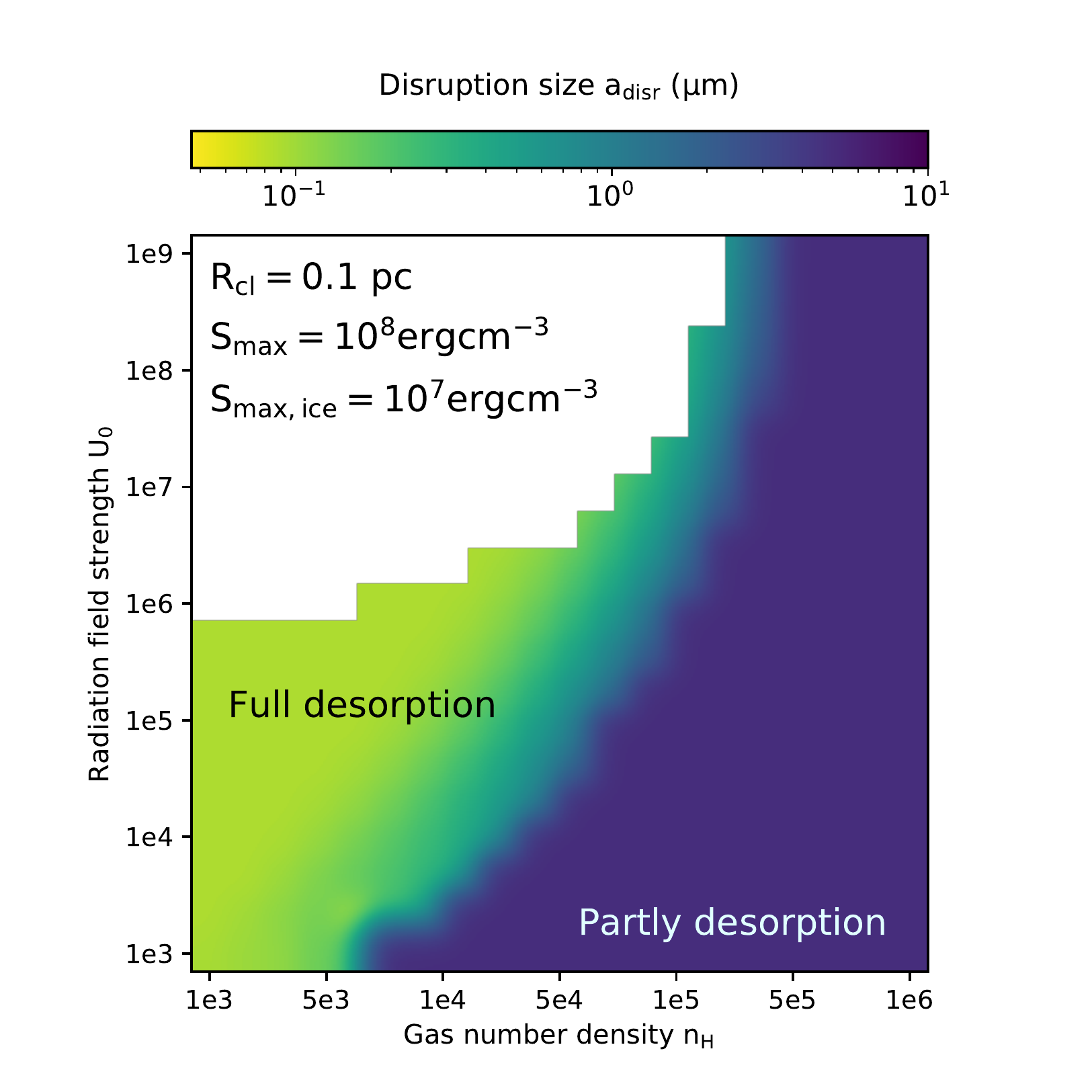} 
        \includegraphics[width=0.49\textwidth]{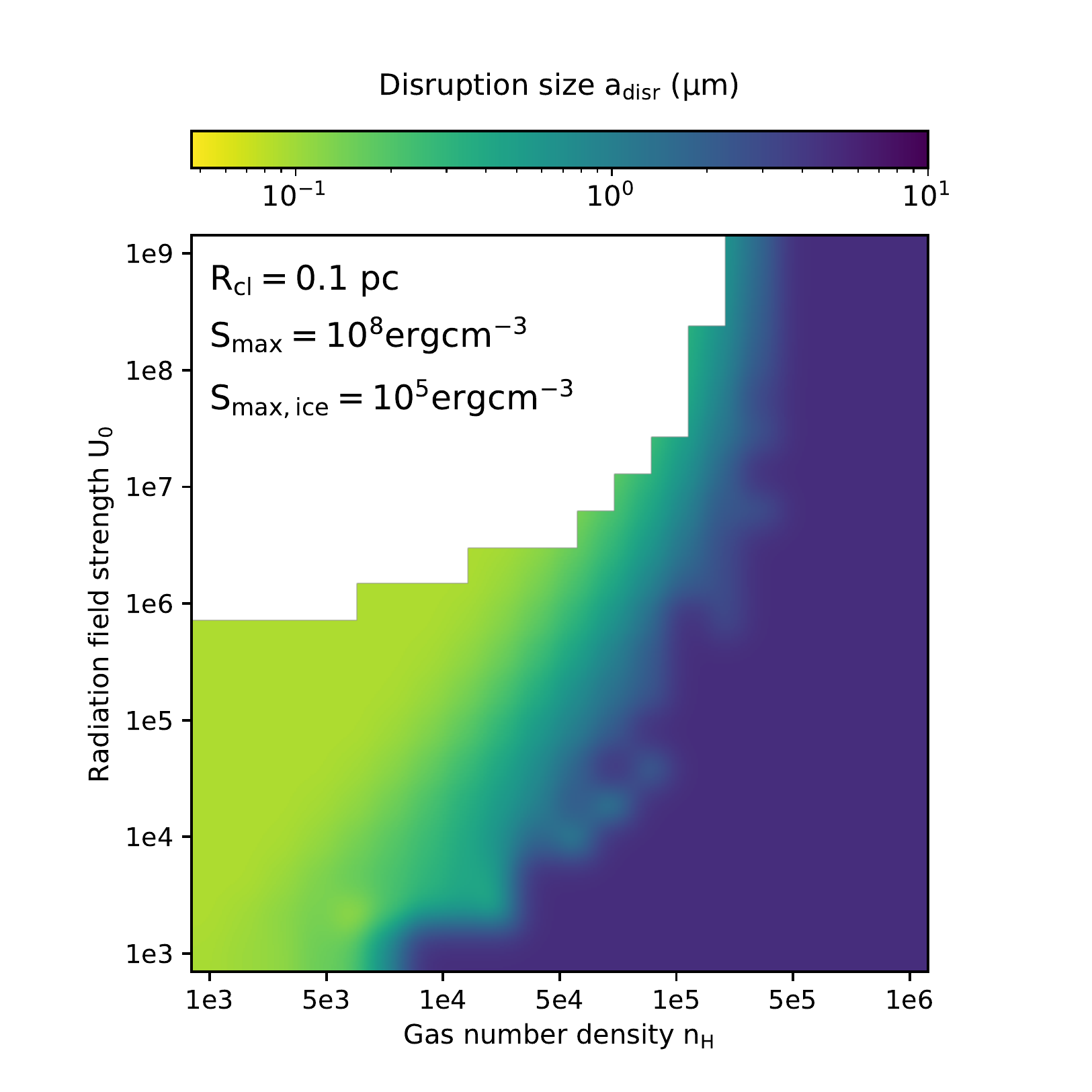}
        \includegraphics[width=0.49\textwidth]{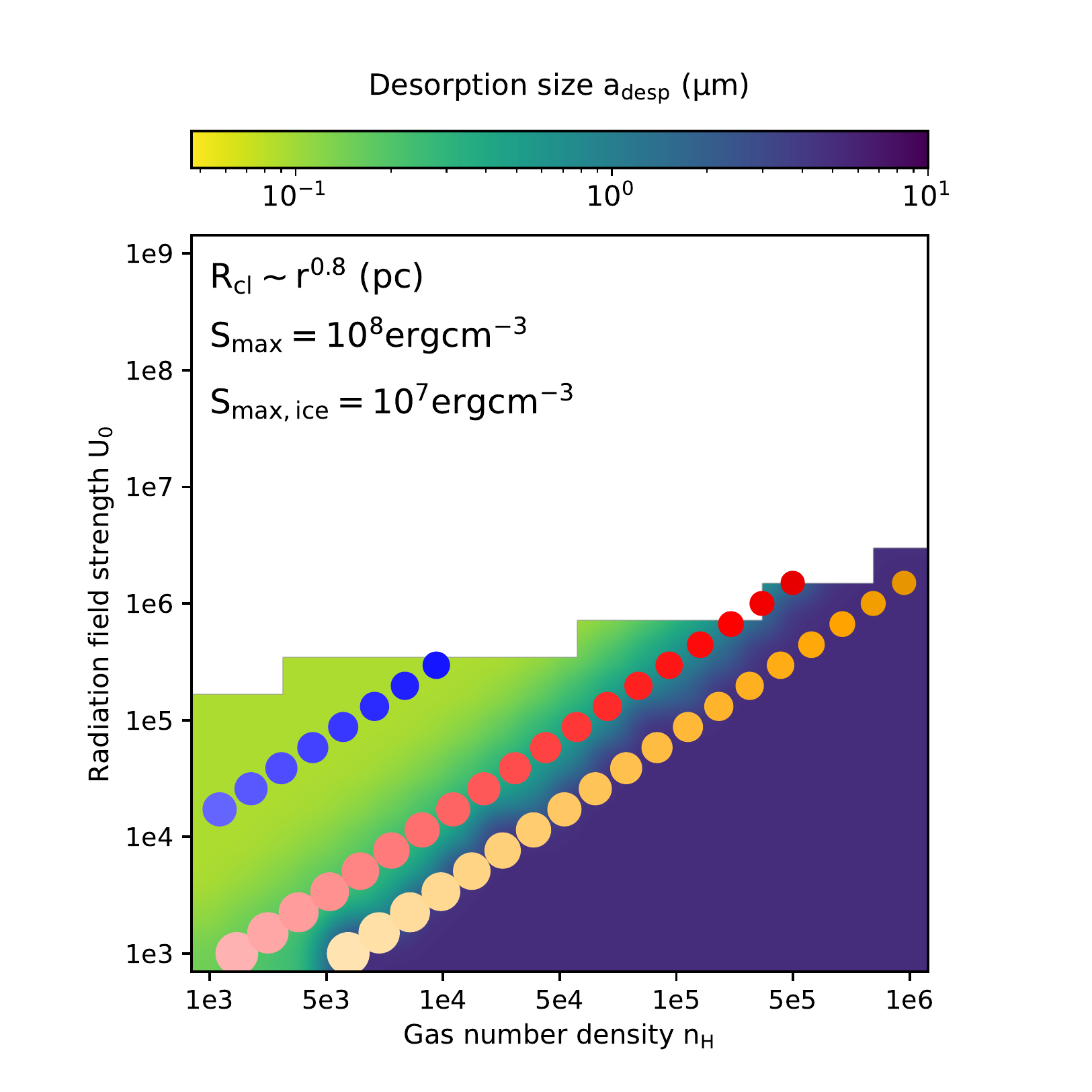}
        \includegraphics[width=0.49\textwidth]{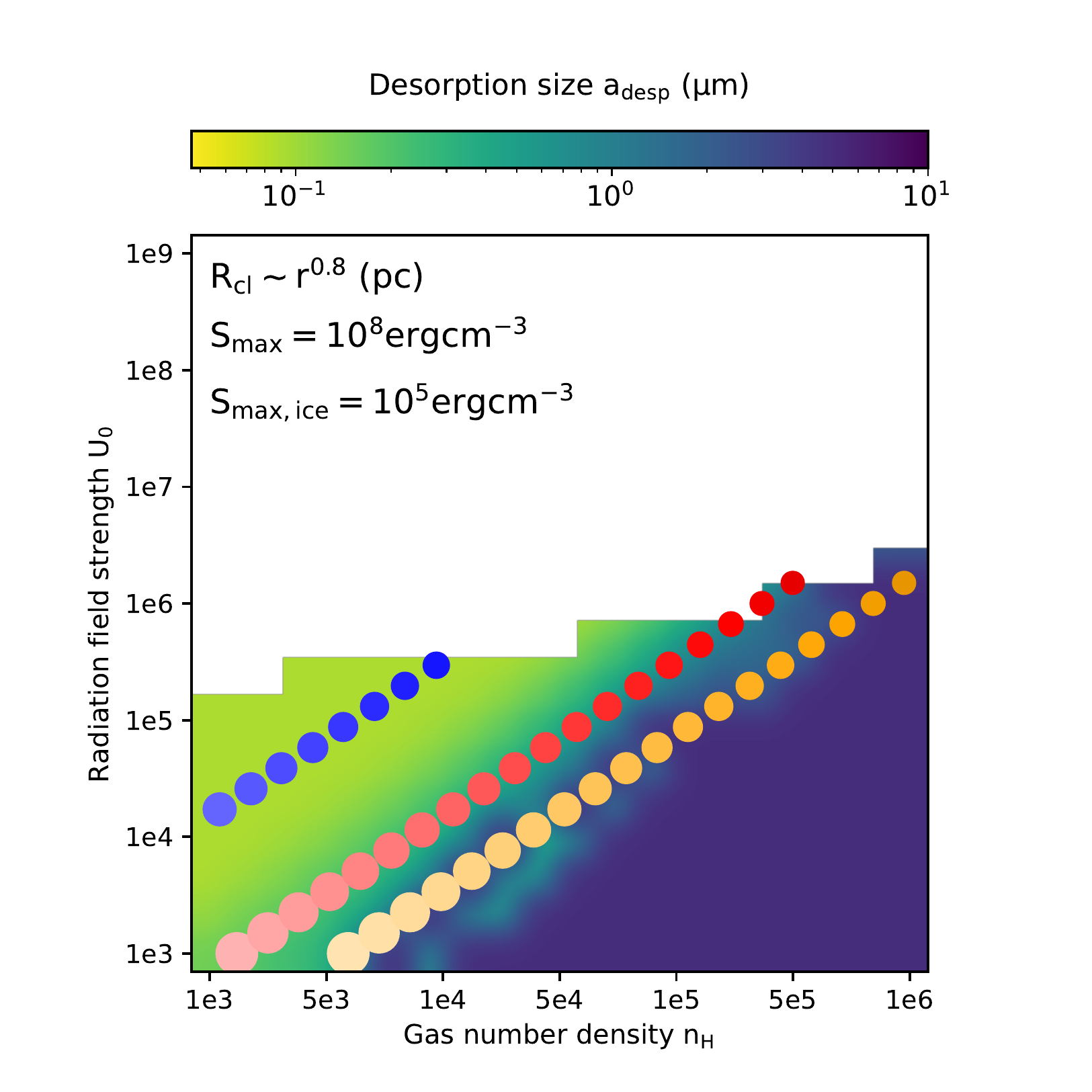}
    \caption{Upper panels: dependence of grain desorption sizes of icy grain mantles at the far side of clumps of size $R_{\rm cl} = 0.1$ pc on the radiation strength $U_{0}$ and the gas density $n_{\rm H}$, assuming $S_{\rm max, ice} = 10^{7}\erg\cm^{-3}$ (left) and $S_{\rm max,ice} = 10^{5}\erg\cm^{-3}$ (right). Assuming that the grain core is fixed at $a_{\rm c} = 0.08\mum$ and has $S_{\rm max} = 10^{8}\erg\cm^{-3}$. The snow line where ice can condense on grain surface is the boundary between white and color region. Green and blue color implies the full desorption of ice mantle inside clumps while darkblue implies the desorption only on the front half of the clumps. Lower panels: similar as the upper panel but account for the variation of clump size with distances given by the upper left panel of Figure \ref{fig:adisr_clump_distance_cloud_size}. The sample of out studied is marked on the map.}
          \label{fig:adesp_clump_distance}
\end{figure*}

Our numerical calculations of the gas temperature (see Equation \ref{eq:Tgas_torus}) show that ice can exist on the composite dust grains beyond $r \sim 1$ pc. We will next study the evolution of icy grains under the rotational desorption effect, assuming that the grain core has $S_{\rm max} = 10^{8}\erg\cm^{-3}$.

\subsubsection{Smooth torus model}\label{sec:adesp_smooth}

The left panel of Figure \ref{fig:a_desp_smooth} shows the dependence of grain desorption size of icy grain mantles with the fixed grain core of $a_{\rm c} = 0.08\mum$ on distances, assuming different gas density profiles. The maximum tensile strength of ice mantles is $S_{\rm max, ice} = 10^{7}\erg\cm^{-3}$ (thick lines), and $S_{\rm max, ice} = 10^{5}\erg\cm^{-3}$ (thin lines). One can see the overlap between $a_{\rm desp}$ and  $a_{\rm desp, max}$ at $a_{\rm max} = 10\mum$ beyond $r \geq 1$ pc for $n_{\rm H, 1} > 10^{4}\cm^{-3}$, indicating no effect of rotational desorption at the parsec scale of torus region. Ice mantles are only detached from the grain core only if the gas density at 1 pc declines below $n_{\rm H, 1} \leq 5\times10^{3}\cm^{-3}$. 

The right panel of Figure \ref{fig:a_desp_smooth} shows the space-varying desorption size for icy dust with the fixed thickness of ice mantles of $\Delta a_{\rm m} = 0.05\mum$. Similar to the left panel, rotational desorption only can affect icy grains at the parsec scale if $n_{\rm H,1} \leq 5\times10^{3}\cm^{-3}$. The desorption happens stronger for lower tensile strengths of ice mantles. In addition, one can see that thin ice mantles are hard to be detached from the core, i.e., higher $a_{\rm desp}$ and lower $a_{\rm desp,max}$, that belongs to the lower tensile stress applying on the contact region between the solid core and the ice mantle (see Equation \ref{eq:tensile_stress}). 

\subsubsection{Clumpy torus model}\label{sec:adesp_clumpy}
The upper panel of Figure \ref{fig:adesp_clumpy_fixcore} shows the variation of $a_{\rm desp}-a_{\rm desp, max}$ within the clump of radius $R_{\rm cl} = 0.05$ pc at $r = 1.5$ pc with different gas density profiles, assuming the fixed grain core of $a_{\rm c} = 0.08\mum$, and $S_{\rm max,ice} = 10^{7}\erg\cm^{-3}$. Icy grain mantles are desorbed stronger near the front side of clumps and weaker in the middle and back side. Interestingly, one can see that rotational desorption now can affect into the presence of icy grain mantles at the parsec scale of torus region thanks to the direct illumination of UV-optical photons from the center region. The detachment happens stronger for the clumps with a lower gas density (i.e., lower $n_{\rm H,1}$). For instance, at the far side of clumps, the desorption size decreases from $10\mum$ (no rotation desorption effect) for $n_{\rm H, 1} = 3\times10^{5}\cm^{-3}$ to $a_{\rm desp} \sim 0.6\mum$ and $0.09\mum$ for $n_{\rm H, 1} = 10^{5}\cm^{-3}$ and $10^{4}\cm^{-3}$, respectively.

The lower panel of Figure \ref{fig:adesp_clumpy_fixcore} shows the similar results as the upper panel, but for $S_{\rm max,ice} =10^{5}\erg\cm^{-3}$. For lower maximum tensile strength of ice mantles, i.e., smaller desorption threshold, rotational desorption can detach all icy grains with size greater than $1\mum$ even in the dense clumps with $n_{\rm H,1} = 3\times10^{5}\cm^{-3}$ at 1.5 pc.  

Figure \ref{fig:adesp_clumpy_fixmantle} shows the variation of grain desorption size inside the clump of $R_{\rm cl} = 0.05$ pc located at 1.5 pc for the different gas density profiles, assuming the fixed thickness of ice mantles of $\Delta a_{\rm m} = 0.05\mum$. Similar to the case of fixed grain core, ice mantles can be detached from the core due to rotational desorption effect. The detachment is stronger for dilute clumps and porous structure of ice mantles. However, the thin ice mantle is less separated from the core due to lower induced tensile stress at the contact region, i.e, higher $a_{\rm desp}$ compared with the results in Figure \ref{fig:adesp_clumpy_fixcore}. For example, with $S_{\rm max,ice} = 10^{7}\erg\cm^{-3}$, ice mantles with thickness $\Delta a_{\rm m} = 0.05\mum$ never be desorbed from the core in dense clumps with $n_{\rm H,1} \geq 10^{5}\cm^{-3}$.

Upper panels of Figure \ref{fig:adesp_clump_distance} shows the dependence of the grain desorption size inside the clump of $R_{\rm cl} = 0.1$ pc with the radiation strength ($U_{0}$) and gas density ($n_{\rm H}$) at the front (near) side of the clump. The results are for icy grains with a fixed $a_{\rm c} = 0.08\mum$ and icy mantles with $S_{\rm max,ice} = 10^{7}\erg\cm^{-3}$ (left panel) and $S_{\rm max,ice} = 10^{5}\erg\cm^{-3}$ (right panel). Empty parts present for the region where water vapour cannot condense into ice, i.e., $T_{\rm dust} > 170$ K. Similar effect as RATD, icy grain mantle inside clumps with low gas density and one with lower maximum tensile strength $S_{\rm max,ice}$ are detached by rotational desorption stronger.

Lower panels show the similar results as the upper panels but accounting for the variation of clump size with distances given by the upper left panel of Figure \ref{fig:adisr_clump_distance_cloud_size}. The sample of our studied cases (upper left panel of Figure \ref{fig:adisr_clump_distance_cloud_size}) are also marked in the map. One can see that if $n_{\rm H,1} < 10^{5}\cm^{-3}$, large icy grain mantle with $a \geq 0.5\mu m$ will be strongly separated from the core even if they locate in large clumps at $r \sim 10$ pc ($U_{0} \sim 10^{3}$). If the clumpy torus is denser with $n_{\rm H, 1} \geq 3\times10^{5}\cm^{-3}$, thick ice mantles up to micron-sized can remain on the grain surface at the far side of clumps due to inefficient rotational desorption. Ice mantle with composite structure can be easier to be detached from the core by rotational desorption, but the difference is not clear (see the right panel). In this case, ice mantles increases the sticking coefficient and facilitate the sticky collisions between icy grains to form larger dust aggregates. However, the increase in the size of composite grains induces the decrease of the tensile strength. Consequently, newly formed large dust aggregate can be disrupted by RATD, which prevents grain growth and blanet formation (see the lower left panel of Figure \ref{fig:adisr_clump_distance_cloud_size}). However, the formation of large grains and further evolution can happen if clumps are denser and bigger than our consider cases.

\subsection{Disruption and desorption timescale} \label{sec:time_RATD}
Dust grains in the AGN torus may experience strong rotational damping due to IR re-emission, i.e., $F_{\rm IR} \ge 1$ (see \citealt{Hoang_2020} for the explanation). Thus, the moment when composite grains of size $a$ are disrupted by RATD can be found by setting $\omega(t)$ given by Equation (\ref{eq:omega(t)}) to equal $\omega_{\rm disr}$ given by Equation (\ref{eq:omega_disr}). The disruption time $t_{\rm disr}$ is given by:
\bea 
t_{\rm disr} = -\tau_{\rm damp}\ln(1 - \frac{\omega_{\rm disr}}{\omega_{\rm RAT}}) ~ \rm s.
\label{eq:time_RATD}
\ena

\begin{figure*}
        \includegraphics[width = 0.5\textwidth]{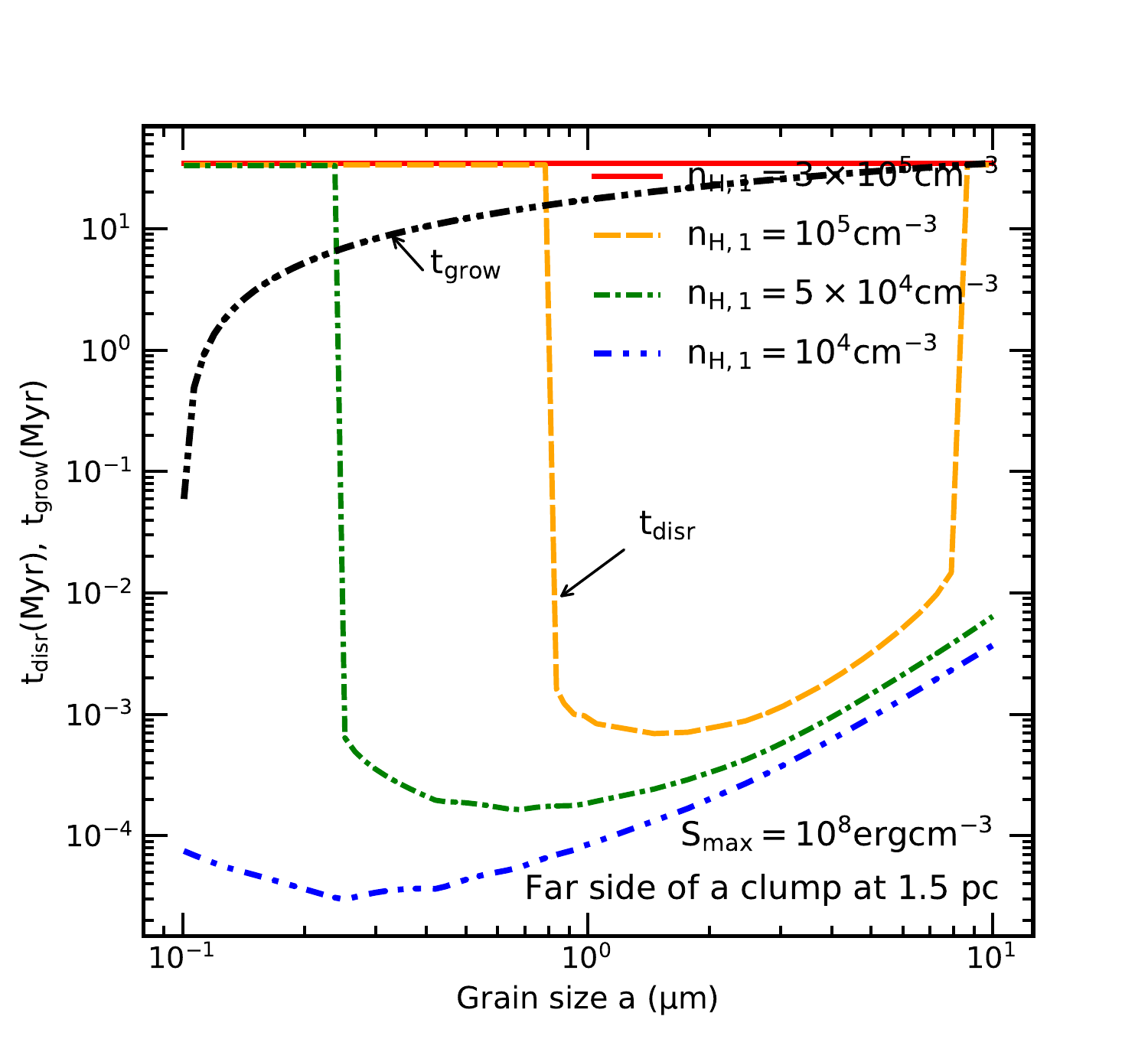}
        \includegraphics[width = 0.5\textwidth]{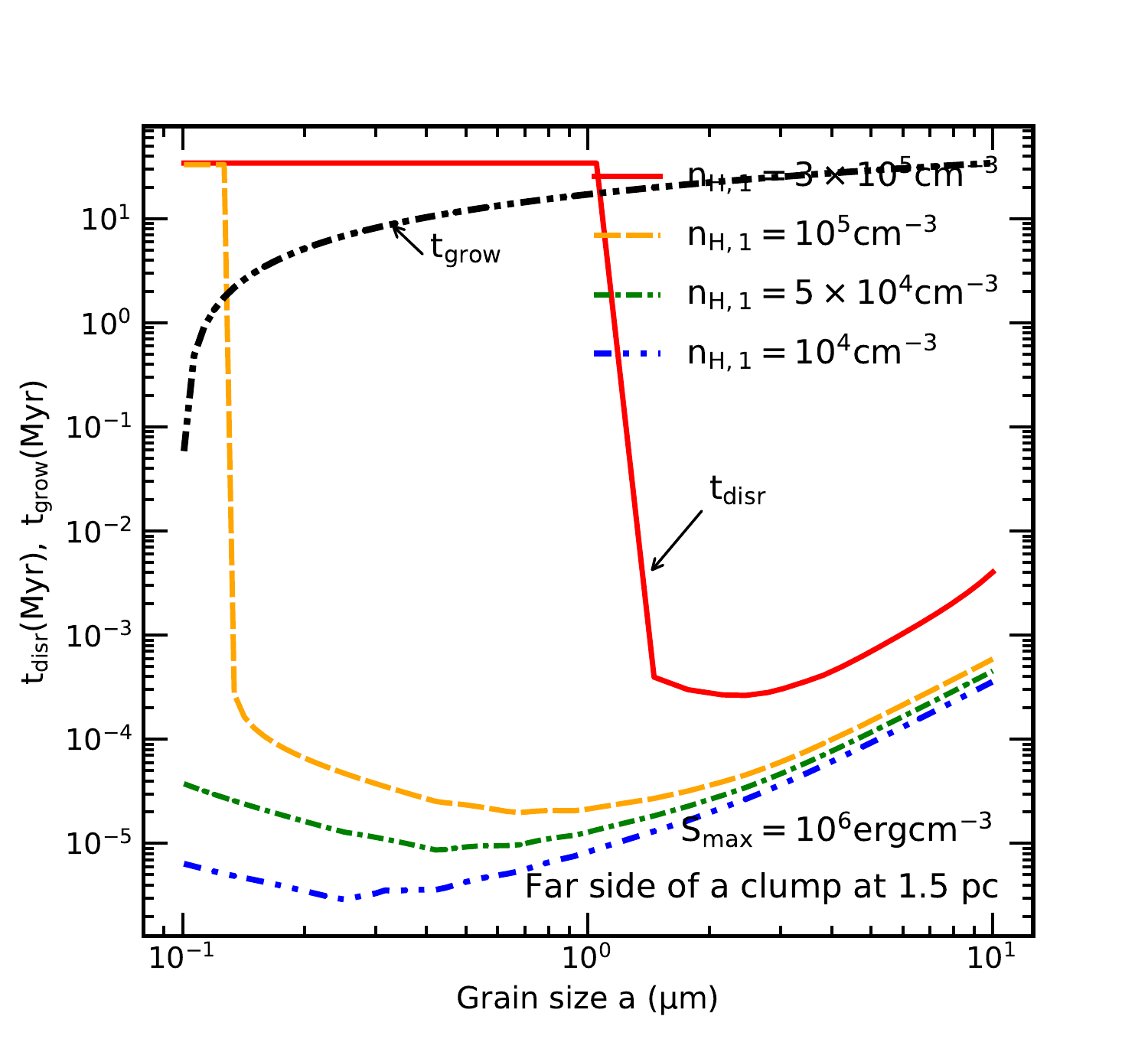}
     \caption{Dependence of disruption timescale $t_{\rm disr}$ on grain sizes for composite dust grains at the far side of the clump of $R_{\rm cl} = 0.05$ pc at $r = 1.5$ pc, assuming different gas density $n_{\rm H, 1}$, $S_{\rm max} = 10^{8}\erg\cm^{-3}$ (left panel) and $S_{\rm max} = 10^{6}\erg\cm^{-3}$ (right panel). The growth timescale (black dashed-dot-dot line) of corresponding grain size is also plotted in the figure for comparing. Dust disruption only happens within $\sim$ kyr, much shorter than the time required for grain growth.} 
     \label{fig:tdisr_size}
\end{figure*}

Similarly, the moment when the ice mantle detaches from the solid grain core by rotational desorption can be found by setting $\omega(t) = \omega_{\rm desp}$ (Equation \ref{eq:omega_desp}). The desorption time $t_{\rm desp}$ is then:
\bea 
t_{\rm desp} = -\tau_{\rm damp}\ln(1 - \frac{\omega_{\rm desp}}{\omega_{\rm RAT}}) ~ \rm s.
\label{eq:time_desp}
\ena
 
\section{Grain Growth vs. Disruption times}\label{sec:timescale}
In Sections \ref{sec:a_disr} and \ref{sec:a_desp}, we study the effect of rotation disruption and desorption on composite and icy grains in the midplane of the torus and find that grain growth and blanet formation can occur in the torus the idealized smooth torus model. However, large composite grains and icy grain mantle of size $a\geq 1\mum$ are efficiently disrupted by RATs for the clumpy torus model, assuming the same gas density distribution adopted in \cite{Wada2019,Wada2021}. Thus, to check whether grain growth can occur in the clumpy torus, we will calculate the timescale of grain growth in the hit-and-stick stage and compare it with the timescale of dust destruction and ice desorption in Sections \ref{sec:time_growth} and \ref{sec:time_compare}.
 
\subsection{Grain Growth Timescale}\label{sec:time_growth}
\cite{Wada2019} considered the possibility that icy grain mantles (of initial size $a_{0} \sim 0.1\mum$) in the torus can grow to larger ones via coagulation, assuming that grain growth follows the ballistic cluster-cluster aggregation (BCCA) model. The mass growth rate $dm/dt$ during the hit-and-stick stage via the BCCA route is given by (\citealt{Wada2019,Wada2021}):
\bea 
\frac{dm}{dt} = \frac{2 \sqrt{2\pi} \Sigma_{\rm d} a^{2} \Delta v}{H_{\rm d}},
\label{eq:dm_dt}
\ena 
where $\Sigma_{\rm d}$ is the surface mass density, $\Delta v$ is the relative velocity between dust aggregates, and $H_{\rm d}$ is the scale height of the dust disk.

The growth rate of grain size of $a$ can be written as
\bea 
\frac{1}{a}\frac{da}{dt} = \frac{3}{8}\sqrt{\frac{\pi}{2}} \eta \sqrt{\alpha} R_{\rm e}^{1/4} \sqrt{\frac{G M_{\rm BH}}{R^{3}}},
\ena 
where $\eta$ is the gas-to-dust mass ratio, $\alpha$ is the dimensionless parameter to determine the kinematic viscosity in the turbulent disk (\citealt{Shakura}), $R_{\rm e}$ is the Reynolds number, $M_{\rm BH}$ is the mass of SMBH, and $R$ is the distance from the AGN center.

The time required to form a dust aggregate of size $a$ is given by (see Appendix \ref{sec:appen} for a detailed derivation):
\bea 
t_{\rm grow}(a) &=& \frac{8}{3 \eta \alpha^{1/2} R_{\rm e}^{1/4}}\sqrt{\frac{2}{\pi}} \rm ln \Bigg(\frac{a}{a_{0}}\Bigg) \Bigg(\frac{G M_{\rm BH}}{R^{3}}\Bigg)^{-1/2} \nonumber \\
&\approx& 3.4489 ~\rm Myr ~\rm ln\Bigg(\frac{a}{a_{0}}\Bigg) \Bigg(\frac{\eta}{0.01}\Bigg)^{-1} \Bigg(\frac{\alpha}{0.02}\Bigg)^{-3/4} \nonumber \\
&\times& \Bigg(\frac{M_{\rm BH}}{10^{6} M_\odot}\Bigg)^{-3/8} \Bigg(\frac{R}{\rm pc}\Bigg)^{9/8}\Bigg(\frac{c_{\rm s}}{1~\rm km s^{-1}}\Bigg)^{-1/4},
\ena
where $a_{\rm 0} = 0.1\mum$ is the radius of icy monomers and $c_{\rm s}$ is the sound velocity.

\begin{figure}
        \includegraphics[width=0.42\textwidth]{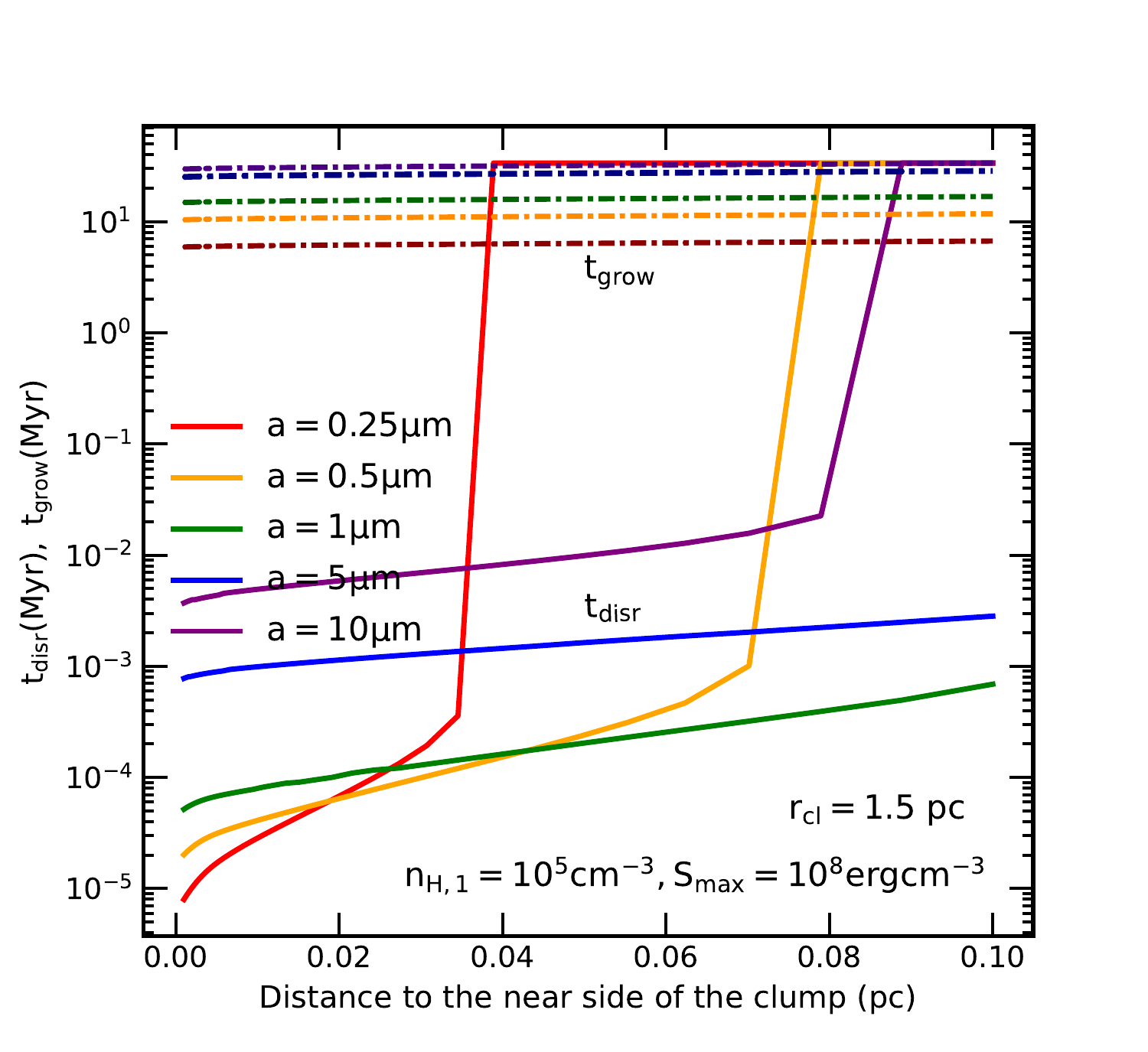}
        \vspace{0.00mm}
        \includegraphics[width=0.42\textwidth]{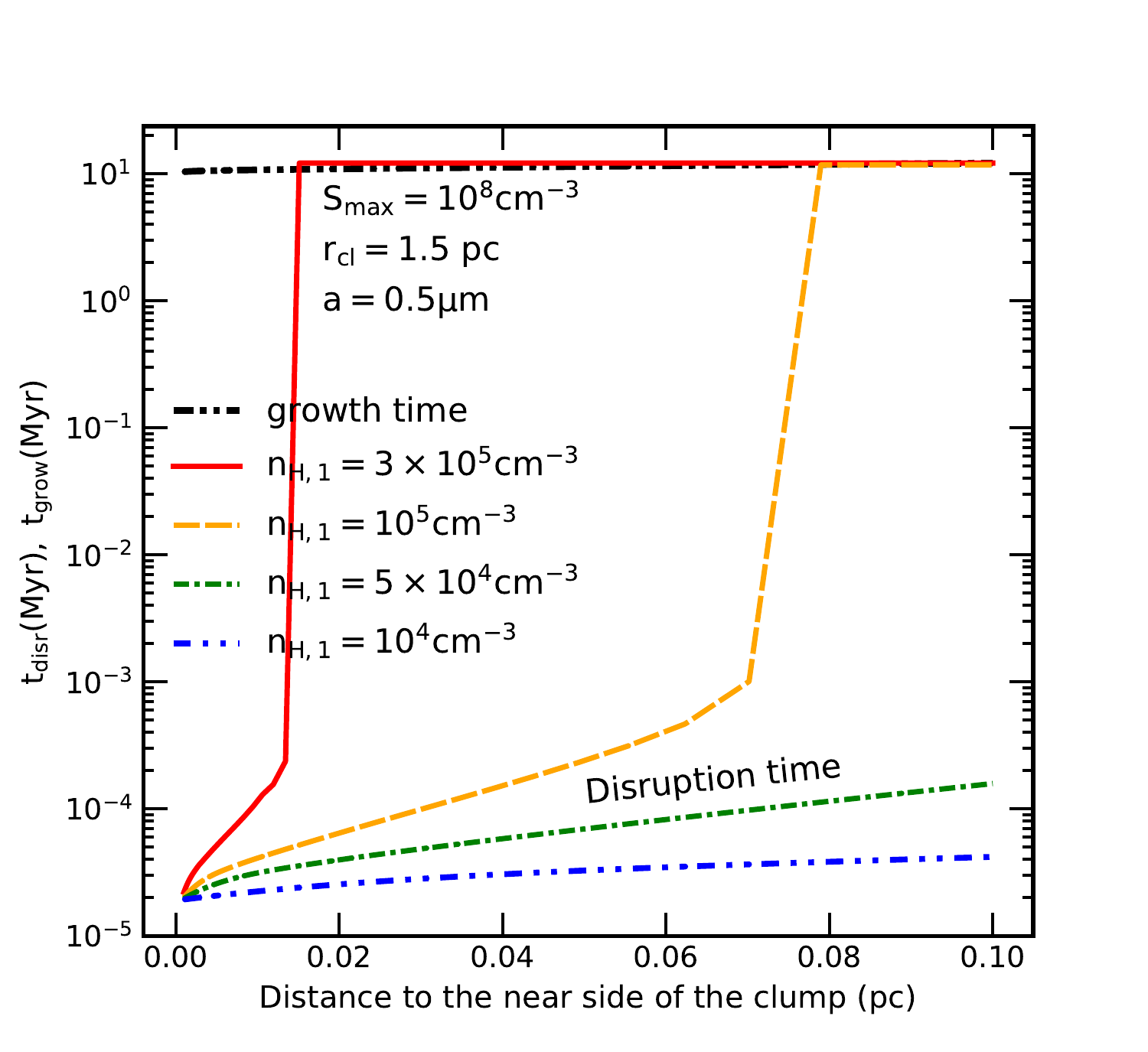}
        \vspace{0.00mm}
        \includegraphics[width=0.42\textwidth]{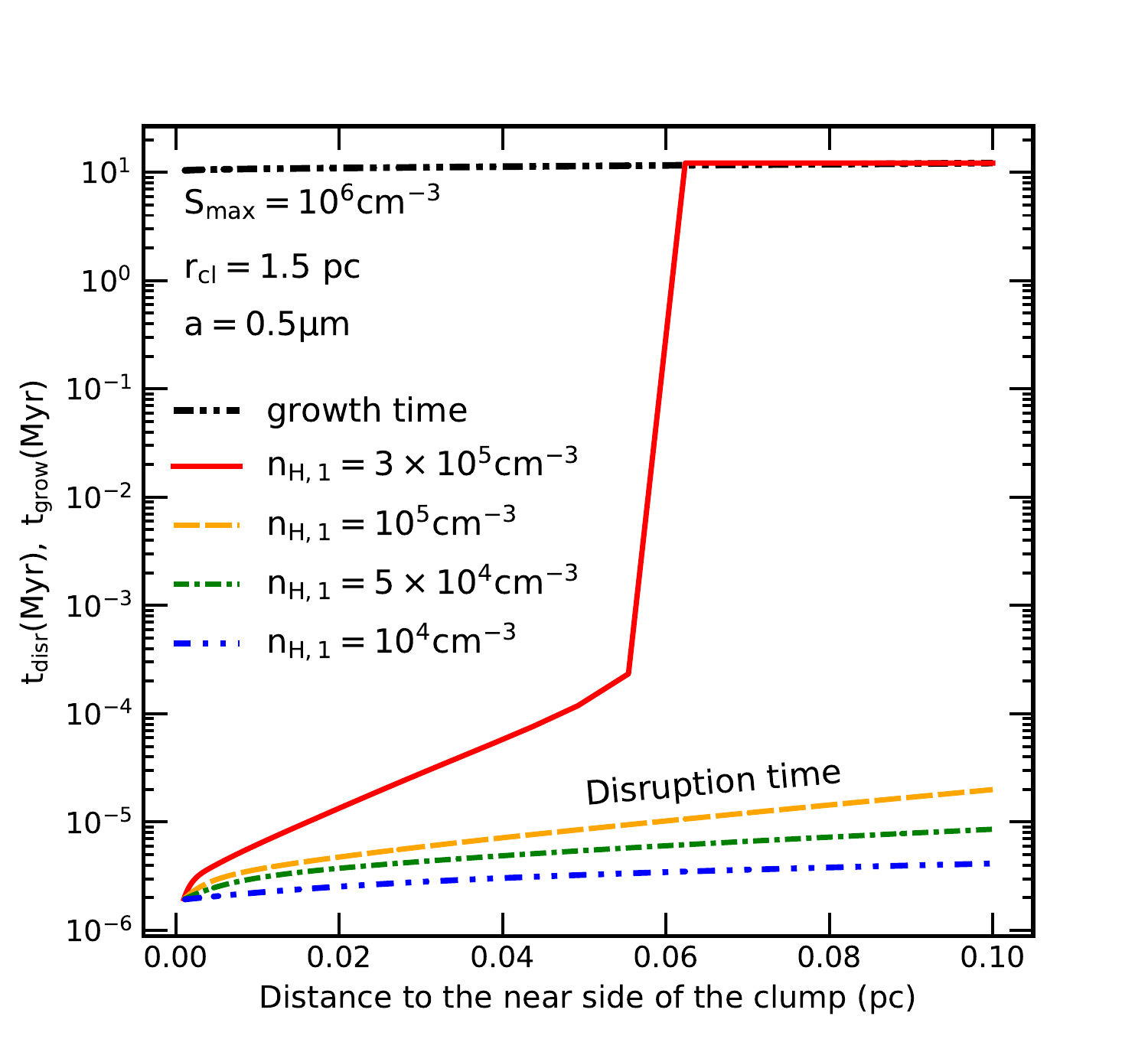}
     \caption{Comparison of grain growth (solid lines) and disruption time (dashed-dot lines) of composite grains as a function of distance within a clump of radius $R_{\rm cl} = 0.05$ pc located at $r = 1.5$ pc from the AGN center. The upper panel shows the timescale for different grain sizes. The central and lower panels is for grains of size $a = 0.5\mum$ with different gas density profiles, $S_{\rm max} = 10^{8}\erg\cm^{-3}$ and $10^{6}\erg\cm^{-3}$, respectively.}
          \label{fig:tdisr_distance}
\end{figure}

\begin{figure*}
        \includegraphics[width = 0.5\textwidth]{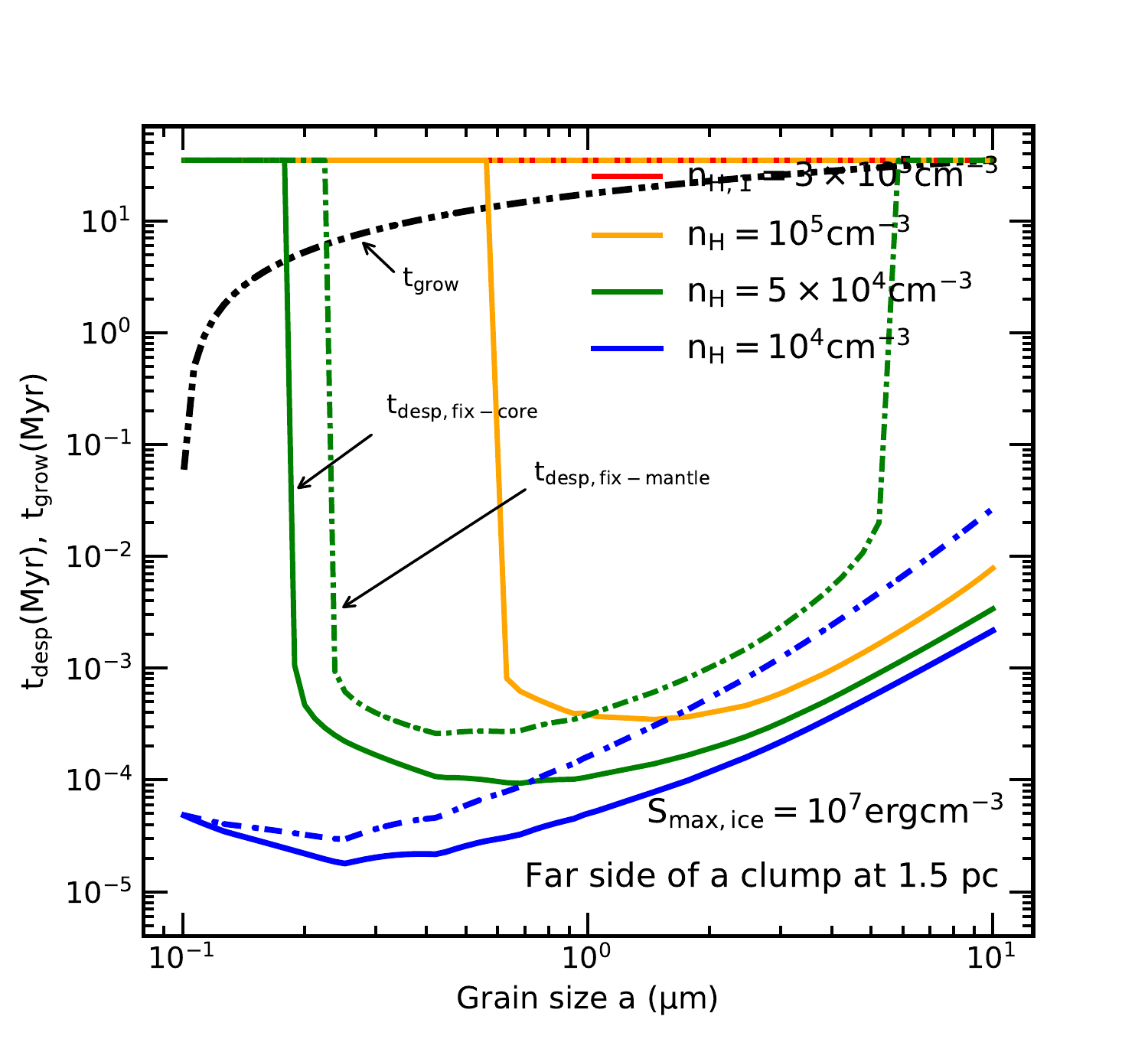}
        \includegraphics[width = 0.5\textwidth]{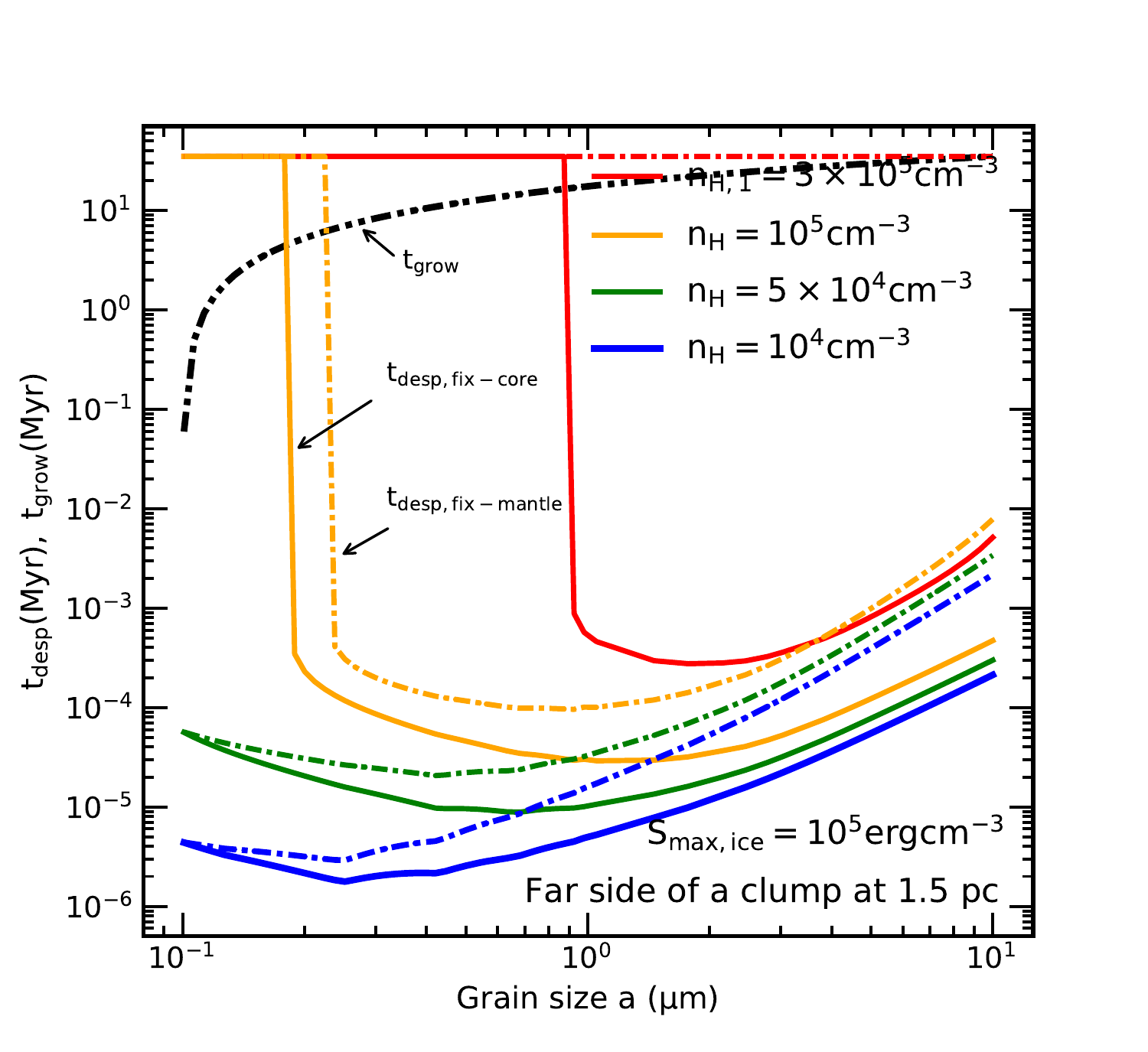}
     \caption{Variation of grain growth (black dashed-dot-dot line) and rotational desorption times of different grain sizes at the far side of the clumps $R_{\rm cl} = 0.05$ pc at $r = 1.5$ pc, assuming different gas density profile, $S_{\rm max, ice} = 10^{7}\erg\cm^{-3}$ (left panel) and $10^{5}\erg\cm^{-3}$ (right panel). The solid lines show the results for icy grains with a fixed grain core of $a_{\rm c} = 0.08\mum$ while the dashed-dot lines is for the case of fixed ice mantle's thickness of $\Delta a_{\rm m} = 0.05\mum$. Similar to RATD, the separation of the ice mantle from the grain core happens quickly just after few kyr.} 
     \label{fig:tdesp_radius}
\end{figure*}

\begin{figure*}
   \begin{minipage}[t]{0.47\textwidth}
     \centering
        \includegraphics[width=1.05\textwidth]{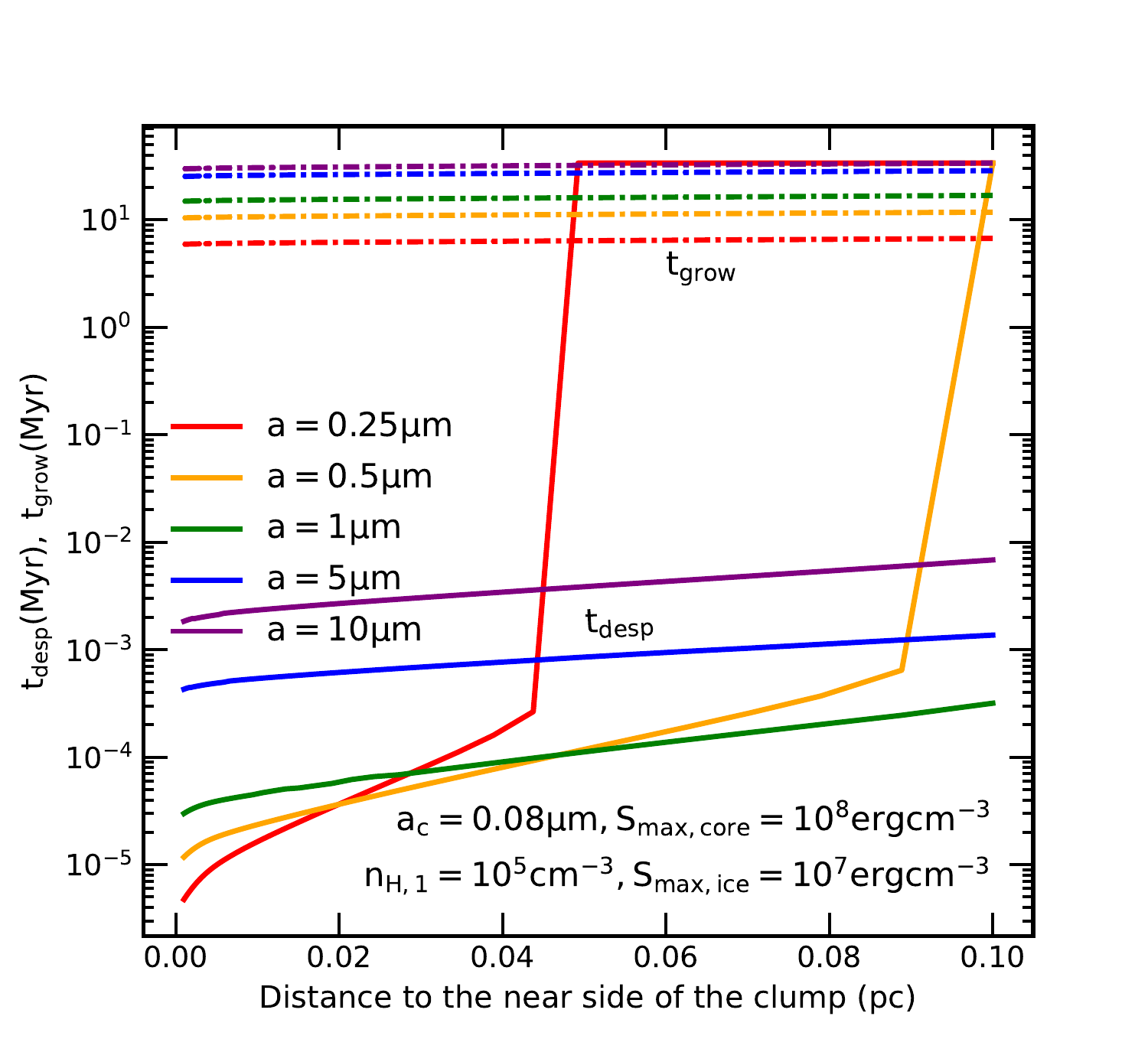}
        \includegraphics[width=1.05\textwidth]{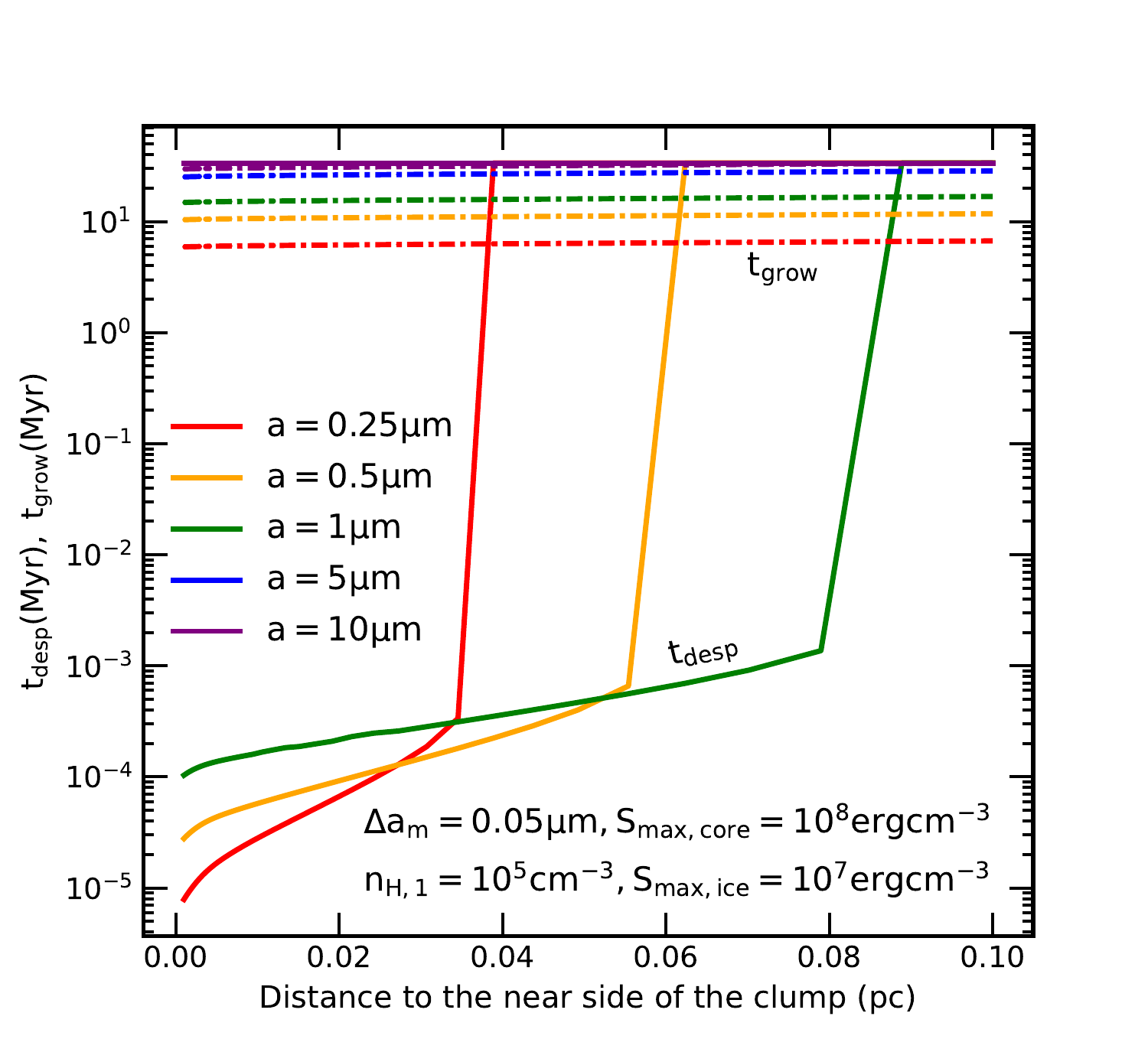}
     \caption{Upper panel: comparison between the variation of rotational desorption (solid lines) and grain growth timescale (dashed-dot lines) of different grain sizes inside the clump at $r = 1.5$ pc, assuming $n_{\rm H, 1} = 10^{5}\cm^{-3}$, icy grains with the fixed grain core of $a_{\rm c} = 0.08\mum$, and ice mantles has $S_{\rm max, ice} = 10^{7}\erg\cm^{-3}$. Lower panel: similar results as the upper panel but for the case of constant ice mantle's thickness of $\Delta a_{\rm m} = 0.05\mum$.}
          \label{fig:tdesp_distance_radius}
         \end{minipage}\hfill
   \begin{minipage}[t]{0.47\textwidth}
     \centering    
    \includegraphics[width=1.05\textwidth]{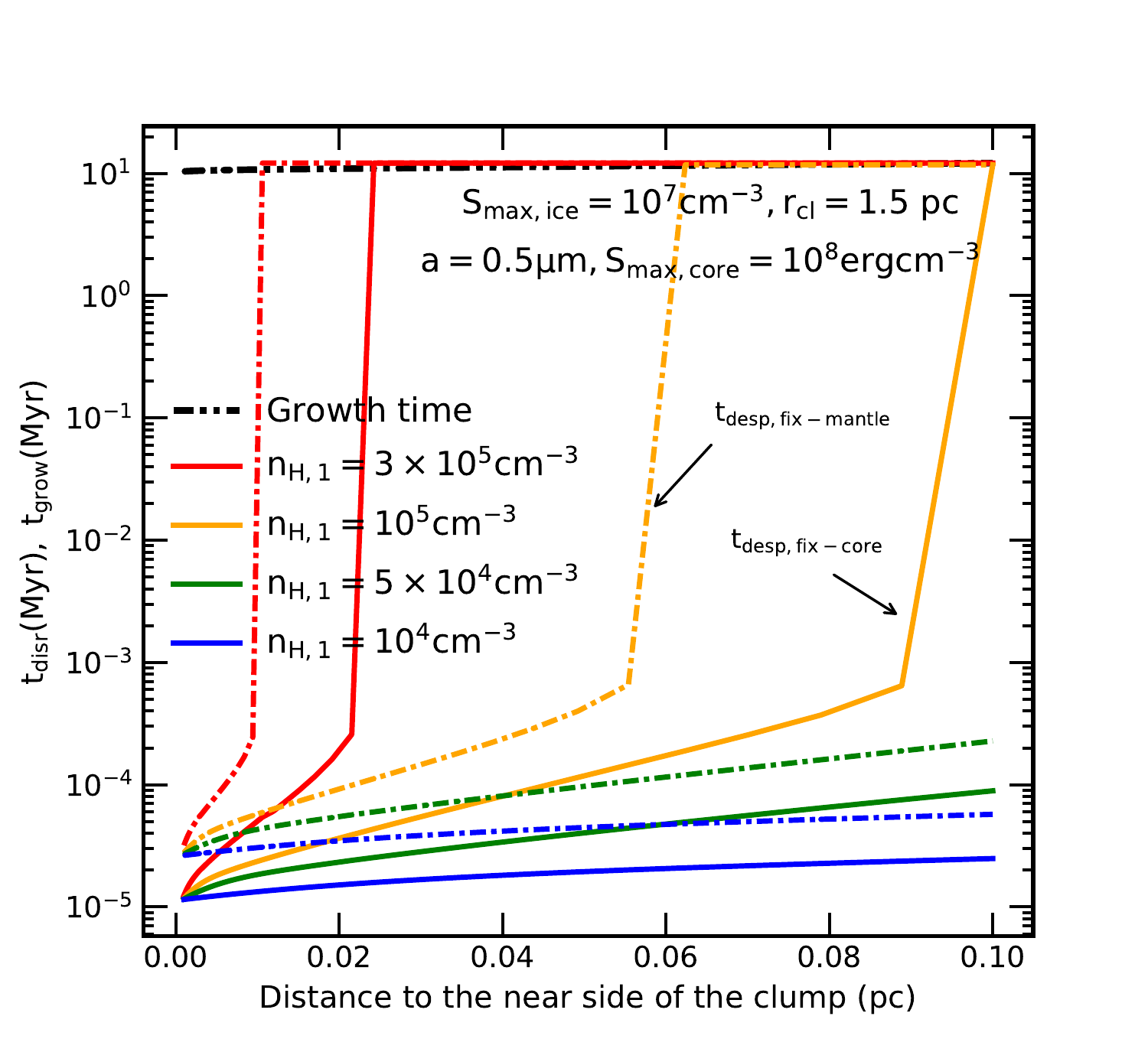}
        \includegraphics[width=1.05\textwidth]{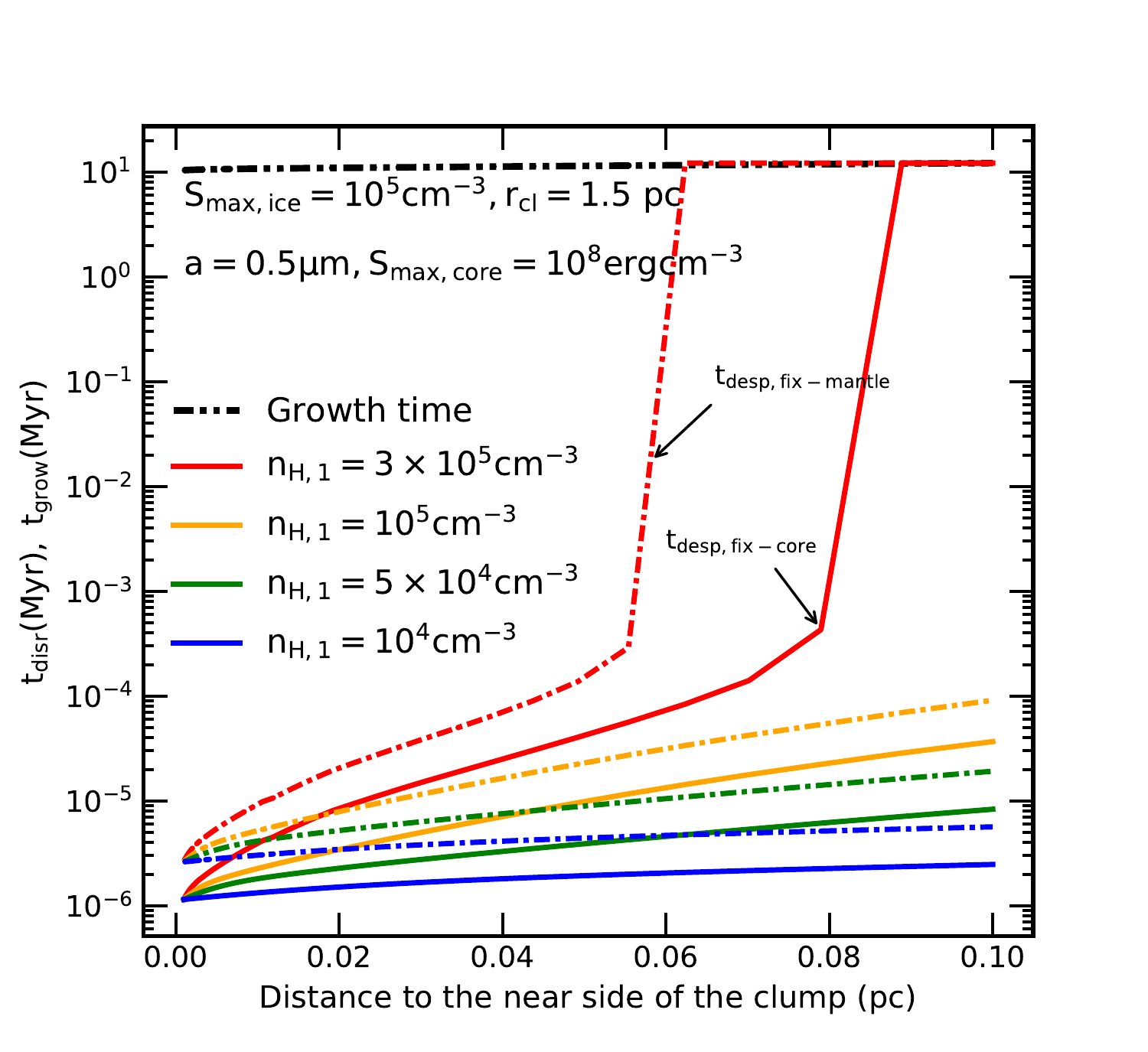}
   \caption{Variation of the rotation desorption timescales for icy grain mantle of size $a = 0.5\mum$ inside the clump at $r = 1.5$ pc with different gas density profile, for ice mantles with $S_{\rm max, ice} = 10^{7}\erg\cm^{-3}$ (upper panel) and $S_{\rm max, ice} = 10^{5}\erg\cm^{-3}$ (lower panel), respectively. The comparison between the removal of ice mantles on the fixed grain core of $a_{\rm c} = 0.08\mum$ (solid lines) and fixed ice mantle's thickness $\Delta a_{\rm m} = 0.05\mum$ (dashed-dot lines) is also presented. } 
   \label{fig:tdesp_distance}
   \end{minipage}
\end{figure*}

 \subsection{Comparison of grain growth to rotational disruption and desorption timescales} \label{sec:time_compare}

\subsubsection{Growth and disruption timescales}\label{sec:tdisr}
The left panel of Figure \ref{fig:tdisr_size} shows the growth (gray lines) and disruption timescales (color lines) of different grain sizes varying from $a = 0.1\mum$ to $a = 10\mum$ at the far side of the clump of radius $R_{\rm cl} = 0.05$ pc at $r = 1.5$ pc, assuming different gas density profiles and $S_{\rm max} = 10^{8}\erg\cm^{-3}$. For $n_{\rm H, 1} = 3\times10^{5}\cm^{-3}$, dust aggregates in the far side of the clump can freely grow to micron-sized grains of $a\geq 10\mum$ due to inefficient RATD (see the left panel of Figure \ref{fig:a_disr_clumpy}), i.e., $t_{\rm disr} > t_{\rm grow}$. By decreasing the gas density, sub-micron grains of $a < 1\mum$ can survive, but larger grains will be quickly disrupted by RATD just after $t_{\rm disr} \sim$ kyr, much faster than grain growth of this size by a factor of $\sim 10^{4}$. The disruption time slightly increases for micron-sized grains, but it is still smaller than the growth time by a factor of $\sim 3000$. By decreasing the gas density, dust grains of size $a > 0.1\mum$ become more difficult to form due to efficient RATD, i.e., smaller $t_{\rm disr}$. The big gap between growth and disruption timescale for grains within disrupted range of $a_{\rm disr} \leq a \leq a_{\rm disr,max}$ implies the significant suppress of grain growth via coagulation for the clumpy torus model.

The right panel of Figure \ref{fig:tdisr_size} shows the similar results as the left panel but for dust grains with a lower strength of $S_{\rm max} = 10^{6}\erg\cm^{-3}$. For the same density $n_{\rm H, 1} = 3\times10^{5}\cm^{-3}$, composite dust grains can freely grow to $a\sim 1\mum$, but further growth will be prevented by the RATD effect, i.e., $t_{\rm disr} < t_{\rm grow}$ by a factor of $\geq 3000$. The dust disruption inside clumps happens stronger (i.e., smaller $a_{\rm disr}$) and quicker (i.e., smaller $t_{\rm disr}$) for more dilute clumps. This feature implies the difficulty for large dust aggregates to survive in the clumpy torus model. 

The upper panel of Figure \ref{fig:tdisr_distance} shows the space-varying growth (faint lines) and disruption (dark lines) timescale for different grain sizes inside the clump at $r = 1.5$ pc, assuming $S_{\rm max} = 10^{8}\erg\cm^{-3}$. One can see that the disruption timescale is different for different grain sizes and different locations inside clumps. But generally, composite dust grains within the disrupted range $a_{\rm disr}-a_{\rm disr,max}$ (the left panel of Figure \ref{fig:a_disr_clumpy}) will be destroyed by RATD before it can grow to larger sizes (i.e., $t_{\rm disr} < t_{\rm grow}$ a factor of $\geq 10^{3}$). Otherwise, they are not affected by RATD and can collide with another to form larger ones (i.e., the presence of grain of sizes $a < 0.5\mum$ and $a\sim 10\mum$) at the far side of clumps (the yellow curve in the left panel of Figure \ref{fig:tdesp_radius}). 

The center panel of Figure \ref{fig:tdisr_distance} compares $t_{\rm grow}$ with $t_{\rm disr}$ for composite grains of size $a = 0.5\mum$ with $S_{\rm max} = 10^{8}\erg\cm^{-3}$ inside the clumps at $r = 1.5$ pc, assuming different gas density profiles, $n_{\rm H, 1}$. By reducing the gas density, grains of size $a = 0.05\mum$ will be destroyed by RATD quicker (i.e., smaller $t_{\rm disr} < 1000$ years) and stronger (i.e., RATD happens to the far face of the clump). Similarly, the timescale for dust disruption by RATD is shorter for dust grains with the lower tensile strength (e.g., porous structures) (see the lower panel with $S_{\rm max} = 10^{6}\erg\cm^{-3}$).
 
\subsubsection{Growth and desorption timescales}\label{sec:tdesp}
Figure \ref{fig:tdesp_radius} shows grain growth (gray line) and desorption (color lines) timescales for different sizes of icy grain mantles located at the far side of clumps at $r = 1.5$ pc, assuming $S_{\rm max,ice}  = 10^{7}\erg\cm^{-3}$ and $10^{5}\erg\cm^{-3}$ for the left and right panel, respectively. The dark and faint lines present for the case of fixed grain core and fix ice mantle's thickness. For the dense clump with $n_{\rm H, 1} = 3\times10^{5}\cm^{-3}$, ice can form and freely cover on all grain sizes without separating due to inefficient rotation desorption. For clumps with $n_{\rm H, 1} = 10^{5}\cm^{-3}$, ice can cover around solid grain core up to the size of $a < 0.8\mum$, but thicker ice mantles (larger size of icy grains) will be quickly detached (i.e., $t_{\rm desp} < t_{\rm grow}$), just after few kyr. By decreasing the gas density and $S_{\rm max,ice}$, more grain core-ice mantle structure will be destroyed (i.e., extended $a_{\rm desp}-a_{\rm desp,max}$) in shorter timescales (i.e., smaller $t_{\rm desp}$). However, thin ice mantles are desorbed by rotational desorption weaker, resulting in the longer desorption timescale and thus, maybe not affected even if reducing $S_{\rm max, ice}$ (i.e., the curve with $n_{\rm H,1} = 3\times10^{5}\cm^{-3}$).

The upper panel of Figure \ref{fig:tdesp_distance_radius} shows the space-varying of growth (faint lines) and desorption (dark lines) timescale for different sizes of icy grain mantles inside the clump of size $R_{\rm cl} = 0.05$ pc at $r = 1.5$ pc, assuming the fixed grain core and $S_{\rm max, ice} = 10^{7}\erg\cm^{-3}$. Similar to the disruption timescale properties, icy grain mantles inside the desorption range $a_{\rm desp} - a_{\rm desp, max}$ (the left panel of Figure \ref{fig:a_disr_clumpy}) will be detached by rotational desorption before they can grow to larger sizes. Otherwise, ice still can cover around the grain core and supports the grain growth process. For example, ice can cover around the grain core of $a\geq 2\mum$ with the thickness of $\Delta a_{\rm m} = 0.05\mum$ at the backside of clumps due to inefficient rotational desorption on removing thin ice mantles.  

Figure \ref{fig:tdesp_distance} shows the space-dependence of growth (gray line) and desorption (color lines) timescale of icy grain mantles of size $a = 0.5\mum$ inside the clump at $r = 1.5$ pc. We consider different gas density profiles and the maximum tensile strength of ice mantles of $S_{\rm max, ice} = 10^{7}\erg\cm^{-3}$ (upper panel) and $S_{\rm max,ice} = 10^{5}\erg\cm^{-3}$ (lower panel). The dark and faint lines present the case of fixed grain core and fixed thickness of ice mantles, respectively. One can see that rotational desorption destroy the grain core-ice mantle structure stronger and quicker in the near face of clumps due to higher radiation flux. In addition, this effect will occur stronger, i.e., lower $t_{\rm desp}$ and larger influenced partions of clumps, with decreasing the gas density, decreasing the maximum tensile strength, and increasing the thickness of ice mantles.
 
In conclusion, in the active region of rotational disruption and desorption, large composite/icy dust grains only take about a few thousand years to rotationally disrupted/desorbed. The timescale is longer for the optically thick regions and for grains with compact structures, but generally be shorter than the growth time of the same size by a factor of $\geq 10^{3}$. This implies that RATD is a strong constraint for dust properties and evolution, which cannot be neglected when studying grain growth. 

\section{Discussion}\label{sec:discuss}
\subsection{Dust and ice evolution under AGN radiation feedback}
Dust in the torus is the key component in the unified model of AGN (\citealt{Antonucci_93}; \citealt{Urry_95}) and responsible for the 'infrared bump' feature in the observed spectral energy distribution (\citealt{Barvainis87}; \citealt{Urry_95}; \citealt{Fritz_06}, \citealt{Nenkova08b}; \citealt{Stalevski_2012}). In addition, dust polarization induced by aligned dust grains is an important tool to study magnetic fields in AGN tori. Understanding dust properties and its interactions with AGN radiation field is the key for constraining the feeding and feedback of AGN to the surrounding environment and probing the origin of the central engine. However, detailed knowledge about dust properties subject to AGN radiation feedback is still poor constrain.

Previously, dust destruction due to AGN radiative feedback was studied by \cite{Barvainis87} via thermal sublimation and by \cite{TazakiRyo} for Coulomb explosions. However, these mechanisms only work in a small region near the AGN and effectively for small grains of $a \leq 0.1\mum$, and the Coulomb explosion effect is ineffective in the optically thick torus (\citealt{TazakiRyo}). Thus, dust grains beyond the sublimation front are hardly affected by AGN radiation feedback because of the high attenuation of X-rays and UV photons. 

Nevertheless, \cite{Hoang_2019b} realized that dust grains subject to strong radiation field can be rotationally disrupted by the RATD effect. Applying this mechanism for the smooth torus model of AGN with a high luminosity of $L_{\rm bol} = 10^{46}\erg~\rm s^{-1}$, \cite{Giang_2021} found the significant removal of large grains of $a\geq1\mum$ up to the boundary region of $\sim 10$ pc, for the wide range of maximum tensile strength of $S_{\rm max} = 10^{7} - 10^{10}\erg\cm^{-3}$ and gas density distribution of $n_{\rm H, 1} = 6\times10^{4} - 4\times10^{-5}\cm^{-3}$. The reduction of the maximum grain size to $a_{\rm max} < 10\mum$ by RATD is more effective in the higher latitude region due to its lower gas density. However, for AGN with a low luminosity of $L_{\rm bol} = 10^{42}\erg~\rm s^{-1}$, RATD is not efficient for the smooth torus model with a high gas density (Section \ref{sec:adisr_smooth}). As a result, dust properties at the parsec scale in the midplane will not be affected by RATD for the gas density at 1 pc is high of $n_{\rm H, 1} > 5\times10^{3}\cm^{-3}$.

In contrast, if dust and gas are concentrated into dense clumps (i.e., clumpy torus model), the attenuation of AGN radiation is reduced, and dust grains can still be spun-up to disruption limit by RATs. Consequently, RATD is strong enough to destroy composite dust grains in both dense clumps near the AGN center and large, dilute clumps at large distance of $r \sim 10$ pc (Section \ref{sec:adisr_clumpy}). The significant influence of RATD in the parsec scale of the torus arises from the fact that RATD can work with the optical-NIR photons that can propagate to a longer distance compared to X-rays and UV photons required for sublimation and Coulomb explosions. Moreover, RATD depends sensitively on the grain structure and can disrupt large grains of composite structures in just a few thousand years. The dust properties modified by RATD would have important effects on the observational properties (emission and polarization), dust dynamics and radiation pressure feedback from AGN (e.g., \citealt{Hoang:2021}). A detailed study of these effects will be presented in a followup paper.

On the other hand, icy grain mantles are expected to exist in the torus beyond the sublimation front (i.e., \enquote{original} snow line. However, rapidly rotating icy grain mantles induced by RATs may be rotationally desorbed by centrifugal force (\citealt{Hoang_Tram}, \citealt{Tung_Hoang}). For the low-luminosity AGN, we find that rotational desorption on icy grain mantles is inefficient for the smooth torus model (Section \ref{sec:adesp_smooth}). However, for the clumpy torus model, ice mantles can be desorbed from the grain core beyond the \enquote{original} snow line, and the amount of icy grain mantles will decrease inside the clumps (Section \ref{sec:adesp_clumpy}). Similar to RATD, rotational desorption works with optical-NIR and only takes a few thousand years to detach icy grain mantles, shorter than the growth time by a factor of $\geq 10^{3}$ (see Section \ref{sec:tdesp}). These effects have important implications for accurately constrain the bundance of water and complex molecules and the efficiency of grain coagulation in the torus.

It is worth noting that, in Section \ref{sec:a_desp}, we studied the effect of rotational desorption and found that this mechanism only can remove thick ice mantles of $\Delta a_{\rm m} > 0.1\mum$. Icy dust aggregates with thin ice mantles still survive beyond the \enquote{original} snow line. However, thin ice mantles can be detached from the grain core when accounting for reducing the binding energy between icy particles due to the spinning effect, which is termed as ro-thermal desorption mechanism (\citealt{Hoang_Tung}). 

\subsection{Dynamical barriers of planet formation around a SMBH} \label{sec:planet}

\cite{Wada2019,Wada2021} studied the formation of blanets beyond the snow line of the midplane of torus region around SMBHs from the hit-and-stick stage to the gravitational instability phase. The authors found that blanets can form if the AGN luminosity is low of $L_{\rm bol} \sim 10^{42}\erg ~\rm s^{-1}$ and the gas density is high of $n_{\rm H, 1} \sim 10^{5}\cm^{-3}$. The authors also studied whether grain shattering due to high-velocity collisions could be a critical barrier for blanet formation in the AGN torus. They found that grain fragmentation is not efficient due to the low relative velocity between icy dust aggregates. Particularly, the grain relative velocity, $\Delta v$, is found to span from $\sim 0.1-0.3 {~ \rm m ~ \rm s}^{-1}$ during the hit-and-stick stage to the maximum velocity of $\sim 57 {~ \rm m ~ \rm s}^{-1}$ at the end of gas compression stage, which is much lower than the critical velocity of shattering of $v_{\rm crit} \sim 80 {~ \rm m ~ \rm s}^{-1}$. The value of $\Delta v$ increases with increasing turbulence, that higher viscosity of gas of $\alpha > 0.04$ results in the fragmentation of dust aggregates at km-size (\citealt{Wada2021}). However, during the hit-and-stick stage considered in this paper, the value of $\Delta v$ is much smaller than the shattering threshold, implying that grain shattering is not an important barrier for blanet formation.

 
In this paper, we study the influence of rotation disruption on composite grains in the circumnuclear region with the same conditions adopted in \cite{Wada2019,Wada2021}. We found that, for the smooth torus model with a high density $n_{\rm H,1} \sim 10^{5}\cm^{-3}$ adopted from \cite{Wada2019}, high attenuation of AGN radiation will reduce the efficiency of RATD (Section \ref{sec:adisr_smooth}) at the parsec scale. Thus, RATD is not the major barrier for grain growth and blanet formation around low-luminosity AGN. However, the RATD mechanism is still a critical barrier of blanet formation around standard AGN.

In contrast, for the clumpy torus model, the presence of dust aggregates of $a\geq 1\mum$ inside clumps is almost prohibited by RATD, up to $r \sim 10$ pc (Section \ref{sec:adisr_clumpy}). Inside big clumps of $R_{\rm cl}\sim 0.1$ pc with a high density $n_{\rm H} \sim 10^{4}\cm^{-3}$, grain growth can occur at the far side of clumps due to inefficient RATD (the left panel of Figure \ref{fig:a_disr_clumpy}, and upper left panel of Figure \ref{fig:adisr_clump_distance}). However, the growth of dust aggregates is accompanied by the decrease of internal density due to the increase of voids between monomers ( \citealt{Dominik_1997}, \citealt{Wada_2008}, \citealt{Suyama_2012}, \citealt{Shen_08}). The increase of porosity $P$ of large dust aggregates thus implies the significant reduction of the maximum tensile strength of $S_{\rm max} \sim 10^{6}(1 - P)a_{0}^{-2}$ with $a_{0}$ the size of monomer (\citealt{Green_1995}, \citealt{Li_1997},  \citealt{Hoang_2019a}), and the enhancement of dust disruption by RATD (see the right panel of Figure \ref{fig:a_disr_clumpy} and the upper right panel of Figure \ref{fig:adisr_clump_distance}). Newly formed dust aggregates are destroyed and the maximum size decreases to the size before growing. Moreover, the clear difference between disruption timescale (few thousand years) and growth timescale (few million years) (Section \ref{sec:tdisr}) sets the first barrier that dust must overcome to continue the blanet formation. 

On the other hand, grain growth is expected to occur more efficiently beyond the snow line where the presence of ice mantles increases the sticking coefficient and prevents the fragmentation during the collision between sub-micron composite dust grains, which facilitates the formation of large, highly porous dust aggregates (\citealt{Chokshi_93}, \citealt{Gundlach_2011}). In Section \ref{sec:adesp_smooth}, we show that rotational desorption is not a major barrier for grain growth at the parsec scales if gas and dust distribute smoothly in space. However, if they concentrate into dense clumps within the diffuse torus, ice mantles will be quickly separated from the grain core (Section \ref{sec:adesp_clumpy}). 
The removal of ice mantles thus may cause the dust grains to bounce off each other rather than sticking together or grain shattering (see \citealt{Chiang:2010}), which would reduce the efficiency of grain growth to large aggregates of $\mum - \rm cm$ size. Therefore, rotational desorption of ice mantles becomes the second barrier for blanet formation due to the short timescale (Section \ref{sec:tdesp}) and the large active region of rotational desorption (Section \ref{sec:adesp_clumpy}).

\subsection{Effects of torus models on rotational disruption and desorption}

Indeed, the real morphology and distribution of dust and gas inside the circumnuclear region is still debated. The smooth torus model generally can reproduce the observed IR spectral energy distribution (SED) of the large sample of AGN and the absorption feature of SiO at $10\mum$ toward type II Seyfert galaxies (\citealt{Fritz_06}, \citealt{Schart}). However,   \cite{Schart} showed that the width of SED produced by the smooth torus model is narrower than observations (\citealt{Granato_1994}, \citealt{Granato_1997}), and this model cannot explain the weak emission/absence of the SiO emission at $10\mum$ in Type 1 Seyfert galaxies (\citealt{Hao_2007}, \citealt{Sieben_2005}). The clumpy torus model proposed by \cite{Nenkova2002} can successfully explain these puzzles, it also can explain the detection of multi temperature components at the same distance in the circumnuclear region  (\cite{Nenkova08}, \cite{Nenkova08b}), \citealt{Dullemond}, \citealt{Honig06}, \citealt{Stalevski_2012}). However, interferometric observations toward the torus region do not clearly show the discrete distribution of clumps inside the diffuse medium as illustrated in the clumpy torus model. They reveal the complex, geometrically thick,  inhomogeneous gas and dust density distribution (\citealt{Tristram_09},  \citealt{Shi_2006}, \citealt{Hicks_2009}, (\citealt{Markow_2014}, \citealt{Izumi_2018}), that matches with the multi gas-phase picture driven by the "radiation-driven fountain" model proposed by \cite{Wada_09} (\citealt{Wada_2016}, \citealt{Schartmann_2014}). This model also can reproduce the dependence of the observed properties of type 1 and 2 Seyfert galaxies on luminosity (\citealt{Burlon_2011}, \citealt{Buchner_2015}, and also the X-ray continuum observed toward AGN (\citealt{Buchner}).

 In addition to the simple adopted model of the torus region, we also only solve the radiative transfer of AGN radiation in one dimension, that basically neglects the asymmetric geometry of AGN radiation emitting from the accretion disk, and the detailed dust scattering and thermal dust emission in three dimensions. Thus, a detailed study with a more realistic model of AGN radiation field and dust/gas distribution in the torus region is required to better constrain the effect of RATD and rotational desorption in the torus region. This is the key for answering the question of blanet formation in low luminosity AGN.

\subsection{Upper limit of RATD for the clumpy torus model}\label{sec:clumpy}
From Figure \ref{fig:adisr_clump_distance}, one can see the decrease of RATD efficiency inside clumps with decreasing the radiation strength and increasing the gas density. With the gas profile adopted from \cite{Wada_16,Wada2019} and the clump size distribution adopted in the lower panel of Figure \ref{fig:adisr_clump_distance}, one can see that rotational disruption and desorption are the main barriers for grain growth and blanet formation.

However, if the clumps are bigger (i.e., higher $\beta_{\rm cl}$ in Equation \ref{eq:R_clump}) and have a higher gas density of $n_{\rm H} > 3\times10^{5}\cm^{-3}$, the RATD efficiency will be similar as in the smooth torus model. Large grains in the near face of clumps can be destroyed by RATD due to higher radiation flux, but they can freely grow and form highly porous dust aggregates, i.e., $\mum - \rm cm$ size, if they are located in the far side of clumps. For example, at $r = 1.5$ pc (corresponds to $U_{0} = 2\times10^{5}$), with $n_{\rm H, 1} = 3\times10^{5}\cm^{-3}$, grain growth via coagulation will be prevented by RATD inside the clumps of size $R_{\rm cl} = 0.1$ pc (the upper panels of Figure \ref{fig:adisr_clump_distance}), but it can occur normally and form blanets if the clumps are larger with $R_{\rm cl} \geq 0.2$ pc or $n_{\rm H,1} > 10^{6}\cm^{-3}$. On the other hand, lower radiation field strength (i.e., lower AGN luminosity) also reduces the RATD effect and support for blanet formation inside the circumnuclear region. 

The presence of more than one clump along the same radial direction is another factor that reduces the RATD efficiency, especially if the line of sight passes through many small but very dense clumps near the center region. Accounting for this effect will narrow the configurations supporting the global effect of RATD inside the torus but broaden the conditions that support grain growth and blanet formation.

\subsection{RATD by IR emission from hot dust grains}
In contrast to other dust destruction mechanisms (i.e., thermal sublimation and Coulomb explosion) that require X-ray and extreme UV radiation (\citealt{Barvainis87}, \citealt{TazakiRyo}), RATD can work with low energy photons at optical-IR \citep{Hoang_2020}, resulting in the significant effect of RATD in the parsec scale inside torus regions. However, in Sections \ref{sec:a_disr} and \ref{sec:a_desp}, we only consider the disruption effect by optical-MIR radiation produced by the accretion disk and neglect the IR emission from the hot dust near the sublimation front.

 NIR-submillimeter can penetrate deeply in dense environments, that the additional contribution from thermal dust emission basically extends the active region of rotational disruption and desorption in the circumnuclear region. In particular, large micron-sized grains are mainly spun up by NIR-submillimeter radiation, that the enhanced IR radiation by thermal dust grains will strengthen the RATD effect on destroying large micron-sized composite grains and the rotational desorption effect on removing large icy grain mantles. Consequently, rotational disruption and desorption can be the main barrier for the blanet formation, even in the smooth torus region or in the large clumps near the outer part of the torus. The detailed modeling accounting for the thermal emission of dust grains should be performed to precisely quantify when rotational disruption and desorption prohibit the blanet formation around low-luminosity AGN. 

\subsection{Efficiency of RATD and blanet formation}
The efficiency of RATD depends on the fraction of grains on high-J attractors, denoted by $f_{\rm high-J}$, which in general depends on the grain properties (shape, size, and magnetic properties), and $0<f_{\rm high-J}\le 1$ (\citealt{Hoang:2020}; \citealt{LH:2021}). An extensive study for numerous grain shapes and compositions by \cite{Herranen_2021} found $f_{\rm high-J}\sim 0.2-0.7$, depending on the grain shapes, sizes, and radiation field. The presence of iron inclusions is found to increase $f_{\rm high-J}$ to unity \citep{Hoang_16}, whereas carbonaceous grains have a lower fraction of $f_{\rm high-J}<1$. Therefore, the RATD mechanism prohibits the formation of blanets out of composite grains with iron inclusions, but it does not completely rule out the formation of blanets made of purely carbonaceous material. However, dust grains in the dense torus are presumably present in the composite dust instead of two separate silicate and carbonaceous grain materials. Therefore, the efficiency of blanet formation via dust coagulation around SMBHs is reduced by the RATD effect, especially in the standard AGN with $L_{\rm bol}>10^{42}L_{\odot}$.

\section{Summary}\label{sec:sum}
Previous studies suggest that planets can form in the torus of low-luminosity AGN around SMBHs. Here we have studied the effect of rotational disruption and desorption of dust and ice by radiative torques due to AGN radiation feedback and discussed their implications for grain growth and planet formation around SMBHs. Our main findings are summarized as follows:

\begin{enumerate}

\item For the simplified smooth model of the torus with the gas density profile of $n_{\rm H} =n_{H,1} (r/1\rm pc)^{-3/2}$ and the gas density at 1 pc $n_{H,1}=10^{4} - 10^{5}\cm^{-3}$, as adopted from \cite{Wada2019}, rotational disruption and desorption are only efficient at distances of $r<1$ pc for low-luminosity AGN of $L_{\rm bol} \sim 10^{42}\erg~\rm s^{-1}$. However, these rotational effects are not efficient for disrupting large grains at large distances of $r>1$ pc due to the high extinction of stellar radiation by intervening dust. Therefore, grain growth and blanet formation can occur around low-luminosity AGN as suggested by \cite{Wada2019}.

\item For the more model that gas and dust concentrate into dense clumps that distributes diffusely inside the circumnuclear region, the attenuation of the AGN radiation is reduced, which enhances the spinning rate of dust grains. As a result, the efficiency of rotational disruption of composite grains and the desorption ice mantles is increased. The sticking coefficient between monomers during the collisions would be reduced due to the removal of ice mantles around the grain core, and the increase in size of dust aggregates is suppressed by RATD to $a_{\rm max} < 10\mum$ even in the large clumps of $R_{\rm cl}\sim 0.2$ pc at $r = 10$ pc, assuming the same gas density distribution adopted from \cite{Wada2019}. The timescale of rotational disruption and desorption is shorter than the growth time a factor of $\geq 10^{3}$, such that rotational disruption and desorption become a major barrier for grain growth and blanet formation in the AGN torus.

\item Grain growth and blanet formation may occur inside large clumps of $R_{\rm cl} >  0.1$ pc with higher gas density than ones adopted for our modeling.  In addition, since the observations support the inhomogeneous, multi-gas phase structure of torus region driven by the "radiation-driven fountain" model, the detailed calculations with more realistic AGN source and torus model, and the additional contribution of IR dust radiation are required to determine whether grain growth and blanet formation can occur in the presence of grain rotational disruption/desorption. 

\end{enumerate}
\acknowledgements
{\bf We thank the referee for comments that helped improve the presentation of our paper.}
T.H. acknowledges the support by the National Research Foundation of Korea (NRF) grants funded by the Korea government (MSIT) through the Mid-career Research Program (2019R1A2C1087045). 

\appendix 
\section{Hit-and-stick timescale of BCCA model}
\label{sec:appen}
In the blanet formation picture suggested by \cite{Wada2019}, icy grain mantles in the midplane of torus first collide and stick together via Ballistic Cluster-Cluster Aggregation (BCCA) model to form larger dust aggregates (\citealt{Okuzumi}). The maximum size at the end of this stage is about cm-sized (\citealt{Wada2019}). Then, cm-sized grains continue to grow further via 2) gas pressure compression, 3) gravitational compression, and form blanets of the Earth-size after 4) gravitational instability phase (\citealt{Gold_1973}). 

During the hit-and-stick stage, the radius of a BCCA cluster with $N$ icy monomers and grain mass $m$ is described as (\citealt{Wada_2008, Wada_09, Wada2019}):
\bea 
a = N^{0.5} a_{0} = \Bigg(\frac{m}{m_{0}}\Bigg)^{0.5} a_{0},
\label{eq:a_d}
\ena
where $a_{0} = 0.1\mum$ is the size of icy monomer at which \cite{Wada2019} starts to track grain growth and $m_{0} = 4/3\pi a_{0}^{3} \rho_{0}$ is the mass of monomer with $\rho_{0} = 1 ~\rm g ~ \rm cm^{-3}$ the mass density of ice. 

The rapid increase in size of newly form dust aggregates is accompanied with the significant decrease of its mass density, which is:
\bea 
\rho_{\rm int} = \rho_{0}\Bigg(\frac{m}{m_{0}}\Bigg)^{(1 - 3/D)},
\label{eq:rho_int}
\ena
where $D\approx1.9$ is the fractal dimension for dust aggregates formed via BCCA model (\citealt{Mukai_1992}, \citealt{Okuzumi_2009}).  
 
From Equation (\ref{eq:a_d}), the mass $m$ of grain size $a$ is given by:
\bea
m = m_{0} \Bigg(\frac{a}{a_{0}}\Bigg)^{2},
\ena 
and its derivative is:
\bea 
dm =  \frac{m_{0}}{a_{0}^{2}} 2 a da = \frac{4/3 \pi a_{0}^{3} \rho_{0}}{ a_{0}^{2}} 2 a da = \frac{4 \pi \rho_{0} a_{0}}{3} 2 a da.
\label{eq:dm}
\ena

The growth rate of dust aggregates during the hit-and-stick stage is (\citealt{Wada2021}):
\bea 
\frac{d m}{dt} = \frac{2 \sqrt{2\pi} \Sigma_{\rm d} a^{2} \Delta v}{H_{\rm d}}.
\label{eq:dm__dt}
\ena 
where $\Sigma_{\rm d} = \Sigma_{\rm g} \eta$ is the dust mass surface density with $\Sigma_{\rm g} = m_{\rm H} N_{\rm H}$ the gas mass surface density and $N_{\rm H}$ the gas column density;  $\eta = 0.01$ the typical dust-to-gas mass ratio in ISM. The second term $\Delta v \approx 1/2 \sqrt{\alpha} ~ c_{\rm s} R_{\rm e}^{1/4} S_{\rm t}$ is the collision velocity between dust aggregates with $\alpha$ the kinematic viscosity of gas due to the turbulence (\citealt{Shakura}). In our study, we adopt the similar value of $\alpha = 0.02$ as in the Blanet paper of \cite{Wada2021}. $c_{\rm s}$ is the sound velocity:
\bea
c_{\rm s} = \sqrt{\frac{k_{\rm B} T_{\rm gas}}{\nu m_{\rm H}}},
\ena
with $k_{\rm B}$ the Boltzmann constant, $T_{\rm gas}$ the gas temperature in the torus given by Equation (\ref{eq:Tgas_torus}), and $\nu = 2$. $R_{\rm e}$ is the Reynolds numbers, that presents the ratio between the turbulent and the kinematic viscosities of gas in the torus: 
\bea 
R_{\rm e} = \frac{\alpha c_{\rm s}^{2}}{\nu \Omega_{\rm K}} &\approx& 1.8\times10^{3} \Bigg(\frac{\alpha}{0.02}\Bigg) \Bigg(\frac{M_{\rm BH}}{10^{6} M_\odot}\Bigg)^{-1/2} \Bigg(\frac{R}{\rm pc}\Bigg)^{3/2} \Bigg(\frac{c_{\rm s}}{1~\rm km s^{-1}}\Bigg),
\label{eq:Re}
\ena
where $\Omega_{\rm K}$ is the orbital frequency of dust in the midplane of torus,
\bea
\Omega_{\rm K} = \sqrt{\frac{G M_{\rm BH}}{R^{3}}},
\ena 
with $M_{\rm BH}$ the mass of the supermassive black hole and $R$ the distance toward AGN center. Assuming the Eddington ratio of 0.01, one gets $M_{\rm BH} \sim 8\times10^{5}M_{\odot}$ for AGN with $L_{\rm bol} = 10^{42}\erg~\rm s^{-1}$.

The last term $S_{\rm t}$ in the equation of $\Delta v$ is the normalized stopping time, i.e, Stoke number, at which dust particle is coupling with ambient gas, i.e., velocity of dust particle reaches the maximum values due to the gas drag, which is:
\bea 
S_{\rm t} = \frac{\pi \rho_{\rm int} a}{2 \Sigma_{\rm g}}.
\ena 

Lastly, the term  $H_{\rm d}$ in Equation (\ref{eq:dm__dt}) is the scale height of the dust disk, which is:
\bea  
H_{\rm d} = \Bigg(1 + \frac{S_{\rm t}}{\alpha} \frac{1+2S_{\rm t}}{1+S_{\rm t}}\Bigg)^{-1/2} H_{\rm g},
\label{eq:Hd}
\ena 
where $H_{\rm g}$ is the scale height of the gas disk, which is:
\bea
H_{\rm g} = \frac{c_{\rm s}}{\Omega_{\rm K}}.
\ena  
 
In the hit-and-stick stage, $S_{\rm t} << 1$ and nearly be a constant with the growth of dust grains (see \citealt{Wada2019,Wada2021}). Thus, we can adopt $H_{\rm d} \approx H_{\rm g}$.

Then, substituting Equation (\ref{eq:dm}) to Equation (\ref{eq:dm__dt}), the increase of grain size $a$ with time $t$ can be described as:

\bea 
\frac{da}{dt} = \frac{3}{4}\sqrt{\frac{2}{\pi}} \frac{\Sigma_{\rm d} \Delta v }{\rho_{0} a_{0} H_{\rm d}} a, 
\ena
or:
\bea 
\frac{1}{a}\frac{da}{dt} = \frac{3}{8}\sqrt{\frac{\pi}{2}} \eta \sqrt{\alpha} R_{\rm e}^{1/4} \sqrt{\frac{G M_{\rm BH}}{R^{3}}}.
\ena 

The growth timescale of size a thus is numerically calculated as:
\bea 
t_{\rm grow}(a) &=& \frac{8}{3 \eta \alpha^{1/2} R_{\rm e}^{1/4}}\sqrt{\frac{2}{\pi}} \rm ln \Bigg(\frac{a}{a_{0}}\Bigg) \Bigg(\frac{G M_{\rm BH}}{R^{3}}\Bigg)^{-1/2} \nonumber \\
&\approx& 3.4489 ~\rm Myr ~\rm ln\Bigg(\frac{a}{a_{0}}\Bigg) \Bigg(\frac{\eta}{0.01}\Bigg)^{-1} \Bigg(\frac{\alpha}{0.02}\Bigg)^{-3/4} \nonumber \\
&\times& \Bigg(\frac{M_{\rm BH}}{10^{6} M_\odot}\Bigg)^{-3/8} \Bigg(\frac{R}{\rm pc}\Bigg)^{9/8}\Bigg(\frac{c_{\rm s}}{1~\rm km s^{-1}}\Bigg)^{-1/4}.
\ena

We note that we derive the timescale of the size growth above from Equation (\ref{eq:dm__dt}), not taking the timescale from \cite{Wada2021}. From Equation (\ref{eq:dm__dt}) and (\ref{eq:a_d}), the mass growth rate is:
\bea 
\frac{dm_{\rm d}}{dt} = \frac{2\sqrt{2 \pi} \Sigma a_{0}^{2} (m_{\rm d}/m_{0}) \Delta v}{H_{\rm d}} =  \frac{2\sqrt{2 \pi} \Sigma a_{0}^{2} m_{\rm d} \Delta v}{H_{\rm d} 4/3 \pi a_{0}^{3} \rho_{\rm 0}} =  \frac{3 \Sigma m_{\rm d} \Delta v}{\sqrt{2 \pi} H_{\rm d} a_{0}\rho_{\rm 0}}
\ena
which is smaller than one in paper of Wada et al. by a factor of 4, and the timescale of grain size growth is longer than their results a factor of 4.
 
\bibliography{reference}
 
\end{document}